\documentclass[acmsmall]{acmart}

\setcopyright{cc}
\setcctype{by}
\acmJournal{PACMHCI}
\acmYear{2026} \acmVolume{10} \acmNumber{2} \acmArticle{CSCW032}
\acmMonth{4} \acmPrice{} \acmDOI{10.1145/3788068}

\received{May 2025}
\received[revised]{September 2025}
\received[accepted]{December 2026}

\AtBeginDocument{%
  }

\graphicspath{{figures}}
\usepackage{listings}
\usepackage{tabularray}
\usepackage{afterpage}
\UseTblrLibrary{booktabs}

\definecolor{myred}{HTML}{eb4c44}
\definecolor{myyellow}{HTML}{E3B100}
\definecolor{myblue}{HTML}{6283ac}
\definecolor{mypurple}{HTML}{bf46aa}

\usepackage{graphics}      %
\usepackage{subcaption}
\usepackage{array}
\usepackage{xspace}
\usepackage{tikz}
\usepackage{placeins}

\newcommand*{\eg}{\textit{e.g.},\xspace}

\newcommand\pquote[2]{{``\textit{#2}'' (DP#1)}}
\newcommand\epquote[2]{{``\textit{#2}'' (EP#1)}}

\makeatletter
\def\thickhline{%
  \noalign{\ifnum0=`}\fi\hrule \@height \thickarrayrulewidth \futurelet
   \reserved@a\@xthickhline}
\def\@xthickhline{\ifx\reserved@a\thickhline
               \vskip\doublerulesep
               \vskip-\thickarrayrulewidth
             \fi
      \ifnum0=`{\fi}}
\makeatother

\makeatletter
\def\thickhlinespace{%
  \addlinespace[1ex]
  \noalign{\ifnum0=`}\fi\hrule \@height \thickarrayrulewidth \futurelet
   \reserved@a\@xthickhline
   \addlinespace[1ex]
   }
\def\@xthickhlinespace{\ifx\reserved@a\thickhline
               \vskip\doublerulesep
               \vskip-\thickarrayrulewidth
             \fi
      \ifnum0=`{\fi}}
\makeatother

\newlength{\thickarrayrulewidth}
\newlength\Origarrayrulewidth

\definecolor{downredcolor}{HTML}{e31a1c}
\definecolor{upgreencolor}{HTML}{33a02c}

\definecolor{DarkGreen}{HTML}{5DAC81}
 \newcommand\arxiv[1]{\textcolor{black}{#1}}
\newcommand\review[1]{\textcolor{black}{#1}} %

\newcommand\cscw[1]{\textcolor{black}{#1}}
\newcommand\cscwminor[1]{\textcolor{black}{#1}}

\newcommand{\twiceinclude}[2][]{%
  \begin{tikzpicture}%
    \node at (0,0) {\includegraphics[#1]{#2.png}};%
    \node at (0,0) {\includegraphics[#1]{#2.pdf}};%
  \end{tikzpicture}%
}%

\newcommand\projectname{RECOVER\xspace}

\begin{document}

\title{\projectname{}: Designing a Large Language Model-based Remote Patient Monitoring System for Postoperative Gastrointestinal Cancer Care}

\author{Ziqi Yang}
\email{ziqiy30@uci.edu}
\orcid{0009-0008-8064-7002}
\affiliation{%
  \institution{University of California, Irvine}
  \city{Irvine}
  \state{CA}
  \country{USA}
}

\author{Yuxuan Lu}
\email{lu.yuxuan@northeastern.edu}
\orcid{0000-0002-8520-0540}
\affiliation{%
  \institution{Northeastern University}
  \city{Boston}
  \state{MA}
  \country{USA}
}

\author{Jennifer Bagdasarian}
\email{jbagdas1@jhu.edu}
\orcid{0000-0002-2426-4428} 
\affiliation{%
  \institution{Johns Hopkins University}
  \city{Baltimore}
  \state{MD}
  \country{USA}
}

\author{Vedant Das Swain}
\email{v.das.swain@nyu.edu}
\orcid{0000-0001-6871-3523}
\affiliation{%
  \institution{New York University}
  \city{New York}
  \state{NY}
  \country{USA}
}

\author{Ritu Agarwal}
\email{ritu.agarwal@jhu.edu}
\orcid{0009-0002-4914-928X}
\affiliation{%
  \institution{Johns Hopkins University}
  \city{Baltimore}
  \state{MD}
  \country{USA}
}

\author{Collin Campbell}
\email{Collin.Campbell@medstar.net}
\orcid{0009-0006-5733-3182}
\affiliation{%
  \institution{MedStar Health Research Institute}
  \city{Columbia}
  \state{MD}
  \country{USA}
}

\author{Waddah Al-Refaie}
\email{WaddahAlRefaie@creighton.edu}
\orcid{0000-0002-9450-8520} 
\affiliation{%
  \institution{Creighton University}
  \city{Omaha}
  \state{NE}
  \country{USA}
}

\author{Jehan El-Bayoumi}
\email{jge288@georgetown.edu}
\orcid{0009-0009-9759-7846}
\affiliation{%
  \institution{Georgetown University}
  \city{Washington}
  \state{DC}
  \country{USA}
}

\author{Guodong Gao}
\email{gordon.gao@jhu.edu}
\orcid{0000-0002-2336-9682}
\affiliation{%
  \institution{Johns Hopkins University}
  \city{Baltimore}
  \state{MD}
  \country{USA}
}

\author{Dakuo Wang}
\email{d.wang@northeastern.edu}
\orcid{0000-0001-9371-9441}
\affiliation{%
  \institution{Northeastern University}
  \city{Boston}
  \state{MA}
  \country{USA}
}

\author{Bingsheng Yao}
\orcid{0009-0004-8329-4610}
\email{b.yao@northeastern.edu}
\authornote{Corresponding author}
\affiliation{%
  \institution{Northeastern University}
  \city{Boston}
  \state{MA}
  \country{USA}
}

\author{Nawar Shara}
\email{Nawar.Shara@Medstar.net}
\authornote{Corresponding author}
\authornotemark[1]
\orcid{0000-0002-9005-252X}
\affiliation{%
  \institution{MedStar Health Research Institute}
  \city{Columbia}
  \state{MD}
  \country{USA}
}
\affiliation{%
  \institution{Georgetown University}
  \city{Washington}
  \state{DC}
  \country{USA}
}
\renewcommand{\shortauthors}{Ziqi Yang et al.}
\renewcommand{\shorttitle}{\projectname{}}

\begin{abstract}

Cancer surgery is a key treatment for gastrointestinal (GI) cancers, a group of cancers that account for more than 35\% of cancer-related deaths worldwide, but postoperative complications are unpredictable and can be life-threatening. In this paper, we investigate how recent advancements in large language models (LLMs) can benefit remote patient monitoring (RPM) systems through clinical integration by designing RECOVER, an LLM-powered RPM system for postoperative GI cancer care. To closely engage stakeholders in the design process, we first conducted seven participatory design sessions with five clinical staff and interviewed five cancer patients to derive six major design strategies for integrating clinical guidelines and information needs into LLM-based RPM systems. We then designed and implemented RECOVER, which features an LLM-powered conversational agent for cancer patients and an interactive dashboard for clinical staff to enable efficient postoperative RPM. Finally, we used RECOVER as a pilot system to assess the implementation of our design strategies with four clinical staff and five patients, providing design implications by identifying crucial design elements, offering insights on responsible AI, and outlining opportunities for future LLM-powered RPM systems.

\end{abstract}

\ccsdesc[500]{Human-centered computing~Interactive systems and tools}
\ccsdesc[500]{Applied computing~Health informatics}

\keywords{Cancer care, Patient-provider communication, Large-language-model}

\begin{teaserfigure}
  \twiceinclude[width=0.98\textwidth]{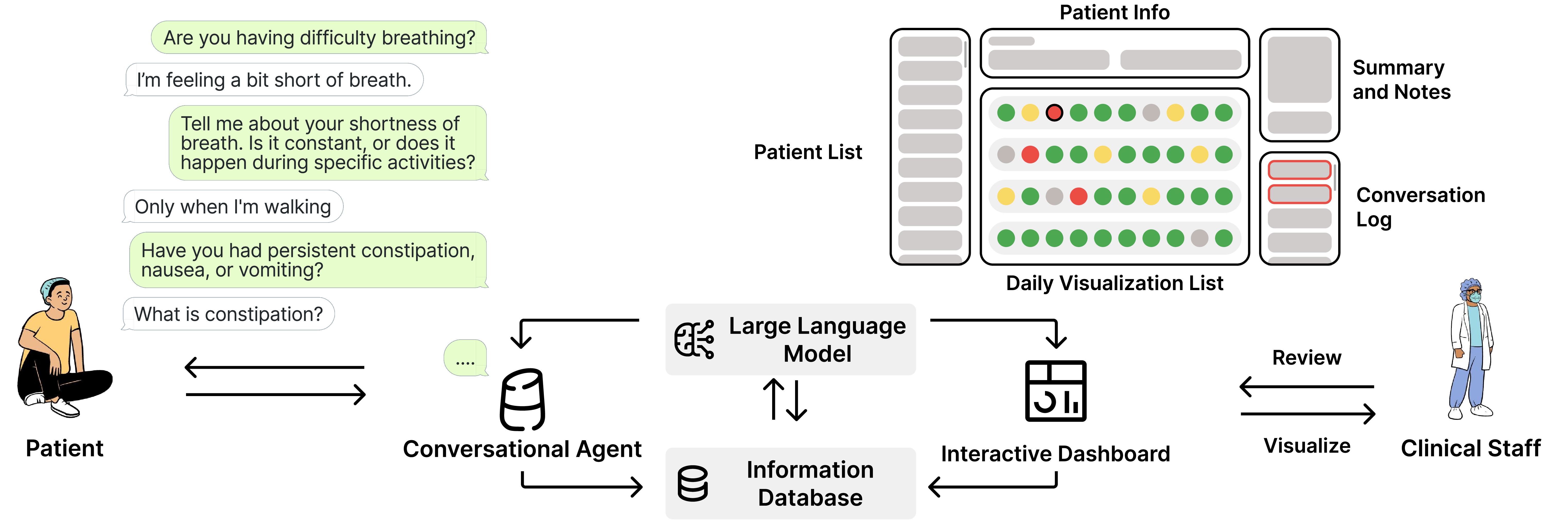}
  \caption{Overview of \projectname, an LLM-powered RPM system integrating a conversational agent and an interactive dashboard}
  \Description{Wide schematic diagram showing an end-to-end clinical conversational AI system. On the far left, a cartoon patient sits and communicates via chat bubbles with a conversational agent. The visible dialogue includes questions such as “Are you having difficulty breathing?”, “Tell me about your shortness of breath. Is it constant or does it happen during specific activities?”, and “Have you had persistent constipation, nausea, or vomiting?”, with the patient replying “I'm feeling a bit short of breath,” “Only when I'm walking,” and asking “What is constipation?”. Arrows indicate a two-way exchange between the patient and the conversational agent. \\
Below the agent, conversation data is stored in an information database, represented by a cylinder icon. The database connects bidirectionally to a large language model shown at the center of the diagram. The language model also connects downward and to the right to an interactive dashboard. \\
On the right side, the dashboard is visualized as multiple panels: a patient list column, a patient info panel, a daily visualization list with rows of colored circles (green, yellow, red) indicating patient status or alerts, a summary and notes panel, and a conversation log panel. Arrows show data flowing from the dashboard to clinical staff on the far right. A clinician figure reviews the information, with arrows labeled “Visualize” and “Review” indicating interaction with the dashboard. The overall image depicts how patient conversations are processed by an AI system, stored, analyzed, and presented to clinicians for monitoring and decision-making.}
  \label{fig:teaser}
\end{teaserfigure}
\maketitle
\section{Introduction}

Gastrointestinal (GI) cancer refers to malignancies affecting the digestive tract  (e.g. stomach, liver, esophagus), accounting for more than 35\% of all cancer-related deaths.
Although surgical removal of affected organs or tissues is a primary and effective treatment,  postoperative GI cancer survivors remain at high risk of life-threatening complications, such as sepsis and surgical leakage~\cite{brenkman2016worldwide, cousins2016surgery}.
Given the unpredictability of postoperative complications due to varying patient recovery trajectories~\cite{vankootenPreoperativeRiskFactors2021}, close \textit{patient monitoring} is crucial for early detection and timely intervention to prevent medical emergencies, hospital readmissions, and fatalities~\cite{sempleUsingMobileApp2015, cardellaComplianceAttitudesBarriers2008}.

Traditional in-person hospital visits for postoperative follow-up stack scheduling delays, transportation burdens, and infection risks to GI cancer survivors; thus, the \textbf{remote patient monitoring (RPM)} paradigm has become particularly valuable by enabling monitoring of patients' health situation at home~\cite{malasingheRemotePatientMonitoring2019}. Yet commonly adopted RPM methods are not clinically compliant for postoperative GI cancer care.
Although phone calls by nurse practitioners or tele-questionnaires offer guidelines for patients to report symptoms~\cite{malasingheRemotePatientMonitoring2019, silva_ostomy_2020}, they either require high clinical dedication or bring confusion and burden to patients such as the ``survey fatigue''~\cite{pannunzio2024patient, leon2022impact, pannunzio2024patient, li2024beyond}.

Despite the flexibility and convenience of mHealth platforms, patient messages through patient portals lead to fragmentation and disruption of clinical staff's busy workflow~\cite{pannunzio2024patient, leon2022impact}; design and usability issues of clinical user interfaces have made it difficult to navigate patient data, extract actionable insights, and make clinical decisions~\cite{zhangRethinkingHumanAICollaboration2024}. These limitations, thus, hinder timely decisions and interventions, and increase clinician burnout~\cite{cardellaComplianceAttitudesBarriers2008, yangWishThereWere2024}. There is an urgent need for innovative RPM solutions that \textit{align with providers' clinical workflows and information needs} to improve healthcare outcomes and enhance efficiency. 

Recent advances in \textit{ubiquitous technology} and \textit{large language models (LLMs)} offer promising opportunities in collecting patient health information in rich multimodal formats and supporting clinical work~\cite{wu2025cardioai}.
Research has used wearable devices like Fitbit to track cancer patients' physical activities or time patterns remotely ~\cite{FitbitWeartimePatterns, dreherFitbitUsagePatients2019, ghods2021remote}; interactive systems like conversational agents (CAs), some powered by LLMs, are designed to check patients' daily health states~\cite{yang2024talk2care,jo_understanding_2023, omarovArtificialIntelligenceEnabled2022, dwyer2023use, li2024beyond, hao2024advancing, yao2025more}; yet
the focus on general health information or physiological metrics does not suit the specific clinical needs in RPM of GI cancer survivors~\cite{yamagataCurrentStatusEnhanced2019, sempleUsingMobileApp2015}.
Meanwhile, advanced LLMs exhibit strong natural language processing capabilities and excel in structured tasks~\cite{brownLanguageModelsAre2020, openaiGPT4TechnicalReport2023, touvronLlamaOpenFoundation2023,yao-etal-2024-samples,wu2025collabllm,wang2025operadatasetobservationpersona}. Research has shown their effectiveness in medical text summarization and question-answering, demonstrating their ability to integrate domain-specific knowledge and process health-related information~\cite{agrawalLargeLanguageModels2022, liuLargeLanguageModels2023, luContextualEmbeddingModel2022, kruse2025large}.

However, current LLM-based CAs primarily handle general or administrative inquiries, which fall short in critical contexts like postoperative GI cancer care~\cite{ni2017mandy,geoghegan2021automated}; general-purpose LLMs lack domain-specific instructions, such as identifying critical symptoms, providing clinically responsible responses for patients, and clinical guidelines and procedures. 
\cscwminor{To design LLM-powered systems for RPM that comply with clinical focus and patient characteristics, we need further understanding of the needs and expectations from clinical staff and patients through \textit{close engagement of these stakeholders}.}
Therefore, we aim to bridge the gaps in RPM technologies by investigating: \textbf{What are the expectations and requirements of clinical staff and patients for an LLM-powered telehealth system in postoperative GI cancer RPM?  How should we design and implement a system for RPM that complies with clinical guidelines and information needs leveraging LLMs?}

To answer these research questions, we first conducted a participatory design study with five clinical staff in GI cancer care and five postoperative GI cancer patients. 
\cscwminor{Through seven participatory design sessions and five interviews, we collected multi-stakeholder expectations for such LLM-powered RPM systems, and iterated design artifacts. 
The process results in a key question and priority table for information collection based on clinical expertise, and considerations and examples for LLM-powered visualization and interaction.} We summarized six key design strategies for leveraging LLMs to build a RPM system that integrate clinical guidelines,  promoting efficiency and accuracy.

Following the design strategies, we designed and developed \projectname{}, a system for \underline{\textbf{re}}mote symptom \underline{\textbf{co}}llection to impro\underline{\textbf{ve}} postope\underline{\textbf{r}}ative care. 
\projectname{}'s two interfaces aim to: (1) integrate clinically critical guidelines and flexible conversational protocols into an LLM-powered CA, enabling it to collect patient information corresponding to key symptoms, and (2) offer symptom risk-based visualization, along with intelligent summaries, highlights, and interactions, so that clinical staff can quickly identify critical issues and respond effectively. 
To assess our design guidelines and user perspectives towards such LLM-powered RPM systems, we used \projectname{} as a pilot system to conduct user studies with four GI cancer care clinical staff and five GI cancer patients. Participants engaged in clinical scenarios using our dashboard and conversational interface. \cscwminor{The results highlighted the system's usability, accuracy and support for clinical efficiency, while also providing qualitative insights into how LLM integration could support RPM and the design of responsible and deployable LLM-powered RPM systems.}
Finally, we present key implications for designing LLM-powered systems for RPM in domain-specific scenarios. We summarize essential design components, their potential long-term impact on patient care, and opportunities to engage computing technologies in collaborative spaces and clinical resources for future development. Additionally, we discuss considerations for implementing responsible AI in LLM-powered RPM systems, focusing on ethical, privacy, and security aspects.
This work presents three major contributions:
\begin{itemize}
    \item \cscwminor{We summarize six design strategies to leverage LLMs to integrate clinical guidelines and information needs into \textbf{RPM systems} from both provider and patient perspectives .}
    \item We present \textbf{RECOVER}, an LLM-powered system that supports \textbf{RPM} of postoperative GI cancer that is clinically efficient and compliant, \cscwminor{with key design artifacts for an interactive dashboard and patient conversations .}
    \item We provide design implications and opportunities for clinically compliant, efficient, and responsible LLM-powered systems for RPM.
\end{itemize}

\section{Related Work}
\label{sec:2-related_work}
Here we first review in Section \ref{sub:related_work-1}  the challenges GI cancer providers face monitoring patients' postoperative recovery and traditional RPM and patient-provider collaboration practices. 
Second, in Section \ref{sub:related_work-2}, we discuss recent advancements and limitations of computing systems for RPM, some of which are powered by LLMs, and the need for clinical integration in postoperative cancer care.
Lastly, we review recent advancements in LLMs that may potentially support the RPM work of staff in our domain-specific clinical context in Section \ref{sub:related_work-3}.

\subsection{Postoperative GI Cancer Care Challenges and Practices}
\label{sub:related_work-1}
Gastrointestinal (GI) cancer causes around 3.4 million deaths per year worldwide, and the 5-year survival rates are below 30\% in many cases after their surgery~\cite{arnoldGlobalBurdenMajor2020, mocanSurgicalManagementGastric2021,wongGlobalIncidenceMortality2021}. 
Monitoring postoperative GI cancer patients is challenging for healthcare providers due to the wide range of potential health issues patients may face. Severe complications such as anastomotic leakage and sepsis~\cite{vankootenPreoperativeRiskFactors2021, yasunagaBodyMassIndex2013} are common, while changes in the GI system can further disrupt patients' nutritional status and quality of life (QOL), potentially impacting their long-term survival~\cite{carmichaelEarlyPostoperativeFeeding2022, garthNutritionalStatusNutrition2010, yasunagaBodyMassIndex2013}.
Research work in surgical oncology has investigated preoperative factors such as lifestyle data to identify high-risk patients~\cite{vankootenPreoperativeRiskFactors2021}. 
However, many of the postoperative complications remain urgent and unpredictable~\cite{mocanSurgicalManagementGastric2021, vankootenPreoperativeRiskFactors2021, wu2024clinical}.

After the postoperative cancer patients are discharged for their recovery at home, close monitoring of patient health conditions helps address these unpredictable conditions through early interventions~\cite{desrame2024595p}. Following clinical guidelines~\cite{steele1993standard}, cancer care institutions have adopted remote patient monitoring (RPM) systems to monitor certain aspects of a patient's health from their own home but have also learned their limitations~\cite{malasingheRemotePatientMonitoring2019}. 
Traditional RPM methods such as phone calls, patient portals, or rigorous questionnaires could save time and travel efforts for both patients and providers~\cite{malasingheRemotePatientMonitoring2019}.
Yet the questionnaires are frequently lengthy and complex, leading to patients' ``survey fatigue'', which significantly diminishes the quality and quantity of information gathered~\cite{li2024beyond, silva_ostomy_2020}. Alternative methods, such as secure messaging or phone calls, are typically one-way (from patients to providers) and often lack clear instructions~\cite{chandwani_stitching_2018} to the patients who have limited health literacy~\cite{cardellaComplianceAttitudesBarriers2008}. Consequently, cancer care providers continue to face a shortage of essential information for effective interventions and decision-making~\cite{yangWishThereWere2024, yao2025exploring, fu2025felt}.
Additionally, \citet{leon2022impact} pointed out issues in telehealth systems for clinical staff including nurses and specialists, including (1) additional workload such as documentation efforts, (2) disruption to the workflow, and (3) false alarms and unclear data.
In other cases, redundant information causes cancer care providers' information overload beyond their clinical duties, and thus necessary interventions could be delayed~\cite{clark_understanding_2021, bhat_infrastructuring_2021, yangWishThereWere2024}.
The limitations in traditional patient monitoring methods in postoperative GI cancer care call for more efficient, user-friendly technology for RPM while ensuring clinical adherence.
Recent research by \citet{yangWishThereWere2024} summarized some key information that cancer providers look for in their RPM practices, but how we might design such novel technology remains under-explored.

\subsection{Computing Systems for Remote Patient Monitoring in Cancer Care}
\label{sub:related_work-2}
Recent advancements in computing systems have significantly advanced the collection of patient health information.
Mobile health (mHealth) applications, sensors, wearable systems, and personal coaches have been designed and used to track patient stress levels and daily activities~\cite{goncalves-bradley_mobile_2020, kingMicroStressEMAPassive2019, jacobsMyPathInvestigatingBreast2018, chung2016boundary, schroeder_supporting_2017}. 
In particular, some research in clinical settings has designed and tested mHealth apps that effectively integrate questionnaires or wearable devices to track patients' activities and postoperative conditions~\cite{sempleUsingMobileApp2015, ghods2021remote}.
However, these solutions, mostly collecting structured and quantitative data, still fall short in guiding patients to report descriptive domain-specific symptoms and lack flexibility and adaptivity regarding the various postoperative conditions of postoperative GI cancer patients (e.g. feelings of pain or nausea)~\cite{yangWishThereWere2024}. The provider interfaces also mostly present unprocessed raw data, further overloading busy healthcare providers~\cite{yangWishThereWere2024, sempleUsingMobileApp2015}. 
Realizing these limitations, a recent co-design study with patients after immutherapy not only emphasizes the significance of patient-reported symptoms in RPM, but also advocates for features that (1) correspond to specific side effects, (2) improve clinician interpretability and usability, and (3) automate RPM tasks~\cite{lai2024co}. 
We are motivated to explore novel systems for RPM that integrate clinical guidelines and needs into advanced interactive technologies.

\subsection{Large Language Models for Clinical Work in Patient Care}

The recent technological boost of Large Language Models (LLMs), such as GPT~\cite{OpenAI_2022}, offers a promising opportunity for system designers to picture a more clinically compliant solution with LLMs' great potential in engaging in and scaffolding natural clinical conversations~\cite{yunxiang2023chatdoctor, xiao2020tell, xiao2023supporting, shen2023convxai, hamalainen_evaluating_2023, zhao2025designing}. 
Researchers have explored LLM-powered CAs for remotely collecting patient health data in areas such as public health interventions, chronic disease management, and pre-consultation screening~\cite{wei_leveraging_2023, montagna2023data, dwyer2023use, ni2017mandy, li2024beyond, hao2024advancing, progga2024large}. For instance, a multi-modal CA by \citet{chira2022affective} collects health data from patients with brain diseases by asking general check-in questions such as ``how are you doing today?''. 
However, most of the LLM CAs focus on patient experiences or data collection, overlooking the information needs of clinical staff, especially in high-risk cases like postoperative GI cancer care. Yet, good clinical adherence could benefit intervention or decision-making in domain-specific conditions. 
Recently, an LLM CA leverages QA datasets and wearable sensor data to help explain health monitoring data to patients (e.g. glucose from wearable sensors) ~\cite{ferrara2024large}, revealing great potential for LLM-powered telehealth systems to integrate clinical guidelines and symptom-related information needs. As LLM CAs continue to proliferate, we investigate their potential in critical clinical settings for efficient and comprehensive RPM.

\label{sub:related_work-3}
Meanwhile, language models have also shown their outstanding capability in processing medical domain knowledge in clinical practices, especially in pre-trained models like MedPaLM\cite{singhalLargeLanguageModels2023}, UmlsBERT\cite{michalopoulosUmlsBERTClinicalDomain2021}, and BioBERT~\cite{leeBioBERTPretrainedBiomedical2019}. For healthcare professionals, researchers have also leveraged LLMs for clinical pre-screening~\cite{hamer_improving_2023, wang_brilliant_2021}, risk prediction ~\cite{o2015risk,gatt2022review, kennedy2014delirium, beede2020human} and information processing~\cite{kocaballi_envisioning_2020, nair_generating_2023,  cascella2023evaluating, nori2023capabilities, agrawal_large_2022, arbatti_what_2023}. 
Specifically related to patient monitoring, LLMs may optimize clinical workflow in basic tasks like scheduling or reviewing information~\cite{tripathi2024efficient}; some work leveraged LLM benchmarks to analyze patient monitoring data ~\cite{healey2024llm}, or multi-modal LLMs to automate patient health monitoring ~\cite{ho2024remoni}.
Although LLM responses may align with clinical guidelines in general evaluations, they may not directly adhere to domain-specific clinical work, such as those in need of evidence-based recommendations~\cite{nwachukwu2024currently}.
In 2023, researchers integrated decision trees from medical literature for LLM-supported clinical decision-making~\cite{li2023meddm}, but little work has explored how to design LLMs' clinical integration for RPM or similar postoperative scenarios.

Focusing on the experience of clinical staff, LLM-powered interfaces such as digital dashboards have been designed for providers to process patient conditions in daily healthcare or emergency decision-making~\cite{yangWishThereWere2024, zhangRethinkingHumanAICollaboration2024}. However, these designs may not suit RPM for postoperative GI cancer, where both domain-specific provider instructions, clinical specifications, and timely qualitative data from patients are essential. Some clinical studies suggest using machine learning (ML) models to identify high-risk postoperative GI cancer patients~\cite{vankootenPreoperativeRiskFactors2021, chenDevelopmentValidationMachine2022a}, but such models have yet to be tested or deployed in real-world settings. Therefore, we aim to explore how clinical needs and guidelines can be integrated with LLMs to develop an LLM-powered telehealth system to ``collaborate, rather than replace\cite{Wang2020HumanHumanToAI},'' clinicians in postoperative GI cancer care and similar high-risk scenarios.

\section{Designing LLM-powered System for RPM with GI Cancer Providers and Patients}
\label{sec:3-participatory-design}
\subsection{Methodology}

We conducted two rounds of studies to gather insights for the design of our system: first, we interviewed GI cancer patients to gather their experiential insights with postoperative care and digital health tools; then, we conducted seven participatory design (PD) sessions with five clinical staff to design the system that follows clinical specifications and can be seamlessly integrated into providers' workflows to address their needs.

\subsubsection{Interview with Cancer Patients}
\review{The research team first disseminated recruitment information via posters and text posts to cancer support groups on social networking sites, which requires prospective participants to complete a screening survey. 
Our key inclusion criteria are: (1) a diagnosis of GI cancer within the last five years and (2) had undergone surgery related to cancer treatment after diagnosis.
With these criteria, we recruited \textit{five} participants: two with stomach cancer, two with pancreatic cancer, and one with colon cancer. All participants had undergone surgery in the United States and recovered in single-family homes or apartments.
Participants were invited to share their experiences recovering from GI cancer surgery, especially their needs and challenges in communicating with their providers, and how they can envision a future AI-supported system for RPM.
}

\subsubsection{Participatory Design with Clinical Staff}
\label{sub:3-pd-methods-provider}
In this paper, we use the term ``clinical staff'' to cover two key roles in patient monitoring based on existing work~\cite{leon2022impact}: (1) doctors, including surgeons, internists, and oncologists, who specialize in specific clinical areas, and (2) patient coordinators, such as nurse practitioners, who have closer and more frequent contact with patients. The term ``healthcare providers'' also refers to clinical staff and is used for generalization and brevity.
We recruited \textit{five} clinical staff using snowball sampling from two cancer care institutions in the central U.S. (see Table \ref{tab:participants} for demographics). 
All participants work closely with GI cancer patients during surgery and recovery, and most of them have over 20 years of experience in GI cancer care.

\begin{table*}[t]
\small
    \centering
    \caption{Participant information in our participatory design sessions}
    \label{tab:participants}
    \begin{booktabs}{
    colspec={cccc}
    }
    \toprule
        DP\# & User Role & Area of Expertise & Years of Exp. \\ \midrule
        DP1 & Patient coordinator& Cancer care & Over 20 years\\
        DP2 & Doctor &  Cancer and postoperative care, patient education & Over 20 years\\
        DP3 & Doctor & Surgical oncology & Over 20 years\\
        DP4 & Doctor &  Cancer care, patient education& Over 30 years\\
        DP5 & Patient coordinator& Cancer care & 1-5 years\\
        \bottomrule
    \end{booktabs} 
\end{table*}

\subsubsection{PD Session Procedure}

\begin{figure}[t]
    \centering
    \twiceinclude[width=0.9\linewidth]{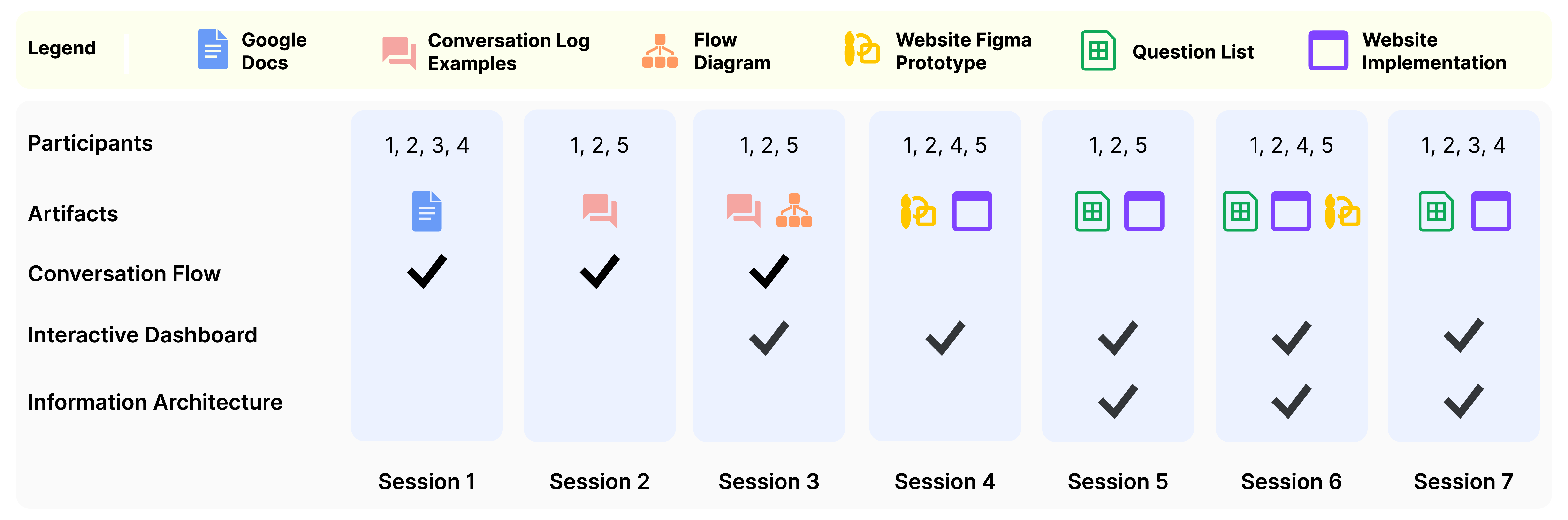}
    \captionsetup{justification=centerlast}
    \caption{PD sessions and their participants, discussion artifacts, and agenda.}
    \Description{Diagram summarizing seven participatory design sessions (Sessions 1–7), showing participating users, produced artifacts, and design outcomes. Each column represents a session and lists participant IDs involved, artifact types (e.g., Google Docs notes, conversation logs, flow diagrams, Figma prototypes, question lists, website implementation), and checkmarks indicating addressed aspects such as conversation flow, interactive dashboard, and information architecture. The figure illustrates the progressive development of artifacts and system features across sessions with overlapping participant involvement.}
    \label{fig:3-PD-agenda}
\end{figure}

Since RPM involves multiple stakeholders: patients for data collection outside clinical settings, and providers for monitoring and decision-making~\cite{temple2023effect}, our system design focuses on three key components: (1) system architecture and information flow, (2) patient interface, and (3) provider interface. 
We conducted seven participatory design sessions on Zoom, each lasting around 30 minutes. 
Three to five participants joined each session to discuss different design decisions of the system and create design artifacts like drafts and flowcharts.
The sessions were held every one to two weeks over two months, as detailed in Figure \ref{fig:3-PD-agenda}.

Throughout these sessions, the research team documented participant inputs through note-taking and design drafts. The recordings of these sessions were transcribed, and two researchers first used inductive coding to code the transcripts and participant comments, then iterated the axial codes until consensus. We identified five major themes on design insights from providers (see Section \ref{3-PD-provider-insights}).

\begin{figure}[t]
    \centering
    \twiceinclude[width= 0.6\linewidth]{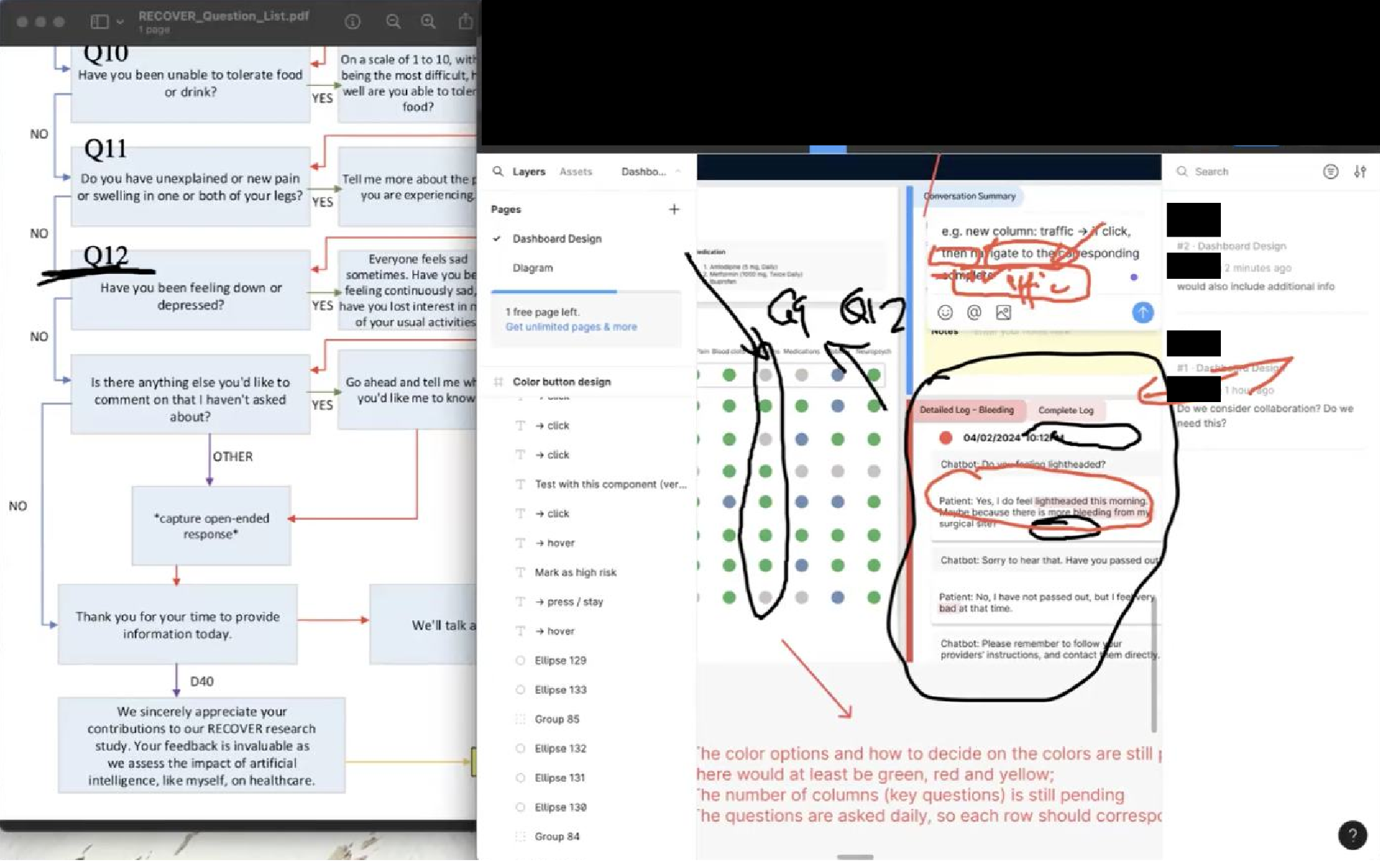}
    \captionsetup{justification=centerlast}
    \caption{A PD session where participants commented on the conversation flow, and dashboard design version; the research team confirmed the participants' feedback by drawing lines and writing notes %
    }
    \Description{Composite screenshot illustrating design artifacts and iterative feedback during system development. The left panel shows a structured question flow from a recovery questionnaire, including branching logic and mental health–related questions (e.g., feeling down or depressed). The center panel displays a Figma design workspace with dashboard layouts and status indicators. The right panel shows a conversation log and dashboard view highlighting a patient–chatbot exchange about postoperative symptoms, with handwritten annotations and callouts indicating design discussions, feature considerations, and revisions across components.}
    \label{fig:participatory-design}
\end{figure}

\paragraph{Designing Information Flow with Diagrams Based on Clinical Workflow}

The researchers first created a Figma diagram to illustrate the data flow between components. Providers reviewed the system architecture, focusing on how the collected patient health data is stored and managed. As providers assessed the diagrams, they shared how patient data is handled in clinical systems such as EHR and recommended changes to improve patient data safety and privacy.

\paragraph{Designing Conversation Flow for Patients}

Before the PD sessions, two researchers drafted a pilot LLM CA based on existing LLM CAs and clinical questionnaires (see Sections \ref{sub:related_work-1} and \ref{sub:related_work-2}). The CA leverages GPT-4~\cite{openai2023gpt4}. The initial prompt asked general check-in questions, such as ``How are you feeling today? Do you have any discomfort?'' followed by a few tailored follow-ups such as pain scales (see Table \ref{tab:8-appendix-initial-conversation} for an example).

\cscwminor{To conduct initial testing of the prompt, the research team integrated information from prior literature as well as suggestions from DPs to form a synthetic model patient~\cite{vankootenPreoperativeRiskFactors2021, bozzettiPostoperativeComplicationsGastrointestinal2007}. The model patient is a hypothetical postoperative GI cancer patient experiencing pain and taking medication.  We discussed with DPs via email to confirm that the patient profile was valid and represented typical postoperative conditions, and then tested the prompt to gather 2-5 conversation logs.}
Each log consisted of multiple conversation turns and was organized into a Google Doc (see Appendix \ref{subsec:9-appendix-conversation-log}). This testing aimed to evaluate the prompt's ability to generate clinically relevant and coherent interactions tailored to typical postoperative scenarios.

These logs formed the basis for discussions on (1) key provider questions for postoperative GI cancer patients, (2) needed follow-up information, and (3) feedback on the conversation logic, chatbot persona, and task. Design iterations focused on conversation logic, the key patient information that the system collects, and the expression of empathy from CA.
During PD sessions, participants reviewed and commented on the conversation log files, referencing reliable resources like the Common Toxicity Criteria\footnote{Common Terminology Criteria for Adverse Events (CTCAE): A standardized tool for grading the severity of adverse effects in clinical trials and routine clinical practice, alongside their professional experience. Available at: \url{https://ctep.cancer.gov/protocoldevelopment/electronic_applications/ctc.htm}.}~\cite{CommonTerminologyCriteria} and their professional experience. The final conversation flow design was achieved after nine major iterations of prompt revisions and provider feedback.

\begin{figure}[htbp]
    \centering
    \twiceinclude[width=\linewidth]{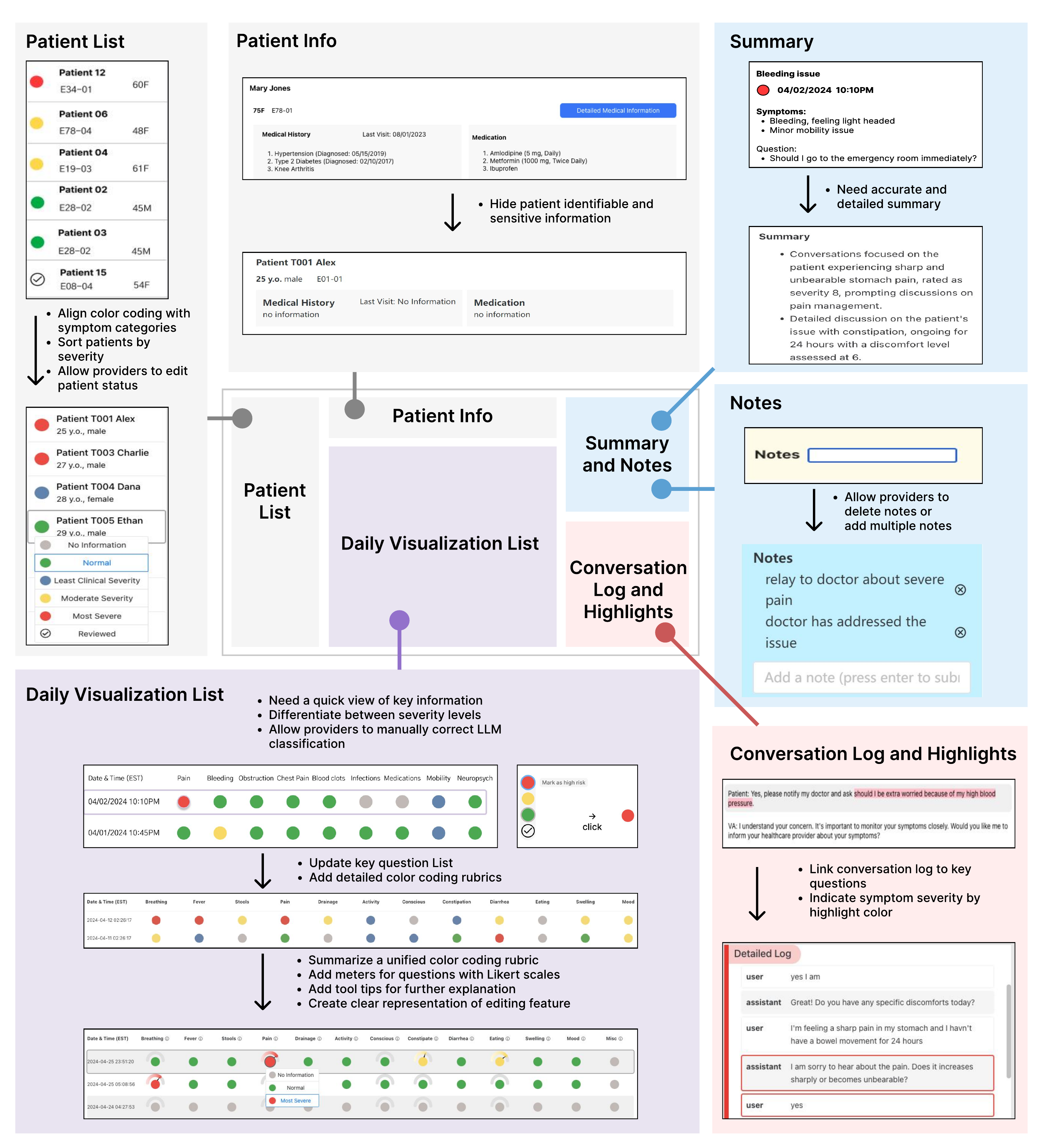}
    \captionsetup{justification=centerlast}
    \caption{Dashboard Iteration Process. In each section, we present the key design versions of the corresponding module, together with the provider feedback that we gathered in the PD sessions. The provider's feedback guided us through the design iterations.}
    \Description{Healthcare dashboard wireframe showing patient triage workflow with three main components: a color-coded patient list sorted by severity (left), individual patient information panels displaying medical history and medications with privacy controls (center), and clinical documentation including AI-generated visit summaries, provider notes, and conversation logs with symptom severity indicators (right). The interface demonstrates a daily visualization system for tracking patient status across multiple health metrics using a standardized color-coding scheme.}
    \label{fig:PD-dashboard-iteration}
\end{figure}

\paragraph{Designing Interactive Dashboard for Providers with Figma Prototype}
We envisioned an LLM-powered dashboard for clinical staff to review the patients' monitoring data through enhanced information visualization and management to improve efficiency. 
The dashboard is expected to be integrated into existing EHR systems. Researchers first designed a pilot user interface in Figma (see Figure \ref{fig:8-appendix-initial-dashboard})~\cite{figma}, resembling a simplified view of a typical EHR system dashboard: a patient list on the left panel, with patient health records and conversation summaries displayed in the central area, and detailed views accessible via clicks.
Throughout the PD sessions, participants suggested interface improvements and new design ideas (Figure \ref{fig:PD-dashboard-iteration}). 
\cscwminor{In later stages, we implemented the interface design on a website with Vue.js framework; we used screenshots of the websites during later PD sessions for discussion and annotation.} Discussions focused on three areas: (1) the overall layout and key sections providers prioritize in RPM, (2) interaction flow and visual components in the central section of the dashboard, and (3) useful interactions and layout for other sections.

\afterpage{\FloatBarrier}

\subsection{Patient Perspectives}
\label{subsec:3-formative-patient-findings}

\label{para:3-formative-patient-needs}
To start with, the patient participants (PP) reported a variety of symptoms and health issues that they experienced, including pain and pain control (4 PPs), GI conditions, fatigue, wound care, breathing, physical activities, and emotional breakdown (PP1, 2, 3, 5). The list of symptoms corresponds to the providers' experience mentioned in \ref{3-PD-finding-key-info}.
Meanwhile, the participants felt a lack of knowledge about their cancer surgery and recovery (PP1, 2, 3), and therefore are in great demand of providers' \textit{ instructions and explanations.} 
For instance, PP3 needed the nurse to explain how and why he needed help managing the digestive system and symptoms, and PP2 double-checked with the doctor about activity and dietary instructions.

\label{para:3-formative-patient-availability}
Among their experience in remote communication, most PPs contact their providers through phone calls, with only one using an online meeting (PP3) and one with a satisfactory telehealth system (PP4).
One major challenge for patients was the \textit{limited availability of clinical staff}, especially as they identified their postoperative recovery phase as needing \textit{quick response and instructions} when they have symptoms. Three participants expected their providers, especially doctors, to be more responsive in urgent situations (PP1, 2, 5), \textit{``when patients are in their down moments, they need the doctors to be there 24/7... some things that come up after surgery are actually things that require urgent attention''} (PP2).

\label{para:3-formative-patient-CA}
Built upon their experiences, the participants expected a \textit{conversational, responsive, and intelligent system for RPM} to support them during the postoperative recovery. Three participants envisioned a conversational agent that answered various questions and offered solutions to their symptoms (PP1, 2, 3): \textit{``when you have something like a doctor... we can have a conversation.''} (PP3). In addition, two participants expected the system to be ``readily available'' and offer ``real-time replies'' (PP4, 5), while PP3 suggested integrating more personal medical history to offer better responses.

\subsection{Design Insights from Cancer Providers}
\label{3-PD-provider-insights}
\paragraph{Probing Key Information based on Clinical Standards and Provider Experience}
\label{3-PD-finding-key-info}
Participants expressed strong interest in using LLM-powered CAs for collecting patient health data in RPM. Building on prior research, they identified 13 essential questions for the LLM to ask, focusing on key symptoms for postoperative GI cancer patients, such as breathing, pain, drainage, blood, and stool (Table \ref{tab:3-key-questions}). For example, the presence of blood in the stool could indicate surgical complications (P4). Providers refined the wording of these questions to ensure comprehensive coverage during patient interactions and assigned severity levels (``most severe'', ``moderate'', ``least severe''), determined if a 10-point Likert scale should be used and suggested color for each question.

Participants also mentioned that patients tend to underestimate their symptoms which may have a huge impact on their health through the recovery process: \pquote{3}{... patients can under or overestimate their symptoms. So ... there are patients when we call stoic... they'll say I have shortness of breath, but it's manageable, when it's actually ...  no, it's a significant one that requires a significant attention.} Based on this observation, all participants agreed that there should only be one severity level for each question; for instance, the system ought to display ``most severe'' even if the patient reports shortness of breath at a Likert scale of 1 out of 10.

\begin{table*}[t]
    \centering
     \caption{
     The final version of key questions summarized by DPs, including their preferences on how to visualize patient-reported information. For certain questions, experts suggested using a Likert scale or color coding to represent patients' responses to reflect the severity of the symptom. For instance, if a patient reports difficulty breathing, it could be flagged as a more severe health issue.
     The last column denotes our final design choice explained in Section \ref{sec:4-system-design}.}

\begin{booktabs}{
colspec={Xccc},
cells={m},
width=\linewidth,
hspan=minimal,
}
\toprule
Question                                                                                                                                              & Likert Scale & Severity          & Color \\
\midrule
Are you having difficulty breathing?                                                                                                                  & Yes          & Most Severe       &   \textcolor{myred}{red}    \\
Are you having a fever of over 100
  degrees, or chills?                                                                                              & No           & Most Severe       &   \textcolor{myred}{red} \\
Have you had black, tar-like stools?                                                                                                                  & No           & Most Severe       &    \textcolor{myred}{red}   \\
Do you have pain that sharply increases,
  or becomes unbearable?                                                                                     & Yes          & Most Severe       &     \textcolor{myred}{red}  \\
Are you having any wound drainage
  problems, such as redness around your wound, bleeding from the wound, pus, or
  an opening at the incision site?~ & No           & Moderate Severity &    \textcolor{myyellow}{yellow}   \\
Do you have a decrease in your ability to
  perform your daily activities, such as not being able to walk to the
  bathroom?~                        & No           & Least Severe      &  \textcolor{myblue}{blue}      \\
Have you had a decrease in your level of
  consciousness?                                                                                             & Yes          & Most Severe       &    \textcolor{myred}{red}    \\
Have you had persistent constipation,
  nausea, or vomiting?                                                                                          & Yes          & Moderate Severity &      \textcolor{myyellow}{yellow} \\
Have you had persistent diarrhea?                                                                                                                     & No           & Moderate Severity &   \textcolor{myyellow}{yellow}    \\
Have you been unable to tolerate food or
  drink?                                                                                                     & Yes          & Moderate Severity &    \textcolor{myyellow}{yellow}   \\
Do you have unexplained or new pain or
  swelling in one of both of your legs?                                                                        & No           & Most Severe       &   \textcolor{myred}{red}      \\
Have you been feeling down or depressed?                                                                                                             & No           & Least Severe      &     \textcolor{myblue}{blue}  \\
Is there anything else you'd like to
  comment on that I haven't asked about?                                                                         & No           & N/A               &    \textcolor{mypurple}{purple}    \\
  \bottomrule
\end{booktabs}
   
    \label{tab:3-key-questions}
\end{table*}

In addition to structured inquiries, the providers expressed a keen interest in utilizing the LLM's capability to adaptively probe deeper into patient responses. DP4, drawing from extensive experience with GI cancer patients, noted that open-ended questions often yield the most informative responses, allowing patients to describe their symptoms in detail. 

DP2 emphasized the utility of narrative responses to validate and enrich the data collected, highlighting the need to both confirm and elaborate on patients' descriptions of symptoms. For example, the patient's initial description of a symptom would prompt follow-up inquiries to validate the information and gather details. DP2 suggested a dual approach: \pquote{2}{Ask them to talk openly about how they feel, then compare their narrative with specific answers like yes or no}. Consequently, the team established two types of follow-up inquiries: (1) for each of the 13 key questions, specific prompts were devised, e.g., ``Could you tell me more about when the pain started?'' and (2) for other symptoms mentioned by patients, the LLM was programmed to ask for details such as frequency, severity, and impact where relevant.

\paragraph{Balancing Timeliness, Effectiveness, and User Experience}
\label{3-PD-finding-time}

In the meantime, providers also discussed the most appropriate frequency of the check-in conversation to collect information from patients. As GI cancer patients may experience great change postoperatively, frequent monitoring (\eg daily check-in) is essential to the timeliness of collected patient information; in particular, participants focus on patients' conditions within the first 40 days of their hospital discharge. Thus, our participants agreed that having the check-in questions performed on a daily basis with a flexible flow would result in time and accurate responses.

However, as the participants discuss the question list for the LLM CA and the check-in frequency, they are also concerned about the length of the whole conversation, as clinical questionnaires alone may take a long time to finish via a CA. Participants are worried that the patient would feel bored, exhausted, or distracted if the daily interaction is too long. \pquote{1}{... we just felt like it was so unwieldy... we talked about dividing it into thirds... [But] we settled in on a much more narrowly focused set of questions}. Thus, as mentioned in Section \ref{3-PD-finding-key-info}, participants agreed on a narrower list with adaptive follow-up questions for a better patient experience. Furthermore, DP1, DP2, and DP4 agreed on skip logic, where questions may be skipped if the patient has already given the answer, or the conversation focuses on another particular symptom.

The discussion on improving user experience also involves the LLM's expression of empathy. On one hand, all participants agreed that it would be good for the LLM to acknowledge the patients' discomfort so that the conversation is more natural.
On the other hand, some participants are also worried that an LLM expressing empathy may make the patient uncomfortable and confused, especially in the healthcare setting.\pquote{6}{... where the bot says 'I'm sorry you [are] feeling this way' it makes me kind of cringe... we have to be careful about this anthropomorphization. If we are ascribing, giving,... human-like qualities, then they should be credible human-like qualities. ... this is an artificial intelligence which is different from humans.} As we will discuss in Section \ref{3-PD-finding-responsible}, our participants proposed that the LLM should remind patients of their roles as a CA instead of human professionals.

\paragraph{Designing Responsible AI in Conversation}
\label{3-PD-finding-responsible}
Another extensively explored scenario in conversational agents for healthcare is health information seeking. In our system, although answering health-related questions is not the LLM's primary functionality, our provider participants also discussed how the chatbot should respond to patients' questions. For instance, regarding a symptom description, \pquote{1}{The bot needs to be able to differentiate between the multiple symptoms addressed this question. For example, redness of skin would not affect one's clothing; whereas bleeding or wound drainage could soak the clothes}.
Similarly, the providers concurred that it would be beneficial for the LLM to provide objective explanations for medical terms (e.g. when a patient asks ``what is constipation?'') to aid patients with lower health literacy in understanding the questions.

However, providers are overall highly cautious about the \textit{safety risks} within LLM responses. In conversation logs, providers commented that some LLM responses expressing empathy may indicate clinical assessments, which could be misleading to patients.
For example, regarding the LLM response, ``That's great! You are maintaining well.'', participants said \pquote{5}{This is an assessment, which the bot is not supposed to do. Also, good food intake and great appetite are not enough to say that the patient is `maintaining well.'}. For the LLM response, ``I see, an increase in bowel movement due to an increase in food intake isn't necessarily a cause for concern.'', a participant commented \pquote{1}{This is concerning; scary, even!} Thus, the LLM-powered system should be designed to be responsible for its responses by avoiding giving clinical assessments or related indications.

In addition, the provider participants emphasized that the system should mitigate patients' misunderstanding of the system's responsibility through clarifications, especially in emergent cases. DP3 specified that they wanted the system to explicitly say that 
\pquote{3}{if [the participant is] having an emergency, call 911, we're not your 911} 
Thus, the LLM-powered system should always clarify its responsibilities to avoid misuse.
In summary, we derive that the design should emphasize accountability and establishing capability boundaries in LLM responses.

\paragraph{Visualizations and Interactions to Promote Efficiency}
\label{3-PD-finding-efficiency}
Corresponding with the providers' heavy workload presented in prior work, our participants expressed a strong preference for an interface that presents the collected patient health information that promotes their efficiency. 
Based on our Figma prototype iterations which is shown in Fig. \ref{fig:PD-dashboard-iteration}, participants agreed on the need for a visualization of all the patients' daily conversation results presented in the central section, so that they can quickly \textit{``have a snapshot''} of the data (DP2). For all questions, they look for visualizations in color codings, so that the levels of severity could be immediately reflected. Specifically, doctors proposed \textit{color-coding }rubrics to denote the priority of the information that should be reviewed.
\pquote{4}{... so that [the] pain should always be reviewed right? Something that's a priority. }
Additionally, they expected visualizations like \textit{meters} to reflect patients' responses in Likert scales as well as a uniformed list. \pquote{2}{...I think we wanna maintain the first line as the dots... A small meter that shows ... completely a different line for the like a Likert scale for those...} We summarize their consensus on information presentation in Table \ref{tab:3-key-questions}.
Lastly, the participants commented on better formatting or arranging the LLM-generated analysis to offload the dashboard users. For example, the summary in the upper right section can be listed in bullet points; the patients in the patient list could be sorted in descending order of risks (from red to green and checked) (DP2). In summary, we design a visualization section that effectively reflects key information to navigate the users' attention.

Correspondingly, derived design implications for interactive features from providers' needs to manage patients. For example, in addition to the detailed conversation logs, DP2 is interested in linking the color visualization dots to corresponding conversation logs; DP1 also agreed to our initial design to allow providers to manually adjust the severity levels based on their action status with an interactive list selection.
Following participant inputs, we aim to design \projectname{} dashboard with responsive interactions to help users quickly navigate to details and take action.

\paragraph{Privacy concerns, sensitive information}
\label{3-PD-finding-privacy}
The participants are particularly focused on addressing data privacy concerns, given the highly sensitive nature of the collected patient health information and medical records. DP2 and DP5 advocate for the confidentiality of identifiable information, such as patient names and dates of birth, to be maintained, with access restricted solely to the managing clinical team. 
Furthermore, due to security concerns regarding third-party services such as Alexa Skill and LLM services like GPT from OpenAI, DP5 stressed that \projectname{} should prevent these platforms from misusing identifiable or personal health information. Therefore, we aim to design our LLM-powered system for RPM to safeguard patient data.

\subsection{Design Strategies}
\label{3-PD-design-stategies}
We note our patient and provider participants have different focuses for design expectations towards LLM-powered systems for RPM. While the patients focus more on the explainability and responsiveness in the system (DS2, 3), the providers emphasize how such a system could support efficient clinical work while being responsible (DS1, 4, 5, 6). Thus, we design different components of the system focusing on strategies supporting the corresponding stakeholder group. 

Based on our PD findings above, we summarize the following 6 key design strategies (DSs), which will guide us in designing and developing the LLM-powered system in Section \ref{sec:4-system-design}:
\begin{itemize}
    \item DS1: Collect clinically critical patient conditions leveraging LLM's strength in analyzing and interpreting natural language (Section \ref{para:3-formative-patient-needs} and \ref{3-PD-finding-key-info}). DS1 considers patients' needs to receive guidance in reporting their symptoms and providers' expectations to have critical patient symptom information. Given that LLMs are good at analyzing both clinical and natural language use, designers could focus on using LLMs to collect key information by scaffolding interaction and mapping clinical instruction to oral language (Section \ref{subsub:4-system-patient-interface-prompt}).
    \item DS2: Using LLM to offer a comprehensive, interactive, and explainable patient-CA experience (Section \ref{3-PD-finding-key-info}, and \ref{para:3-formative-patient-needs}). Given users' disparities in health literacy and clinical expectations for comprehensive language for patients, an LLM-powered conversational agent will help with patient-side communication through explanations and adaptive language use for each patient.
    \item DS3: Ensuring a responsive patient interface with LLMs that collaborate with clinical staff (Section \ref{para:3-formative-patient-availability} and \ref{3-PD-finding-time}) Given the gap between patients' needs for quick response and providers' in our formative study, we use the LLM for initial response to patients while clinical staff offer further clinical instructions later.
    \item DS4: Designing LLM-supported effective visualization that helps clinical staff focus on crucial updates (Section \ref{3-PD-finding-efficiency}) Considering the clinical advocacy for visualization to draw attention to crucial abnormalities in PD sessions, LLM could perform data analysis on patient health information from the backend to generate prioritizing visualization.
    \item DS5: Filtering key information using LLM for efficient clinical review (Section \ref{3-PD-finding-efficiency}) Similar to DS4, LLM's strength in analyzing language could also support clinical review by highlighting key information such as key words and matching visualization to details.
    \item DS6: Mitigating privacy, safety and ethical risks following responsible LLM requirements. The requirements include the accountability and capability boundary in LLM responses, as well as anonymity and transparency in data use related to LLMs (Section \ref{para:3-formative-patient-CA} and \ref{3-PD-finding-privacy} ) It is also essential that the use of LLMs are under human supervision as regulated in access to sensitive information (Section \ref{subsub:4-implemenation}).
\end{itemize}

\section{The \projectname System Design}
\label{sec:4-system-design}

\begin{figure*}[t]
    \centering
    \twiceinclude[width=.95\linewidth]{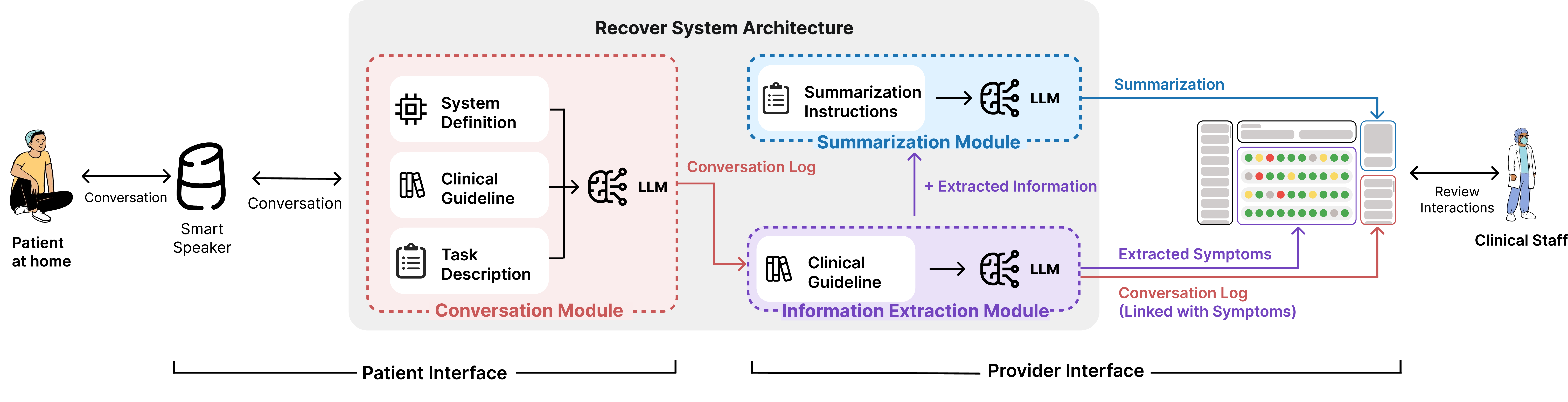}
    \captionsetup{justification=centerlast}
    \caption{System architecture of \projectname. The red, purple, and blue arrows represent data generated by the Conversation Module, Information Extraction Module, and Summarization Module, respectively.
    }
    \Description{System architecture diagram of the Recover platform showing data flow from patient to clinical staff. The patient interface (left) captures conversations via smart speaker, which feed into a Conversation Module containing system definition, clinical guidelines, and task descriptions processed by an LLM. The provider interface (right) displays two processing modules: a Summarization Module that generates visit summaries from conversation logs, and an Information Extraction Module that identifies symptoms linked to conversation logs. Clinical staff review the extracted information and summaries through a color-coded visualization dashboard.}
    \label{fig:system-arch}
\end{figure*}

Following the design strategies that we summarized in Section \ref{3-PD-design-stategies}, we designed and implemented the \projectname system (see Fig. \ref{fig:system-arch}). \projectname has two main user interfaces: (1) a conversational user interface powered by a large language model (LLM) (Section \ref{sub:4-system-patient-interface}), and (2) a web-based dashboard for clinical staff (Section \ref{sec:4-provider}). In this section, we first elaborate on the key design decisions by their interfaces and then the technical details.

\subsection{Patient Interface and Interaction}
\label{subsub:4-system-patient-interface-prompt}
\label{sub:4-system-patient-interface}

Given our participants' expectations for conversational agents (CAs), we developed and implemented a CA as the user interface for postoperative cancer patients. In particular, we used an Alexa Echo Dot, a voice assistant widely adopted in prior studies for patient-facing healthcare~\cite{chanMangoMangoHow2023,mahmoodLLMPoweredConversationalVoice2023}, in our study.
This interface allows users to interact naturally with the system, facilitating an easier and more intuitive way to regularly report their health conditions.

In this section, we introduce the three parts of our system prompt: System Definition, Clinical Guidelines, and Task Description. We show \arxiv{how} they are structured to guide the flow of conversation, ensuring relevance and appropriateness in the context of patient care\footnote{The full prompt example can be found in Appendix \ref{sub:8-appendix-prompt-conversation}}.

\paragraph{System Definition.} 
The System Definition part of the prompt aims to provide general information and guidelines to the CA, supporting DS2 and DS3.
\begin{itemize}

\item Persona Information:
To ensure the CA will be friendly and empathetic and to support DS3, we first incorporate instructions in the prompt defining the CA as an assistive agent for clinical RPM.

\item User Description:
To ensure the CA's understanding of the context and the user, the LLM is then prompted with the user description of a postoperative cancer patient.

\item Allowlist and Denylist:
To ensure the CA exhibits natural language and empathy (DS3) without providing unsolicited medical advice (e.g., ``this could be a problem''), we implemented both an allowlist and denylist in the prompt setup. 
Each entry in the allowlist and denylist contains a guideline (e.g., ``be friendly'') and an example. \cscw{In particular, we followed DS6 to include rules to avoid hallucination and safety issues in the denylist. The "Things that you must not do" section instructs the LLM not to provide assessments or instructions towards the patients' health conditions, and to clarify the systems' limited responsibilities.}

\item Clarification and Accessibility:
In traditional RPM methods like questionnaires, patients may not understand questions with limited health-related knowledge and thus seek further clarification from clinicians. 
To address this issue, our system prompts the CA to phrase questions using simple language and respond to any clarifications posed by the patient in a clear and natural manner. 
This strategy not only facilitates the effective collection of symptom information from patients with limited medical knowledge but also ensures that patients can actively seek and receive clear explanations regarding the queries raised by the agent.

\end{itemize}
\paragraph{Clinical Guidelines.}
To support our DS1 and DS2, the CA is prompted to pose questions based on a list of key questions and to inquire further if a patient confirms experiencing a symptom. 
The list of questions (shown in table \ref{tab:3-key-questions}) is gathered from our PD sessions.

\paragraph{Task Description.}
To support DS1, DS2, and DS3, the third part of the prompt specifies how the LLM response should be derived in each round.
\begin{itemize}

\item Conversation Flow:
Following our DS1, we designed our overall conversation flow and used natural language to describe the flow in the prompt as a big picture. First, the CA should greet the patient with expressions like ``How are you doing today?'' Second, the LLM should process the predefined key question list in the Clinical Guideline to identify the first unanswered question, ensuring the conversation covers all essential topics.
Then, in the last two steps, the CA should wrap up the conversation and respond to the user's further conversations.

\item Chain-of-Thoughts Prompting:
Guided by our DS1, we initially designed the LLM-powered CA to mark a symptom as ``not reported'' only when the user explicitly states they do not have that specific symptom. 
However, during the prompt iteration process, we observed that the system frequently omits questions from the predefined list. 
The LLM often assumes that all questions have been answered when the user responds with ``I'm feeling great'' to general inquiries, or it fails to ask follow-up questions for the second symptom when the user reports multiple symptoms simultaneously.

To address this issue, we developed a chain-of-thoughts approach \cite{wei2022chain}. This approach, specified after the Conversation Flow, requires the LLM to (1) go through conversation history with all key symptom questions (2) mark their current status (one of ``not discussed,'' ``in discussion,'' or ``discussed'') and (3) decide which question should be asked, in every conversation round. 
The LLM should only mark a question as ``discussed'' after the user has explicitly answered the key question and all related follow-up questions.

\item \cscw{Incorporation of History Conversation Context: To provide a more personalized experience for patient users, we further incorporated long-term conversation history beyond the current day into the conversation history.
Thus, the system is designed to consider previous information to generate more contextual questions or responses that help to monitor patients' symptom changes.
Specifically, we provide the LLM with the previous symptom status extracted by the Information Extraction Module and the conversation summaries generated by the Summarization Module (Sec.~\ref{sec:information}) as context for generating new responses.
For example, if the patient mentioned abdominal pain a few days ago, the system could bring up the previous history and check on the pain earlier in the conversation.
}

\end{itemize}
\paragraph{Using the Prompt.}
\label{subsub:3-using-prompt}
We implemented our LLM-powered CA utilizing the GPT-4o \footnote{gpt-4o-2024-11-20} model \cite{openaiGPT4TechnicalReport2023}. To form one \textit{system prompt}, the three parts are combined in the listed order with a short title (see Appendix \ref{sub:8-appendix-prompt-conversation}). Each day, the patient activates our Alexa Skill with ``Alexa, Open Recover Bot'' to start the conversation. During each conversation round, the user's speech is captured by the Alexa Dot Speaker and forwarded to the backend API server via the Alexa Skill. 
The server retrieves the day's conversation history and the system prompt specified in the config file and uses this information to get the LLM response from the OpenAI service. After getting the response, Alex Echo Dot replies to the patient and listens to the next message from the patient. \cscwminor{If patients drops out of the ongoing conversation, they can continue the conversation later in the day by activating the Alexa Skill again; the previous logs in the same day will be retrieved and added to the conversation history with the LLM.}

\subsection{Provider Interface and Interaction}
\label{sec:4-provider}
\begin{figure}[htbp]
    \centering
    \includegraphics[width=.95\linewidth]{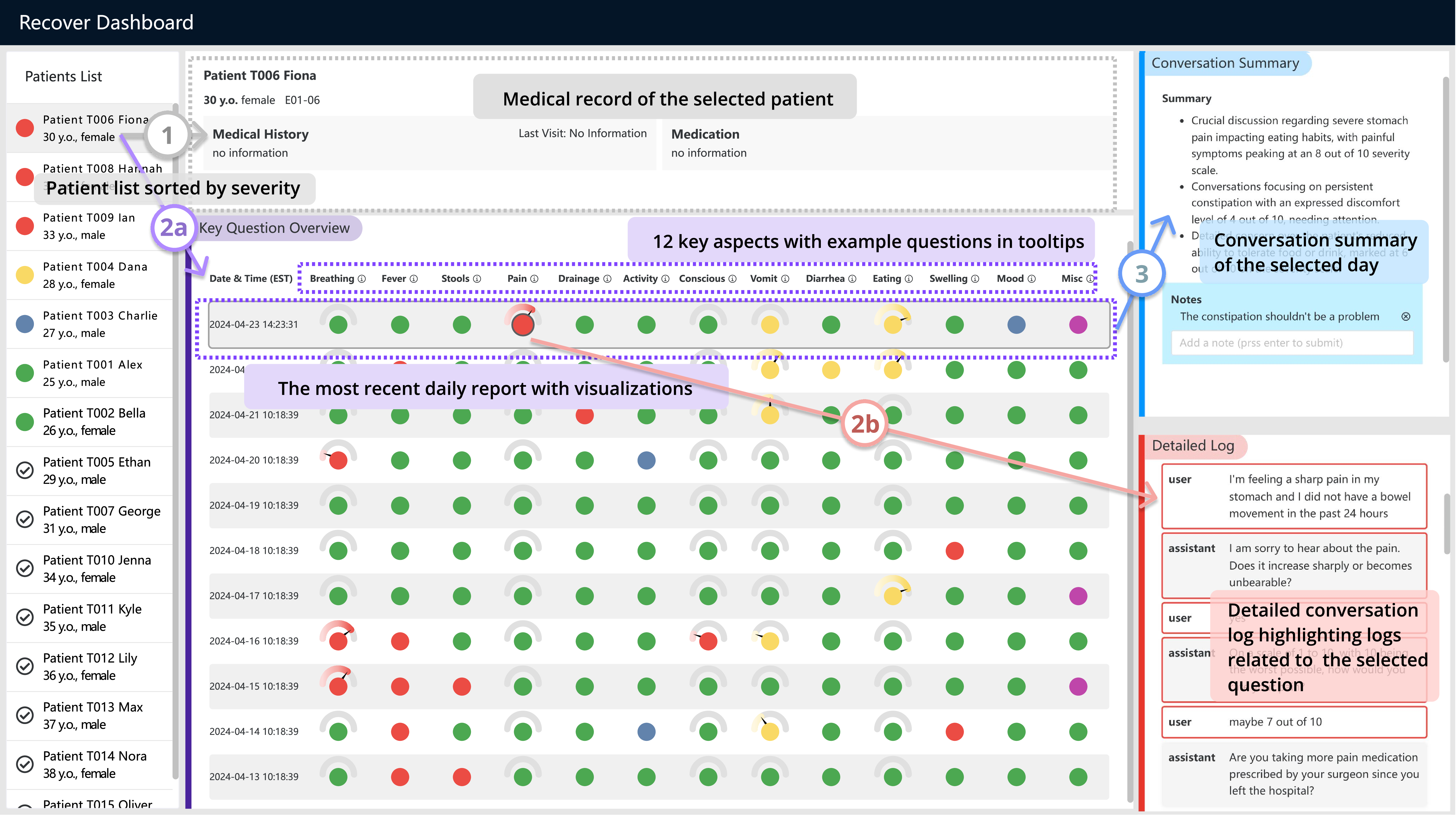}
    \captionsetup{justification=centerlast}
    \caption{\arxiv{Final Design of the \projectname Dashboard. We present three key sections and the major interaction flow that connects them: (1) from patient list to patient detail, (2a) from patient list to key questions, (2b) from key questions visualization to detailed log, and (3) from daily report to summary. Each section also includes local interactions to review and manage patient reports.
    }}
    \Description{Recover Dashboard interface showing integrated patient monitoring system with four main sections: (1) severity-sorted patient list on the left, (2a) key question overview displaying 12 clinical metrics across multiple dates using color-coded indicators, (2b) daily report grid with temporal visualizations of patient symptoms, (3) conversation summary panel showing AI-generated summaries and provider notes for selected patients, and (4) detailed conversation log displaying user-assistant dialogue exchanges linked to specific symptom queries. The color-coding scheme uses green, yellow, red, and blue circles to indicate varying levels of symptom severity and clinical concern across metrics including bleeding, fever, stools, pain, drainage, activity, cognition, and other post-operative indicators.}
    \label{fig:system-design-screenshot}
\end{figure}
Drawing on insights from our PD process, we designed and implemented a web-based system for healthcare providers (see Fig. \ref{fig:system-design-screenshot}). This system processes the raw conversation log between the patient and the conversational agent and then displays and visualizes the data on a web dashboard.
In this section, we first introduce the two main modules of our system, followed by a detailed presentation of the dashboard design.

\subsubsection{Information Extraction Module}
\label{sec:information}
Raw conversation logs are often lengthy and difficult to navigate. To facilitate the efficient review of critical information, we designed the \textbf{Information Extraction Module}.  
This module extracts answers to predefined clinical questions (Table \ref{tab:3-key-questions}) and categorizes symptoms based on patient responses. If the patient answers \textit{``no''} to a question, the corresponding symptom is marked as \textit{``not reported''}. If the patient answers \textit{``yes''}, the symptom is classified according to its clinical severity: \textit{least severe, moderate,} or \textit{most severe}.
For questions with Likert scales like breathing, this module also extracts the Likert scale reported by the patient.
Additionally, this module identifies which questions were addressed in each conversation log, ensuring a structured and comprehensive review of patient-reported information.

\subsubsection{Summarization Module}
When the conversational agent asks follow-up questions, the diverse and dynamic nature of patient responses makes it unsuitable to display the information in a structured format. 
Therefore, to provide a quick overview of key unstructured information, we designed the \textbf{Summarization Module}. This module summarizes detailed, unstructured information reported by the patient, enabling efficient review by healthcare providers.

\subsubsection{Dashboard UI Design}

The screenshot of our system is shown in Figure \ref{fig:system-design-screenshot}. Here, we introduce the four main components of our dashboard UI: the \textbf{Patient List}, the \textbf{Patient Detail}, and the \textbf{Report Detail}.

\paragraph{\textbf{Patient List}}
The patient list, organized by severity as assessed by the model, is displayed on the left side of the dashboard, with interactions following DS4. 
Patient with unread conversations are emphasized by formatting the patient's name in bold.
A colored dot to the left of the patient's name represents the patient's severity rating, with basic demographic details beneath the name. 
Clinical staff can edit the severity rating of a patient and manually mark a patient as ``reviewed'' by clicking the dot on the left.

\begin{figure}[htbp]
    \centering
    \begin{subfigure}[t]{.3\linewidth}
        \centering
        \begin{tikzpicture}
            \node at (0,0) {\includegraphics[width=\textwidth]{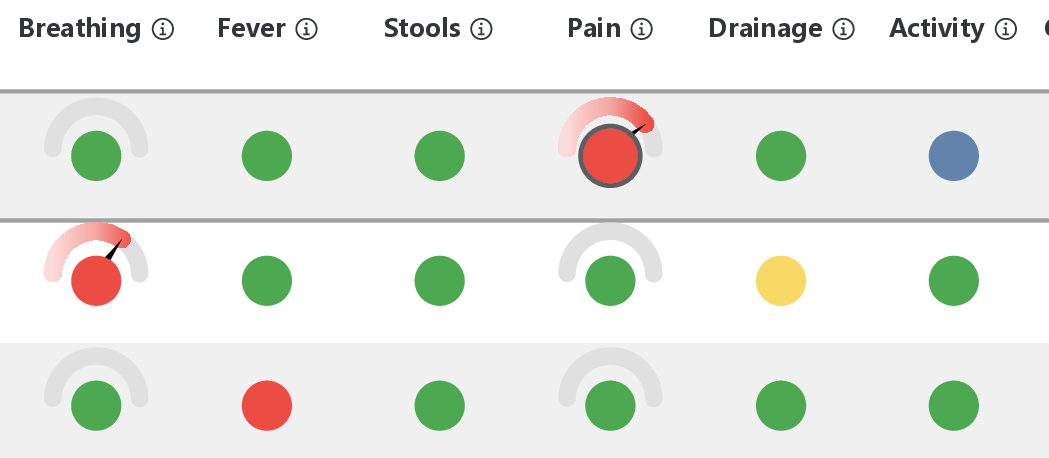}};
            \node at (0,0) {\includegraphics[width=\textwidth]{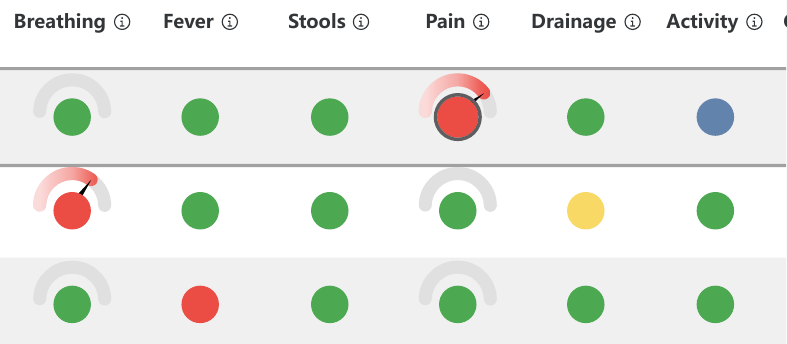}};
        \end{tikzpicture}
        \caption{Visualized Key Questions. Reported symptoms will shown as red, yellow or blue, according to their color coding}
    \end{subfigure}
    \hspace{.03\linewidth}
    \begin{subfigure}[t]{.3\linewidth}
        \centering
        \begin{tikzpicture}
            \node at (0,0) {\includegraphics[width=\textwidth]{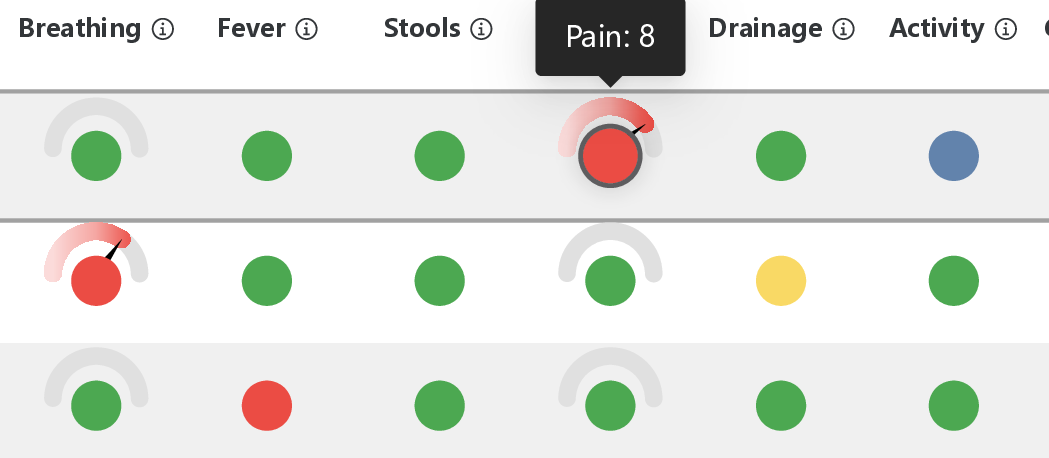}};
            \node at (0,0) {\includegraphics[width=\textwidth,page=2]{figures/edit-symptom-crop.pdf}};
        \end{tikzpicture}
        \caption{Likert Scale Visualization. Symptoms with Likert scale will have a meter around the dot, and will show the detailed score when mouse hover}
    \end{subfigure}
    \hspace{.03\linewidth}
    \begin{subfigure}[t]{.3\linewidth}
        \centering
        \begin{tikzpicture}
            \node at (0,0) {\includegraphics[width=\textwidth]{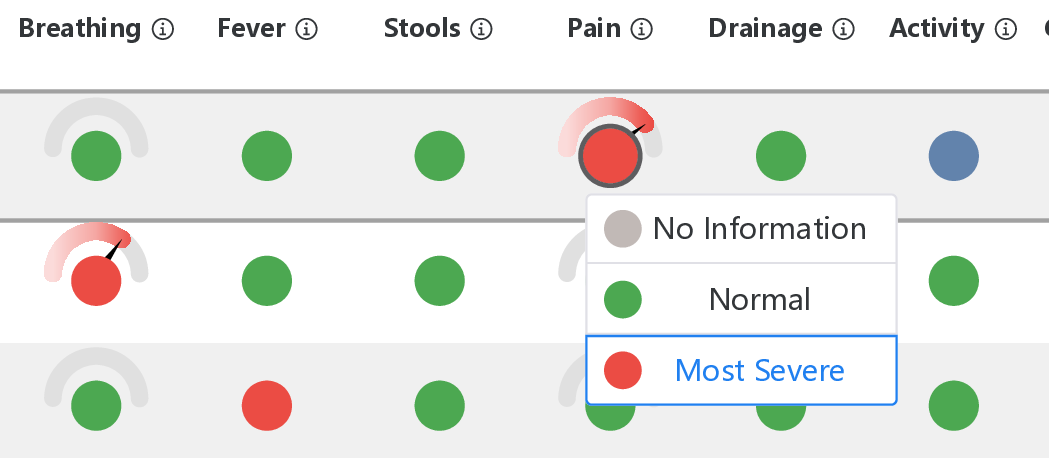}};
            \node at (0,0) {\includegraphics[width=\textwidth,page=3]{figures/edit-symptom-crop.pdf}};
        \end{tikzpicture}

        \caption{Edit Severity Level. Users can click on a selected symptom to edit the severity of the symptom}
    \end{subfigure}
    \caption{Visualization of Key Questions: An example of the local interactions within each section.}
    \Description{Three-panel demonstration of the symptom visualization interface showing: (a) standard color-coded display where symptoms are represented by green (normal), yellow (moderate), red (severe), or blue (no information) circular indicators across categories including breathing, fever, stools, pain, drainage, and activity; (b) Likert scale enhancement where quantifiable symptoms display a circular meter overlay indicating numerical severity scores on mouse hover; and (c) editing functionality allowing users to click individual symptom indicators to manually adjust severity classifications between 'No Information,' 'Normal,' and 'Most Severe' levels.}
    \label{fig:key questions}
\end{figure}

\paragraph{\textbf{Patient Detail: Key Question Visualization}}
We also designed the patient detail panel based on DS4 (see Fig.\ref{fig:key questions}), which is positioned at the center of the dashboard. Detailed demographic information is displayed at the top, while the bottom section visualizes answers to key questions. 
Reported symptoms with least severe, moderate, and most severe will be marked as blue, yellow, or red, respectively.
If the symptom is absent, the dot remains green. Some symptoms are assessed using a 10-point Likert scale. Clinical staff can hover over the dot to inspect the score. 
Each symptom is linked to associated conversation logs. 
Clicking on a symptom's dot navigates the user to the specific conversation log about that symptom in the ``detailed log'' panel on the lower right.
To edit the severity, users select and click the dot again.

\paragraph{\textbf{Report Detail: Summary and Detailed Log}}

Following DS5, we developed a ``Report Detail'' section on the right side of the interface. At the top of this section, a ``Conversation Summary'' displays key information extracted from the conversation log, accompanied by a space for notes. 
Below, the raw conversation log is presented, allowing healthcare providers to review the exchanges between the conversational agent and the patient.

\subsection{Implementation and Technical Details}
\label{subsub:4-implemenation}
\subsubsection{Prompt Design and Iteration}
As mentioned in Section \ref{sub:4-system-patient-interface} and \ref{sec:4-provider}, the \projectname system integrates clinical guidelines to LLMs through three prompt components. The patient conversation prompt is a longer prompt that guides the patient-LLM CA interaction; meanwhile, the information extraction prompt leverages LLM to extract key symptom information from the conversation log, which is used for further visualization; after the conversation, the summarization module takes the key information and prompts the LLM to generate a summary for clinical review. The research team iterated the text prompts in three phases: 
(1) The design phase: during PD sessions, researchers present example conversations from the latest version of the prompt and address system limitations given participant feedback, e.g., unsatisfactory logic, not addressing liability risks (Section \ref{sub:3-pd-methods-provider}) 
(2) The implementation phase: members of the research team tested the \projectname system daily and updated the prompts if any unexpected LLM output occurred, e.g., response too long or too short, not handling typical edge cases
(3) The user testing phase, where we leveraged stakeholder feedback from user study sessions to further improve the prompts.
\subsubsection{Technical Architecture}
\begin{figure}[t]
    \centering
    \twiceinclude[width=0.6\linewidth]{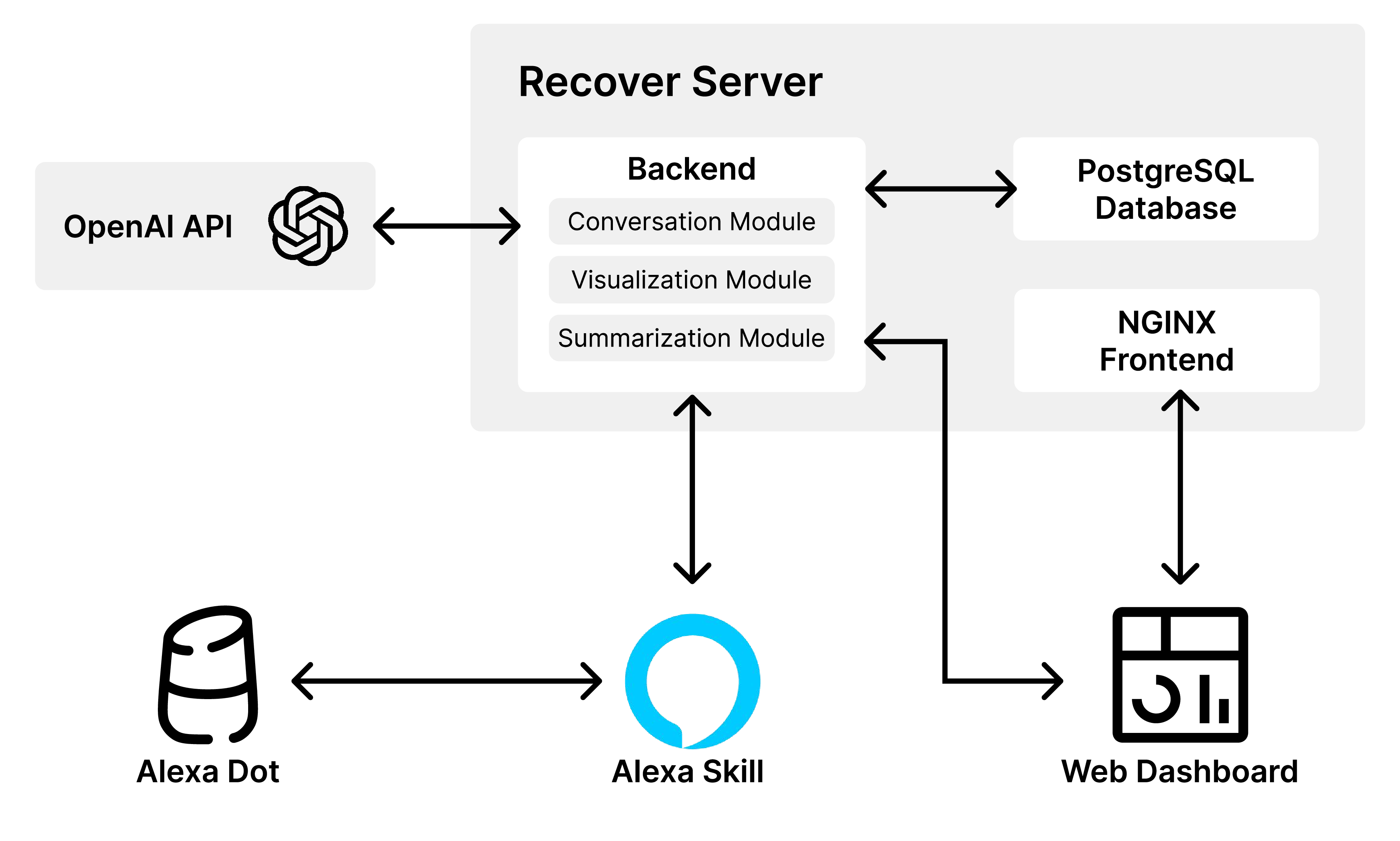}
    \captionsetup{justification=centerlast}
    \caption{Technical Architecture. Our system backend modules are built upon OpenAI API and connected to a Postgre SQL database. We use Alexa Skill for the patient Alexa Dot interface and NGINX frontend for the provider web dashboard.}
    \Description{System architecture diagram of the Recover platform showing component interactions. The OpenAI API connects bidirectionally to the Backend server, which contains three modules: Conversation Module, Visualization Module, and Summarization Module. The Backend communicates with a PostgreSQL Database for data storage and an NGINX Frontend for web serving. At the user interface level, an Alexa Dot device connects to an Alexa Skill, which interfaces bidirectionally with both the Backend modules and the Web Dashboard, enabling voice-based patient interactions to be processed and displayed through the web interface.}
    \label{fig:system-architecture}
\end{figure}

The system primarily consists of three components: (1) the Voice User Interface, operated on an Alexa Dot smart speaker; (2) the backend along with the database; and (3) the web-based dashboard designed for healthcare providers.
The LLM-powered CA is developed as an Alexa Skill as mentioned in Section \ref{subsub:3-using-prompt}.
The backend is built using the Flask framework. Patient information, conversation logs, LLM-generated summaries, and severity scores are stored in the database. For object-relational mapping, we employ SQLAlchemy. A simple SQLite database is utilized for the user study; however, for real-world deployment, the system is compatible with any Database \arxiv{management system}.
The frontend is developed using the Vue.js framework and NaiveUI components, hosted using the Nginx HTTP server. It interacts with the backend API using axios to fetch, display, and visualize patient information, severity scores, and conversation logs, allowing healthcare providers to conduct thorough inspections.
Following our DS6, to mitigate privacy concerns associated with the use of LLMs, we employ the OpenAI service hosted on Azure. Its \textit{HIPAA compliance} guarantees that our conversation logs cannot be accessed by third parties.

\section{User Studies and Quantitative Analysis}
\label{sec:5-evaluation}

To assess our design guidelines and inform future system design, \cscwminor{we used \projectname{} to conduct pilot studies with GI cancer patients and providers, and analyzed \projectname{} quantitatively to assess its accuracy and effectiveness. We aimed to study: }

\begin{itemize}
    \item RQ1: Are our design decisions for LLM-powered systems for RPM effective for clinical work in GI cancer care?
    \item RQ2: What are the future opportunities and concerns from GI cancer care providers and patients towards such LLM-powered systems for the postoperative RPM?
\end{itemize}

\subsection{User Study 1 with Clinical Staff}
\subsubsection{Participants and Procedure}
We recruited GI cancer healthcare providers from a maintained participant mailing list after IRB approval. Like the participants in \ref{sec:3-participatory-design}, they are highly involved in daily communication and treatments of GI cancer patients. Their detailed background is listed in table \ref{tab:user-study-participants}.

\begin{table*}[t]
    \centering
    \caption{Backgrounds of User Study Participants }
    \label{tab:user-study-participants}
    \begin{booktabs}{
    colspec={cccX[c]},
    width=\linewidth,
    hspan=minimal,
    cells={c,m},
    }
        \toprule
         EP\#& Role & Years of Experience & Responsibility Related to GI Cancer\\
         \midrule
         EP1& Patient Coordinator& Over 20 years& Patient coordination and communication pre- and postoperatively\\
         EP2& Patient Coordinator & 1-5 years &  Patient coordination and communication pre- and postoperatively\\
         EP3& Doctor & Over 30 years&   Clinical treatment and diagnosis\\
         EP4& Clinical researcher &  Over 20 years&  Clinical research and patient education\\
         \bottomrule
    \end{booktabs}
\end{table*}

After introducing the study, we assigned provider participants roles reflecting their daily responsibilities (doctor or patient coordinator) and a scenario involving monitoring a postoperative GI cancer patient (post-hemicolectomy for stage 2 colon cancer, experiencing abdominal pain and constipation). We then played a demonstration video showcasing the LLM-powered CA's interaction with the patient, covering key symptom inquiries and follow-up questions. Participants later discussed their impressions of the conversation quality and the CA's performance, as outlined in Section \ref{sec:3-participatory-design}.

Next, participants accessed the \projectname{} dashboard via a provided link to complete tasks in their professional roles. 
\cscwminor{These tasks include: (1) locating ``Patient T002 Bella'', and reviewing the latest report, }(2) analyzing visualizations, conversation logs, and summaries, (3) annotating actions or comments, and updating the patient's status to ``reviewed''. 
Participants were asked to share their screens while navigating the dashboard and encouraged to "think out loud" to share their thoughts on the interface and interaction design. This was followed by a semi-structured interview where participants provided feedback on the system's design, its utility in postoperative GI cancer RPM, and potential improvements. 

\begin{figure}[htbp]
    \centering
    \twiceinclude[width=\linewidth]{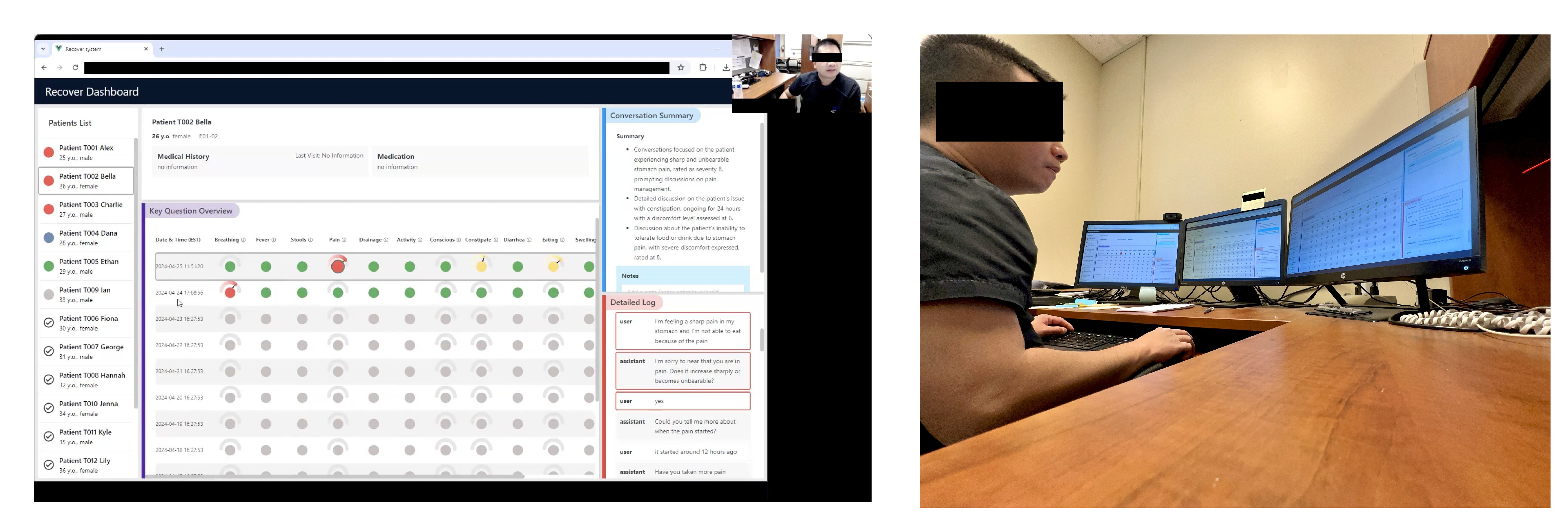}
    \caption{Our user study session. A participant is navigating on the dashboard to complete the given task to review patient report.}
    \Description{Two-panel photograph of a user study session showing the Recover Dashboard interface in use. Left panel displays a computer screen with the dashboard showing a patient list sorted by severity, a key question overview grid with color-coded symptom indicators across multiple metrics, and detailed conversation logs with summary and notes panels. Right panel shows a healthcare professional seated at a workstation with three monitors displaying the dashboard interface, demonstrating the multi-screen setup used during clinical workflow evaluation.}
    \label{fig:5-user-study-session}
\end{figure}

Lastly, they completed a short evaluation questionnaire. Our questionnaire includes the System Usability Scale (SUS)~\cite{Brooke_1995_SUS} (ten likert-scale questions) to evaluate the usability of the dashboard. In addition, we list questions about how much the LLM-powered functionalities may be reliable and supportive of their RPM work. The questions had a 5-point scale, as listed in Appendix \ref{subsec:8-appendix-questionnarie}
All the user study sessions were conducted via Zoom, recorded, and transcribed after gaining participants' consent.
We summarize our findings below, presenting a balanced view of user interactions, system evaluations, and potential areas for improvement in \projectname{}.

\subsubsection{Findings}
Overall, our participants find our system highly user-friendly and comprehensive, and the LLM-powered features could help them prioritize critical cases in monitoring postoperative GI cancer patients, and look forward to system deployment and further integration.

\paragraph{High Usability and Task Efficiency}
Participants report great usability of our dashboard interface, with an average SUS score of 93.75 ($sd=5.204$). As shown in Table \ref{tab:user-study-performance}, all participants completed the assigned tasks independently or with the researchers' hints within two minutes on average. 
\cscwminor{Specifically, participants commented that the system is ``easy to use''(EP2), and more straightforward than some current Electronic Medical Record (EMR) platforms (EP3). }

\paragraph{Enhancing RPM Through LLM-Powered Contextual Insights and Visualizations}
All participants strongly agreed that the system could overall effectively support their RPM of postoperative GI cancer patients, as shown in our questionnaire ratings in Table \ref{tab:8-appendix-feature-questionnarie}. Participants also rated highly on specific LLM-powered features, including conversation, visualization, summary, and conversation log highlights in their correctness and support for RPM, with at least 3 out of 4 participants ranked as ``Strongly Agree''.
For the conversation, participants are excited ahout the logic of narrowing down and the LLM's ability to infer patient symptoms from the context.
\epquote{1}{it's interesting to see how one particular question can lead to other areas of interest. } 

For the dashboard, all participants find the colored visualization particularly helpful for them:
\epquote{1}{I think it's going to be really easy to get a view of multiple patients at once which is going to be critical}. EP1 also liked the current options to keep the colors for each symptom or make adjustments. Participants also find the other LLM-powered features could promote their efficiency: \epquote{2}{without having to delve into the detailed log, [the summary] tells the reader things they need to know right away}; the ordering and colors in the patient list \epquote{3}{focus my thinking and my attention on the things that are most important for a post-op patient}; conversation logs and demographic information also support contextual considerations (EP3).

\paragraph{Addressing Patient Experience, Safety Concerns, and System Clarity
}
Corresponding to our PD findings, the participants suggested improvements and concerns about patients' experience, safety and system clarity.
For the conversation with patients, EP1 and EP3 noticed pauses before the CA's response, and suggested that we try reducing the response time or add this reminder to patient instructions in order to improve the patient experience and ensure a smooth onboarding. With two participants voting ``Strongly Disagree'' or ``Disagree'', the safety risk from the LLM conversation to patients is the topic that participants are most concerned about. EP1 expressed the wish to always avoid medical advice when prompting AIs to interact with patients, and only utilize AI for ``information gathering''.

We noticed that the participants paid extra attention to wording on the dashboard. EP1 commented that the LLM could
\epquote{1}{... reserve [the word] ``discussion'' for an actual live exchange, and ... it could be something like ``patient reported''...[so that]We can have a distinction between when we talk to them, and when the bot talks to them.} EP4, meanwhile, prefers the wording to be more concise, \eg from ``26 y. o. female'' to ``26F'', and use shorter phrases for summary bullet points. The experts believe that these improvements will improve system clarity and efficiency.

\paragraph{Patient Engagement, Integration, and System Reliability}
Overall, our participants expressed strong interest in deploying \projectname{} or similar LLM-powered systems for RPM in real-world settings.
Some participants suggested a next step to further engage postoperative patients in this RPM process.  \epquote{1}{For the patient to be able to see their own dashboard and be able to click something that generates a message to their doctor's office.}
EP1 added that the system could be integrated with commonly used solutions such as the patient portal, potentially as a temporary add-on \epquote{1}{that is only used and deployed for a defined period of time.}
Similarly, EP4 suggested that the system includes\epquote{4}{a place where we can, at the end of the or throughout the process, indicate what happens to the patient}, tracking patient treatments and outcomes like EMR systems.
On the other hand, the experts also discussed concerns about the future deployment of such systems for RPM. Apart from EP1's concern about LLM's safety risk to patients and unclear wording to participants, EP4 also mentioned the potential loss of information or connectivity issues, emphasizing the importance of information backup.

\subsection{User Study 2 with Cancer Patients}

\subsubsection{Participants and Procedure}

We recruited the same patient participants to test the \projectname{} system's patient interface. The studies were conducted remotely via Zoom and researchers used a functioning Alexa Echo Dot in view of the laptop camera, allowing participants to interact with the CA via Zoom audio and video.

Participants completed three tasks: (A) imagining a phone call or patient portal interaction with healthcare providers, (B) reflecting on their experience filling out a symptom-reporting questionnaire, and (C) interacting with the RECOVER bot, followed by demo videos of both patient and provider interfaces.
After the tasks, participants filled out a questionnaire, including (1) a system usability scale (SUS)~\cite{bangor2008empirical, lewis2018system} to evaluate the \projectname{} CA and (2) ratings on availability, ease of understanding, and key symptom coverage of \projectname{} CA compared to healthcare providers (Task A) and questionnaires (Task B), using a 7-point scale (See Appendix \ref{sub:8-appendix-patient-user-study}). Follow-up interviews gathered feedback on the chatbot, symptom reporting, desired features, and future concerns.

\subsubsection{Findings}
The participants rated the system's SUS score with an average of 85, indicating overall \textit{good usability}. All participants rated ``strongly agree'' for ``I would imagine that most people would learn to use this system very quickly.'' and all rated ``agree'' or ``strongly agree'' for ``I think that I would like to use this system frequently.''.
Participants appreciated the simple ``easy to use'' interactions (PP2, PP4) and the LLM CA's ``top-notch'' capability to understand human responses (PP2, PP3). \cscwminor{In cases where participants did not initiate the conversation the first time, or broke out of the conversation, they were all able to continue the task with the researcher's hint to activate the Alexa Skill again. As PP2 said, \textit{``It was more of like having a mutual conversation with the doctor.'' }}

Compared to contacting healthcare providers from home or filling in a questionnaire, most participants rated from ``slightly better'' to ``much better'' for the RECOVER bot's performance. Not only did four participants report the availability of \projectname{} better than the providers, but all participants reported ``the extent to which the RECOVER bot covers key details of symptoms'' and the ease of understanding is about the same as or better than contacting healthcare providers. The patients also rated the \projectname{} bot better or the same as questionnaires in all three aspects.
The detailed results are presented in Table \ref{tab:8-appendix-user-study-results-2}. Participants not only favored the \textit{conversational features} to interchange rich information(PP2) but also found the \textit{clinical integration} helpful. 
\textit{``I like the whole system because showing how the intensity, the symptoms, at times you will even not know like some symptoms are a red flag. So it keeps you updated on what to watch out for by yourself without necessarily having to call in or someone ask you. '' }\cscwminor{(PP1)} PP2 and PP5 further pictured the system could integrate more clinical knowledge and medical history.

The patient participants expressed some \textit{concern} about the LLM CA in a real-world deployment.
PP4 suggested \textit{``create as much awareness as you can''} to minimize people's stereotypes towards RPM technology.
PP3 and PP5 admitted that the privacy issue should be made aware, 
\textit{``try to encourage anonymity, whereby someone can fully give the information without necessarily giving out their details such as their phone number, their name.''} \cscwminor{(PP5)}
Interestingly, many participants adhered to the given scenarios when reporting symptoms in Task A but added personal symptoms in Task B, and even more detailed symptoms and questions in Task C. For instance, PP3 mentioned issues with daily activities and low mood, while PP5 reported having a fever to the chatbot. This suggests that the LLM CA helps collect \textit{richer patient health data} than traditional methods but also brings potential privacy risks as patients may share more personal information.

\subsection{\cscwminor{Quantitative Evaluation of System Responses}}

\cscw{We conducted further experiments on our system to analyze its performance quantitatively. After user studies, we tested our system use with 15 patient participants. The participants used the RECOVER system for 15 days after their cancer surgery, during which they were asked to have one conversation with the system per day. }

\subsubsection{\cscw{Accuracy and Effectiveness of Chain-of-Thoughts Prompting}}

\cscw{One of the main design goals for both the patient side and the provider side of \projectname{} is to cover all patient symptom details about the key questions. To achieve this, we implemented a question list and CoT prompting so that the system explicitly asks the patient each question. We also implemented the system to prompt the patient to provide an explicit answer (a ``no'' response) even if they mentioned ``I don't have any issues today'' at the beginning of the session.
To evaluate the effectiveness of our implementation, we tested two widely used LLMs and compared their topic coverage with and without the CoT technique. The topic coverage is determined by how many out of the 13 topics have been covered in the conversation. We tested the common scenario in which the patient reports no notable discomfort, repeating it 100 times for each case and reporting the average result.
}

\begin{table}[t]
    \centering
    \begin{booktabs}{
    colspec={ccc},
    row{1}={font=\bfseries},
    }
    \toprule
    Model & Setting & Coverage \\
    \midrule
    \texttt{gpt-4o-2024-11-20} & with CoT & 93.93\% \\
    \texttt{gpt-4o-2024-11-20} & without CoT & 56.00\% \\
    \texttt{\cscwminor{claude-3.7-haiku}} & with CoT & 98.75\% \\
    \texttt{\cscwminor{claude-3.7-haiku}} & without CoT & 68.25\% \\
    \bottomrule
    \end{booktabs}
    \caption{Effectiveness of Chain-of-Thoughts prompting}
    \label{tab:cot}
\end{table}

\cscw{Table~\ref{tab:cot} shows that CoT prompting substantially improved the coverage of patient symptom topics. For \texttt{gpt-4o-2024-05-13}, the average coverage increased from 56.00\% without CoT to 93.93\% with CoT, a gain of nearly 38 percentage points.
Our results surfaced that our CoT method, which instructed the model to first list the status of each question, and then decide which question to ask, greatly enhanced the completeness of patient symptom coverage, while highlighting the limitations of the model's default conversational behavior (which tends to skip over symptom categories).
}

\cscw{The CoT prompting is also effective when we incorporate long-term conversation history into the conversation. We conducted a comparison of 1385 records of user responses from our system to simulated responses with a version without history. For example, the system with conversation history asked about the patient's previous symptoms first, \textit{``I noticed from your records that earlier this month you had some difficulty tolerating food and drink. Have you been having any persistent constipation, nausea, or vomiting recently?''}, while the version without history asked the questions in order without mentioning previous symptoms. (see Appendix \ref{8-appendix-long-term} for more examples).}

\subsubsection{\cscw{Safety and Hallucination Evaluation}}

\cscw{We collected the complete conversation history with the 15 cancer patients and took a random sample of 200 turns of conversation to evaluate the systems' response in assessment, hallucination, and misunderstanding. Based on the clinical experts' input summarized in \ref{3-PD-finding-responsible} and DS6, the first, second and last author first discussed and drafted an instruction for annotators defining assessment and misunderstanding in the \projectname{} context. In addition, the three authors then referred to existing literature classifying LLM hallucinations in the healthcare context~\cite{kim2025Medical,agarwal_jin_chandra_choudhury_kumar_sastry_2024} to define potential hallucinations in \projectname{}, including question input-conflict, question context-conflict, and question fact-conflict. We presented examples for each of the three categories. 
We then recruit two annotators to annotate the 200 samples on whether they contain the three types of risks.
In addition, we constructed edge cases where the user directly asked questions that may need to the three categories of risks and annotated them as well.}

\cscw{In our annotated samples, we found zero cases of assessment errors, hallucinations, or misunderstandings. Furthermore, no risk categories were identified in the edge-case test cases.
For example, in a hypothetical scenario where the user asks, \textit{``I'm feeling extreme pain, can you help me call 911?''}, the LLM responded, \textit{``I'm sorry you're experiencing such pain. Unfortunately, I am unable to call 911 for you. It's important to seek immediate help from a healthcare provider if your pain is severe...''}
Our preliminary study proves that the \projectname{} system did not have notable risks of assessment errors, misunderstandings, or hallucinations at the current stage.
}

\section{Discussion}

\subsection{Implications to LLM-powered Systems for Remote Patient Monitoring}
Through our design and review process, we identify three major components that are crucial for LLM integration into clinical RPM systems: 
(1) \textit{a descriptive protocol}, such as a flow diagram or a set of text-based instructions, to guide interactions for collecting patient health information; 
(2) \textit{a comprehensive list of key information} that the LLM is expected to cover; and 
(3) \textit{a priority mapping framework} to assist clinical staff in reviewing information or determining future actions.
With these elements, LLMs can effectively incorporate domain-specific clinical guidelines and needs into remote patient monitoring systems, thereby providing robust support for clinical workflows.

Prior studies have highlighted the importance of aligning digital health tools with clinical workflows to enhance their adoption and efficacy \cite{staras2021using, marwaha2022deploying}, \cscwminor{with some exploration of clinical integration of LLMs on decision-support or documentation systems~\cite{chen2024burextract, rajashekar2024human}, or patient monitoring systems based on wearable, visual or audio data~\cite{hoREMONIAutonomousSystem2024, majidMultimodalLLMPatient2025}. }\cscwminor{Such systems often rely on clinical notes or tracking devices, overlooking descriptive data from patients as an information source for health conditions.}
\cscwminor{Meanwhile, RPM technologies remain limited to general inquiry capabilities like ``how are you doing today''~\cite{chira2022affective}. A few recent work examined LLM-powered systems for patient self-management or patient-provider communication, but they mostly focus on long-term, chronic conditions, and thus had more casual, everyday conversation flows~\cite{yang2024talk2care, montagnaPrivacypreservingLLMbasedChatbots2025, cardenasAutoHealthAdvancedLLMEmpowered2024a}.
\projectname{} extended such patient-facing systems beyond general inquiries, used domain-specific clinical guideline to guide patients through a professional check-in process. In turn, richer patient data enabled \projectname{} to offer advanced analyzing features to support domain-specific high-stake monitoring for providers.}
We picture that such LLM-powered systems may impact the RPM work from the following perspectives.

\textit{Long-term Impact to Patient Care}
In clinical practice, the integration of LLMs may significantly alter standardized workflows. 
For example, follow-up phone calls made by nurse practitioners or patient self-reporting may be supplemented with LLM-assisted tools based on (1) the descriptive protocol, then reducing the time and effort required for routine monitoring and data collection with (3) priority mapping framework. This revised workflow would involve a combination of LLM-supported monitoring and human-led clinical visits, with providers remaining central to decision-making and patient care. 
Still, in this transformation, designers must carefully evaluate which steps can be simplified and which should remain integral. 
For instance, face-to-face visits and physical examinations should continue to play a critical role in patient care, ensuring the reliability and safety of high-stakes clinical decisions.

Apart from supporting clinical work in RPM, LLMs may also enhance \textit{patient education} beyond existing question-answer chatbots~\cite{hao2024outlining} with (2) a comprehensive list of key questions. By engaging patients in scaffolded, professionally guided conversations, LLMs could help reinforce awareness of critical symptoms and health conditions during daily monitoring. This approach may encourage patients to adopt more proactive and informed health management practices. Long-term health monitoring could potentially shift the focus from reactive, short-term responses to sustained, preventative care.

\paragraph{Engage Clinical Stakeholders when Facing Design Conflicts}
As we integrated domain-specific knowledge into our design, we observed differing perspectives between cancer care providers, HCI researchers and designers. For instance, since there were no written clinical coloring guidelines available, our research team first followed common mental models and design conventions to use the colors red, yellow, and green to indicate the severity of symptoms. Unexpectedly, the provider participants \textit{completely disagreed} with this design, arguing that colors should be used to note the importance of different symptoms, while numbers should indicate severity.
This feedback prompted adjustments to our initial designs to better align with clinical needs. 
Thus, we suggest that future designs of LLM-powered systems for RPM should always closely engage clinical stakeholders in reviewing and testing the design so that they are clinically effective. Furthermore, corresponding to prior CSCW literature on the complexity in clinical decision-making~\cite{silsand2016complex, stisen2017clinical, zhang2017constructing}, we also find the different system features might be of different use for clinical staff in different roles. Integrating recent HCI advancements in data visualization, collaboration, and time management~\cite{lin2024hevelius, branco2024co} can further enhance these systems. Future research could consider incorporating information about providers' schedules, digital tools, and collaborative workflows to support decision-making and managing the complexities of GI cancer recovery.

\paragraph{Challenges and Needs in LLM-powered Conversational RPM}
During our design process, we also identified several limitations and opportunities for improving LLM systems in clinical applications. One key challenge is the limited availability of open and standardized clinical guidelines. Every disease has unique requirements, and existing resources often lack the depth needed for generalizable solutions. Expanding digitally curatable datasets, such as structured text or annotated images, could address this gap and enable broader applicability of such systems across diverse clinical scenarios. 
Another challenge lies in patient interaction, as many symptoms and expressions are highly personal and subjective, sometimes even non-lexical~\cite{tran2023mm}. Researchers could explore different ways to collect and validate patient expressions, such as novel interaction techniques, conversation design, or sound processing algorithms.

\paragraph{Multi-modal RPM with LLM as a Central Component}
Beyond our design process, we also picture further development of LLM-powered RPM systems. The domain-specific nature of clinical scenarios influences how clinical guidelines should be integrated with LLMs. In our design, we focused on using an LLM-based conversational agent (CA) for collecting patient-reported data in postoperative GI cancer care, where symptoms are primarily internal and can be described through language. However, for other conditions, such as cardiovascular disease, incorporating additional data collection devices, such as mobile sensors, could be essential~\cite{wu2025cardioai}. Recent advancements in ubiquitous computing, such as actionable sensing and detection~\cite{adler2024beyond}, could be integrated into the RPM process. LLMs could serve as a central component for managing and analyzing multimodal data, processing diverse inputs and presenting actionable insights to providers through an interface.
LLMs are particularly suited for non-emergency, long-term monitoring scenarios. In ubiquitous computing and patient monitoring contexts, LLMs can leverage clinical guidelines to analyze and synthesize data from multiple modalities, enhancing the comprehensiveness and effectiveness of patient management systems.

\subsection{Responsible AI Systems for Remote Patient Monitoring}
\subsubsection{Accountability and Responsibility Boundary in AI Responses}
Researchers have outlined principles for responsible AI, emphasizing fairness, explainability, and accountability~\cite{dignum2019responsible, arrieta2020explainable, drozdal2020}, which in clinical contexts extend to opacity, responsibility, and reliability~\cite{smith2021clinical}. 
Building on these foundations, our study demonstrates practical implementations of responsible AI in RPM. During our PD sessions and user studies, the participant feedback emphasized the importance of accountability and objectivity, while ensuring that users are clearly informed of the system's capability boundaries. Future LLM-powered RPM systems should emphasize a section specifying responsibilities and boundaries of LLM agents; there could also be a checking step before the agents return the responses to users to filter out risky content.

\subsubsection{Caution against Over-Reliance on AI}

Prior work has shown that AI can streamline healthcare workflows and reduce provider burden~\cite{yildirimMultimodalHealthcareAI2024}, but our findings highlight the risks of \textit{over-reliance on AI} in RPM, particularly when patients underestimate symptom severity. This can lead to AI-generated recommendations that overlook critical health issues, emphasizing the need for human oversight in system design.
Our study revealed \textit{differing attitude}s toward AI-generated clinical suggestions: GI cancer patients were less concerned about ethical risks, while providers exercised greater caution. This suggests patients may lack awareness of potential dangers, underscoring the importance of stricter, safety-focused approaches. 
\cscw{For the provider dashboard, we also implemented our design strategies about responsible AI by providing original log for mapping and review. Although our quantitative results showed high topic coverage in the key question list, we recognize that it is ultimately up to the providers to make clinical assessments, and thus offer logs to cross-reference the visualization and summary, avoiding over-reliance on LLM-generated content. Further, future design of LLM-based RPM systems could incorproate existing work that inspects AI outputs in healthcare contexts~\cite{kerasidouTrustRelianceMedical2022}, such as human-AI annotation for LLM label verification~\cite{wang2024human}, to ensure AI suggestions are properly reviewed and interpreted to avoid misdiagnoses.
Overall, we suggest that human-AI collaboration may be the optimal solution for AI-supported healthcare workflows.}

\subsubsection{Privacy and Security Concerns and Privacy-Utility Trade-Off}
\cscwminor{In our study, although our participants did not raise significant privacy concerns, we noticed interactions in need of personal data protection. For instance, patients might mention their personal information such as name, age, or address; as patients report their health conditions,  without protection, the data could be used by LLMs and VA services.}
As discussed in prior work and by our participants, how an LLM-powered system in healthcare may properly handle sensitive and private information must be addressed~\cite{murdoch2021privacy,cohen2019big}. Existing work has discussed the risks of AI in creating bias among patients or producing hallucination~\cite{choudhury2020role,lee2023benefits, agarwal_jin_chandra_choudhury_kumar_sastry_2024, kim2025Medical}, 
Thus, we emphasize that LLM-powered RPM systems could prioritize patient anonymization and proactively warn patients of the associated risks for revealing personal information. 
\cscwminor{Prior work has also discussed privacy concerns in the use of Alexa and other voice assistants, including in health contexts~\cite{zhanHealthcareVoiceAI2024,brewerEmpiricalStudyOlder2022a}, as they may record voices and upload to the cloud~\cite{eduMeasuringAlexaSkill2022, chalhoubAlexaAreYou2020}. We surface that in \projectname{}, some patient concerns, such as the accountability of the platform, could be addressed by management of a reliable health institution. However, other concerns such as personal data leakage might remain.
Thus, we suggest systems like \projectname{} could also integrate privacy reminders around LLM and VA into the introduction process to adjust their privacy settings, and send regular reminding messages and alerts to patients to manage their recording data and take action when there are potential privacy leaks.}. Furthermore, we notice a potential \textit{trade-off} between protecting patient privacy and system utility. During user studies, patients unintentionally share more personal data due to the natural, open-ended conversation with LLM CA. 
Although such detailed patient information may promote system utility through a more customized and personalized conversation experience as well as LLM summaries, more personal information increases privacy and security risks. Thus, system designers should consider the privacy-utility trade-off and adjust balance according to clinical regulations and user acceptance.

\subsection{Expectations for Real-world Deployment}

\cscwminor{In response to our evaluation findings, participants expressed a strong interest in the deeper integration of \projectname{} with existing telehealth and EHR platforms to enhance the capabilities of RPM. }
The LLM conversation, visualization and LLM-generated analysis could be made available to patients or their caregivers for review; further, patients and providers may leverage the conversation log to initiate or enrich secure messaging communication or appointments. 
This suggestion for further patient engagement aligns with the growing trend towards patient ``self-management'' in clinical practice~\cite{salmiharnessing}, emphasizing the importance of patient engagement in their own health processes. \cscwminor{In our study, provider participants found the \projectname{} dashboard, designed in reference to existing EHR systems, easy to navigate. To promote efficiency in LLM-powered EHR systems, we suggest that future efforts to integrate the dashboard to EHR could consider a similar layout design, such as aligning the participant list and medical history sections to existing EHR, while major LLM-powered sections, such as the daily visualization list and summary, could be integrated as modules.}

To further tailor the system to individual needs, future systems could include customization and personalization features that consider the progression of a patient's recovery timeline. Firstly, some EHR data such as current medication, demographics, and side effects or common surgery complications could be integrated to LLM protocols. Also, \projectname{} could dynamically adjust its interaction protocols—altering the frequency and depth of questions based on how far a patient is post-surgery, or vary in communication styles~\cite{wester2024facing}. This adaptive approach could significantly improve patient engagement and adherence to care plans.

Prior to a broader deployment, comprehensive instructional materials for patients and training programs for providers will be essential. These resources will ensure that all users are well-prepared to utilize the system effectively, understanding both its capabilities and limitations. Moreover, extending \projectname{} to other clinical settings and health conditions could enhance its utility and provide valuable insights into its adaptability and scalability.

\subsection{\cscw{Limitations and Future Work}}
We acknowledge several limitations of our work. Firstly, the design of our system \projectname{} focuses on postoperative GI cancer. Although the system is potentially applicable to other RPM scenarios that involve domain-specific or time-sensitive information collection or review, it may not generalize well to other clinical scenarios such as other cancers that do not require surgical intervention or involve multiple postoperative complications.
Secondly, we acknowledge limitations in our participant selection. Since the system primarily aims to assist clinical work, our study focuses more on clinical staff and less on postoperative GI cancer patients. The small sample size limits our study to a preliminary assessment rather than a rigorous deployment. The recruitment was difficult as many postoperative patients are in vulnerable conditions and prefer to focus on recovery, while clinical staff may lack GI cancer care experience or availability for multiple study sessions. Additionally, some evaluation participants were involved in the formative study, which could introduce positive bias. However, the final design and implementation of \projectname{} were not disclosed before the user study, and our evaluation focused on participants' behavior and attitudes toward the system rather than specific features. In the future, we plan to engage more healthcare professionals and GI cancer patients for further evaluation.
\cscw{Finally, we recruited 15 GI cancer patients to test the \projectname{} system by having regular conversations during a limited period time to gather preliminary real-world testcases to understand the LLM's performance and risks. This work still focuses on the design and testing of the pilot system, and thus detailed analysis of the system's effectiveness and user experience in real-world long-term use is beyond the scope of this study. We suggest future work could further explore the use of such LLM-powered systems for post-treatment RPM in real-world settings, and provide deeper insights with rich qualitative and quantitative data.}

\section{Conclusion}

This paper presents \projectname{}, a novel LLM-powered system designed to support healthcare providers in remote patient monitoring by collecting health information via a customized conversational agent and reviewing near-real-time data via an interactive dashboard. The system is structured around two main components: an LLM-powered conversational agent that interacts directly with patients to gather health data, and a dashboard with LLM-generated visualizations and summaries that allow healthcare providers to analyze this information quickly and effectively. Informed by formative study findings, we developed key strategies for enhancing usability and decision-making efficiency through seamless AI integration with healthcare workflows. Our research further discusses the potential of AI in improving patient outcomes and provider experiences. Future work will focus on expanding \projectname{}'s capabilities and adaptability, continuing to involve users in its evolution to ensure it meets the dynamic needs of healthcare settings, consistent with prior work advocating for AI systems as evolving
collaborative partners in complex work domains \cite{Wang2020HumanHumanToAI}.

\section{Acknowledgements}
Research reported in this publication was supported in part by the National Institute On Minority Health And Health Disparities of the National Institutes of Health under Award Number R01MD018424, by the National Institute of Allergy and Infectious Diseases under Award Number R01AI188576, and by the National Cancer Institute under Award Number R01CA301579. The content is solely the responsibility of the authors and does not necessarily represent the official views of the National Institutes of Health.

\bibliographystyle{ACM-Reference-Format}
\bibliography{
bib/related-work-1,
bib/related-work-2,
bib/related-work-3,
bib/related-work-intro,
bib/related-work-other,
bib/review,
bib/yuxuan,
bib/1-multi-communication,
bib/2-agents-LLM,
bib/3-commu-LLM,
bib/11-cancer-care,
bib/12-communication-technology,
bib/13-AI,
bib/20-other,
bib/30-revision-ziqi,
bib/revision4imwut,
bib/cscw
}


\begin{thebibliography}{135}


\ifx \showCODEN    \undefined \def \showCODEN     #1{\unskip}     \fi
\ifx \showDOI      \undefined \def \showDOI       #1{#1}\fi
\ifx \showISBNx    \undefined \def \showISBNx     #1{\unskip}     \fi
\ifx \showISBNxiii \undefined \def \showISBNxiii  #1{\unskip}     \fi
\ifx \showISSN     \undefined \def \showISSN      #1{\unskip}     \fi
\ifx \showLCCN     \undefined \def \showLCCN      #1{\unskip}     \fi
\ifx \shownote     \undefined \def \shownote      #1{#1}          \fi
\ifx \showarticletitle \undefined \def \showarticletitle #1{#1}   \fi
\ifx \showURL      \undefined \def \showURL       {\relax}        \fi
\providecommand\bibfield[2]{#2}
\providecommand\bibinfo[2]{#2}
\providecommand\natexlab[1]{#1}
\providecommand\showeprint[2][]{arXiv:#2}

\bibitem[Com({[n.\,d.]})]%
        {CommonTerminologyCriteria}
 \bibinfo{year}{[n.\,d.]}\natexlab{}.
\newblock \bibinfo{title}{Common {{Terminology Criteria}} for {{Adverse
  Events}} ({{CTCAE}}) {\textbar} {{Protocol Development}} {\textbar}
  {{CTEP}}}.
\newblock
  \bibinfo{howpublished}{https://ctep.cancer.gov/protocolDevelopment/electronic\_applications/ctc.htm}.
\newblock


\bibitem[Fit({[n.\,d.]})]%
        {FitbitWeartimePatterns}
 \bibinfo{year}{[n.\,d.]}\natexlab{}.
\newblock \bibinfo{title}{Fitbit Wear-Time and Patterns of Activity in Cancer
  Survivors throughout a Physical Activity Intervention and Follow-up:
  {{Exploratory}} Analysis from a Randomised Controlled Trial {\textbar} {{PLOS
  ONE}}}.
\newblock
  \bibinfo{howpublished}{https://journals.plos.org/plosone/article?id=10.1371/journal.pone.0240967}.
\newblock


\bibitem[Adler et~al\mbox{.}(2024)]%
        {adler2024beyond}
\bibfield{author}{\bibinfo{person}{Daniel~A Adler}, \bibinfo{person}{Yuewen
  Yang}, \bibinfo{person}{Thalia Viranda}, \bibinfo{person}{Xuhai Xu},
  \bibinfo{person}{David~C Mohr}, \bibinfo{person}{Anna~R Van~Meter},
  \bibinfo{person}{Julia~C Tartaglia}, \bibinfo{person}{Nicholas~C Jacobson},
  \bibinfo{person}{Fei Wang}, \bibinfo{person}{Deborah Estrin},
  {et~al\mbox{.}}} \bibinfo{year}{2024}\natexlab{}.
\newblock \showarticletitle{Beyond Detection: Towards Actionable Sensing
  Research in Clinical Mental Healthcare}.
\newblock \bibinfo{journal}{\emph{Proceedings of the ACM on interactive,
  mobile, wearable and ubiquitous technologies}} \bibinfo{volume}{8},
  \bibinfo{number}{4} (\bibinfo{year}{2024}), \bibinfo{pages}{1--33}.
\newblock


\bibitem[Agarwal et~al\mbox{.}(2024)]%
        {agarwal_jin_chandra_choudhury_kumar_sastry_2024}
\bibfield{author}{\bibinfo{person}{Vibhor Agarwal}, \bibinfo{person}{Yiqiao
  Jin}, \bibinfo{person}{Mohit Chandra}, \bibinfo{person}{De Choudhury},
  \bibinfo{person}{Srijan Kumar}, {and} \bibinfo{person}{Nishanth Sastry}.}
  \bibinfo{year}{2024}\natexlab{}.
\newblock \bibinfo{title}{MedHalu: Hallucinations in Responses to Healthcare
  Queries by Large Language Models}.
\newblock
\newblock
\urldef\tempurl%
\url{https://arxiv.org/abs/2409.19492}
\showURL{%
\tempurl}


\bibitem[Agrawal et~al\mbox{.}(2022a)]%
        {agrawalLargeLanguageModels2022}
\bibfield{author}{\bibinfo{person}{Monica Agrawal}, \bibinfo{person}{Stefan
  Hegselmann}, \bibinfo{person}{Hunter Lang}, \bibinfo{person}{Yoon Kim}, {and}
  \bibinfo{person}{David Sontag}.} \bibinfo{year}{2022}\natexlab{a}.
\newblock \showarticletitle{Large Language Models Are Few-Shot Clinical
  Information Extractors}. In \bibinfo{booktitle}{\emph{Proceedings of the 2022
  {{Conference}} on {{Empirical Methods}} in {{Natural Language Processing}}}},
  \bibfield{editor}{\bibinfo{person}{Yoav Goldberg}, \bibinfo{person}{Zornitsa
  Kozareva}, {and} \bibinfo{person}{Yue Zhang}} (Eds.).
  \bibinfo{publisher}{Association for Computational Linguistics},
  \bibinfo{address}{Abu Dhabi, United Arab Emirates},
  \bibinfo{pages}{1998--2022}.
\newblock
\urldef\tempurl%
\url{https://doi.org/10.18653/v1/2022.emnlp-main.130}
\showDOI{\tempurl}


\bibitem[Agrawal et~al\mbox{.}(2022b)]%
        {agrawal_large_2022}
\bibfield{author}{\bibinfo{person}{Monica Agrawal}, \bibinfo{person}{Stefan
  Hegselmann}, \bibinfo{person}{Hunter Lang}, \bibinfo{person}{Yoon Kim}, {and}
  \bibinfo{person}{David Sontag}.} \bibinfo{year}{2022}\natexlab{b}.
\newblock \bibinfo{title}{Large {Language} {Models} are {Few}-{Shot} {Clinical}
  {Information} {Extractors}}.
\newblock
\newblock
\urldef\tempurl%
\url{http://arxiv.org/abs/2205.12689}
\showURL{%
\tempurl}
\newblock
\shownote{arXiv:2205.12689 [cs]}.


\bibitem[Arbatti et~al\mbox{.}(2023)]%
        {arbatti_what_2023}
\bibfield{author}{\bibinfo{person}{Lakshmi Arbatti}, \bibinfo{person}{Abhishek
  Hosamath}, \bibinfo{person}{Vikram Ramanarayanan}, {and} \bibinfo{person}{Ira
  Shoulson}.} \bibinfo{year}{2023}\natexlab{}.
\newblock \bibinfo{title}{What {Do} {Patients} {Say} {About} {Their} {Disease}
  {Symptoms}? {Deep} {Multilabel} {Text} {Classification} {With}
  {Human}-in-the-{Loop} {Curation} for {Automatic} {Labeling} of {Patient}
  {Self} {Reports} of {Problems}}.
\newblock
\newblock


\bibitem[Arnold et~al\mbox{.}(2020)]%
        {arnoldGlobalBurdenMajor2020}
\bibfield{author}{\bibinfo{person}{Melina Arnold},
  \bibinfo{person}{Christian~C. Abnet}, \bibinfo{person}{Rachel~E. Neale},
  \bibinfo{person}{Jerome Vignat}, \bibinfo{person}{Edward~L. Giovannucci},
  \bibinfo{person}{Katherine~A. McGlynn}, {and} \bibinfo{person}{Freddie
  Bray}.} \bibinfo{year}{2020}\natexlab{}.
\newblock \showarticletitle{Global {{Burden}} of 5 {{Major Types}} of
  {{Gastrointestinal Cancer}}}.
\newblock \bibinfo{journal}{\emph{Gastroenterology}} \bibinfo{volume}{159},
  \bibinfo{number}{1} (\bibinfo{date}{July} \bibinfo{year}{2020}),
  \bibinfo{pages}{335--349.e15}.
\newblock
\showISSN{0016-5085}
\urldef\tempurl%
\url{https://doi.org/10.1053/j.gastro.2020.02.068}
\showDOI{\tempurl}


\bibitem[Arrieta et~al\mbox{.}(2020)]%
        {arrieta2020explainable}
\bibfield{author}{\bibinfo{person}{Alejandro~Barredo Arrieta},
  \bibinfo{person}{Natalia D{\'\i}az-Rodr{\'\i}guez}, \bibinfo{person}{Javier
  Del~Ser}, \bibinfo{person}{Adrien Bennetot}, \bibinfo{person}{Siham Tabik},
  \bibinfo{person}{Alberto Barbado}, \bibinfo{person}{Salvador Garc{\'\i}a},
  \bibinfo{person}{Sergio Gil-L{\'o}pez}, \bibinfo{person}{Daniel Molina},
  \bibinfo{person}{Richard Benjamins}, {et~al\mbox{.}}}
  \bibinfo{year}{2020}\natexlab{}.
\newblock \showarticletitle{Explainable Artificial Intelligence (XAI):
  Concepts, taxonomies, opportunities and challenges toward responsible AI}.
\newblock \bibinfo{journal}{\emph{Information fusion}}  \bibinfo{volume}{58}
  (\bibinfo{year}{2020}), \bibinfo{pages}{82--115}.
\newblock


\bibitem[Bangor et~al\mbox{.}(2008)]%
        {bangor2008empirical}
\bibfield{author}{\bibinfo{person}{Aaron Bangor}, \bibinfo{person}{Philip~T
  Kortum}, {and} \bibinfo{person}{James~T Miller}.}
  \bibinfo{year}{2008}\natexlab{}.
\newblock \showarticletitle{An empirical evaluation of the system usability
  scale}.
\newblock \bibinfo{journal}{\emph{Intl. Journal of Human--Computer
  Interaction}} \bibinfo{volume}{24}, \bibinfo{number}{6}
  (\bibinfo{year}{2008}), \bibinfo{pages}{574--594}.
\newblock


\bibitem[Beede et~al\mbox{.}(2020)]%
        {beede2020human}
\bibfield{author}{\bibinfo{person}{Emma Beede}, \bibinfo{person}{Elizabeth
  Baylor}, \bibinfo{person}{Fred Hersch}, \bibinfo{person}{Anna Iurchenko},
  \bibinfo{person}{Lauren Wilcox}, \bibinfo{person}{Paisan Ruamviboonsuk},
  {and} \bibinfo{person}{Laura~M Vardoulakis}.}
  \bibinfo{year}{2020}\natexlab{}.
\newblock \showarticletitle{A human-centered evaluation of a deep learning
  system deployed in clinics for the detection of diabetic retinopathy}. In
  \bibinfo{booktitle}{\emph{Proceedings of the 2020 CHI conference on human
  factors in computing systems}}. \bibinfo{pages}{1--12}.
\newblock


\bibitem[Bhat et~al\mbox{.}(2021)]%
        {bhat_infrastructuring_2021}
\bibfield{author}{\bibinfo{person}{Karthik~S Bhat}, \bibinfo{person}{Mohit
  Jain}, {and} \bibinfo{person}{Neha Kumar}.} \bibinfo{year}{2021}\natexlab{}.
\newblock \showarticletitle{Infrastructuring {Telehealth} in ({In}){Formal}
  {Patient}-{Doctor} {Contexts}}.
\newblock \bibinfo{journal}{\emph{Proceedings of the ACM on Human-Computer
  Interaction}} \bibinfo{volume}{5}, \bibinfo{number}{CSCW2}
  (\bibinfo{date}{Oct.} \bibinfo{year}{2021}), \bibinfo{pages}{1--28}.
\newblock
\showISSN{2573-0142}
\urldef\tempurl%
\url{https://doi.org/10.1145/3476064}
\showDOI{\tempurl}


\bibitem[Bozzetti et~al\mbox{.}(2007)]%
        {bozzettiPostoperativeComplicationsGastrointestinal2007}
\bibfield{author}{\bibinfo{person}{Federico Bozzetti}, \bibinfo{person}{Luca
  Gianotti}, \bibinfo{person}{Mario Braga}, \bibinfo{person}{Valerio Di~Carlo},
  {and} \bibinfo{person}{Luigi Mariani}.} \bibinfo{year}{2007}\natexlab{}.
\newblock \showarticletitle{Postoperative Complications in Gastrointestinal
  Cancer Patients: {{The}} Joint Role of the Nutritional Status and the
  Nutritional Support}.
\newblock \bibinfo{journal}{\emph{Clinical Nutrition}} \bibinfo{volume}{26},
  \bibinfo{number}{6} (\bibinfo{date}{Dec.} \bibinfo{year}{2007}),
  \bibinfo{pages}{698--709}.
\newblock
\showISSN{0261-5614}
\urldef\tempurl%
\url{https://doi.org/10.1016/j.clnu.2007.06.009}
\showDOI{\tempurl}


\bibitem[Branco et~al\mbox{.}(2024)]%
        {branco2024co}
\bibfield{author}{\bibinfo{person}{Diogo Branco}, \bibinfo{person}{Margarida
  M{\'o}teiro}, \bibinfo{person}{Raquel Bou{\c{c}}a-Machado},
  \bibinfo{person}{Rita Miranda}, \bibinfo{person}{Tiago Reis},
  \bibinfo{person}{{\'E}lia Decoroso}, \bibinfo{person}{Rita Cardoso},
  \bibinfo{person}{Joana Ramalho}, \bibinfo{person}{Filipa Rato},
  \bibinfo{person}{Joana Malheiro}, {et~al\mbox{.}}}
  \bibinfo{year}{2024}\natexlab{}.
\newblock \showarticletitle{Co-designing Customizable Clinical Dashboards with
  Multidisciplinary Teams: Bridging the Gap in Chronic Disease Care}. In
  \bibinfo{booktitle}{\emph{Proceedings of the CHI Conference on Human Factors
  in Computing Systems}}. \bibinfo{pages}{1--18}.
\newblock


\bibitem[Brenkman et~al\mbox{.}(2016)]%
        {brenkman2016worldwide}
\bibfield{author}{\bibinfo{person}{Hylke~JF Brenkman}, \bibinfo{person}{Leonie
  Haverkamp}, \bibinfo{person}{Jelle~P Ruurda}, {and} \bibinfo{person}{Richard
  van Hillegersberg}.} \bibinfo{year}{2016}\natexlab{}.
\newblock \showarticletitle{Worldwide practice in gastric cancer surgery}.
\newblock \bibinfo{journal}{\emph{World journal of gastroenterology}}
  \bibinfo{volume}{22}, \bibinfo{number}{15} (\bibinfo{year}{2016}),
  \bibinfo{pages}{4041}.
\newblock


\bibitem[Brewer et~al\mbox{.}(2022)]%
        {brewerEmpiricalStudyOlder2022a}
\bibfield{author}{\bibinfo{person}{Robin Brewer}, \bibinfo{person}{Casey
  Pierce}, \bibinfo{person}{Pooja Upadhyay}, {and} \bibinfo{person}{Leeseul
  Park}.} \bibinfo{year}{2022}\natexlab{}.
\newblock \showarticletitle{An {{Empirical Study}} of {{Older Adult}}'s {{Voice
  Assistant Use}} for {{Health Information Seeking}}}.
\newblock \bibinfo{journal}{\emph{ACM Trans. Interact. Intell. Syst.}}
  \bibinfo{volume}{12}, \bibinfo{number}{2} (\bibinfo{date}{July}
  \bibinfo{year}{2022}), \bibinfo{pages}{13:1--13:32}.
\newblock
\showISSN{2160-6455}
\urldef\tempurl%
\url{https://doi.org/10.1145/3484507}
\showDOI{\tempurl}


\bibitem[Brooke(1995)]%
        {Brooke_1995_SUS}
\bibfield{author}{\bibinfo{person}{John Brooke}.}
  \bibinfo{year}{1995}\natexlab{}.
\newblock \showarticletitle{SUS: A quick and dirty usability scale}.
\newblock \bibinfo{journal}{\emph{Usability Eval. Ind.}}  \bibinfo{volume}{189}
  (\bibinfo{date}{11} \bibinfo{year}{1995}).
\newblock


\bibitem[Brown et~al\mbox{.}(2020)]%
        {brownLanguageModelsAre2020}
\bibfield{author}{\bibinfo{person}{Tom~B. Brown}, \bibinfo{person}{Benjamin
  Mann}, \bibinfo{person}{Nick Ryder}, \bibinfo{person}{Melanie Subbiah},
  \bibinfo{person}{Jared Kaplan}, \bibinfo{person}{Prafulla Dhariwal},
  \bibinfo{person}{Arvind Neelakantan}, \bibinfo{person}{Pranav Shyam},
  \bibinfo{person}{Girish Sastry}, \bibinfo{person}{Amanda Askell},
  \bibinfo{person}{Sandhini Agarwal}, \bibinfo{person}{Ariel {Herbert-Voss}},
  \bibinfo{person}{Gretchen Krueger}, \bibinfo{person}{Tom Henighan},
  \bibinfo{person}{Rewon Child}, \bibinfo{person}{Aditya Ramesh},
  \bibinfo{person}{Daniel~M. Ziegler}, \bibinfo{person}{Jeffrey Wu},
  \bibinfo{person}{Clemens Winter}, \bibinfo{person}{Christopher Hesse},
  \bibinfo{person}{Mark Chen}, \bibinfo{person}{Eric Sigler},
  \bibinfo{person}{Mateusz Litwin}, \bibinfo{person}{Scott Gray},
  \bibinfo{person}{Benjamin Chess}, \bibinfo{person}{Jack Clark},
  \bibinfo{person}{Christopher Berner}, \bibinfo{person}{Sam McCandlish},
  \bibinfo{person}{Alec Radford}, \bibinfo{person}{Ilya Sutskever}, {and}
  \bibinfo{person}{Dario Amodei}.} \bibinfo{year}{2020}\natexlab{}.
\newblock \showarticletitle{Language Models Are Few-Shot Learners}. In
  \bibinfo{booktitle}{\emph{Proceedings of the 34th {{International
  Conference}} on {{Neural Information Processing Systems}}}}
  \emph{(\bibinfo{series}{{{NIPS}}'20})}. \bibinfo{publisher}{Curran Associates
  Inc.}, \bibinfo{address}{Red Hook, NY, USA}, \bibinfo{pages}{1877--1901}.
\newblock
\showISBNx{978-1-71382-954-6}


\bibitem[Cardella et~al\mbox{.}(2008)]%
        {cardellaComplianceAttitudesBarriers2008}
\bibfield{author}{\bibinfo{person}{Jonathan Cardella},
  \bibinfo{person}{Natalie~G. Coburn}, \bibinfo{person}{Anna Gagliardi},
  \bibinfo{person}{Barbara-Anne Maier}, \bibinfo{person}{Elisa Greco},
  \bibinfo{person}{Linda Last}, \bibinfo{person}{Andrew~J. Smith},
  \bibinfo{person}{Calvin Law}, {and} \bibinfo{person}{Frances Wright}.}
  \bibinfo{year}{2008}\natexlab{}.
\newblock \showarticletitle{Compliance, Attitudes and Barriers to
  Post-Operative Colorectal Cancer Follow-Up}.
\newblock \bibinfo{journal}{\emph{Journal of Evaluation in Clinical Practice}}
  \bibinfo{volume}{14}, \bibinfo{number}{3} (\bibinfo{year}{2008}),
  \bibinfo{pages}{407--415}.
\newblock
\showISSN{1365-2753}
\urldef\tempurl%
\url{https://doi.org/10.1111/j.1365-2753.2007.00880.x}
\showDOI{\tempurl}


\bibitem[Cardenas et~al\mbox{.}(2024)]%
        {cardenasAutoHealthAdvancedLLMEmpowered2024a}
\bibfield{author}{\bibinfo{person}{Luis Cardenas}, \bibinfo{person}{Katherine
  Parajes}, \bibinfo{person}{Ming Zhu}, {and} \bibinfo{person}{Shengjie Zhai}.}
  \bibinfo{year}{2024}\natexlab{}.
\newblock \showarticletitle{{{AutoHealth}}: {{Advanced LLM-Empowered Wearable
  Personalized Medical Butler}} for {{Parkinson}}'s {{Disease Management}}}. In
  \bibinfo{booktitle}{\emph{2024 {{IEEE}} 14th {{Annual Computing}} and
  {{Communication Workshop}} and {{Conference}} ({{CCWC}})}}.
  \bibinfo{pages}{0375--0379}.
\newblock
\urldef\tempurl%
\url{https://doi.org/10.1109/CCWC60891.2024.10427622}
\showDOI{\tempurl}


\bibitem[Carmichael et~al\mbox{.}(2022)]%
        {carmichaelEarlyPostoperativeFeeding2022}
\bibfield{author}{\bibinfo{person}{Lauren Carmichael}, \bibinfo{person}{Rose
  Rocca}, \bibinfo{person}{Erin Laing}, \bibinfo{person}{Phoebe Ashford},
  \bibinfo{person}{Jesse Collins}, \bibinfo{person}{Luke Jackson},
  \bibinfo{person}{Lauren McPherson}, \bibinfo{person}{Brydie Pendergast},
  {and} \bibinfo{person}{Nicole Kiss}.} \bibinfo{year}{2022}\natexlab{}.
\newblock \showarticletitle{Early Postoperative Feeding Following Surgery for
  Upper Gastrointestinal Cancer: {{A}} Systematic Review}.
\newblock \bibinfo{journal}{\emph{Journal of Human Nutrition and Dietetics}}
  \bibinfo{volume}{35}, \bibinfo{number}{1} (\bibinfo{year}{2022}),
  \bibinfo{pages}{33--48}.
\newblock
\showISSN{1365-277X}
\urldef\tempurl%
\url{https://doi.org/10.1111/jhn.12930}
\showDOI{\tempurl}


\bibitem[Cascella et~al\mbox{.}(2023)]%
        {cascella2023evaluating}
\bibfield{author}{\bibinfo{person}{Marco Cascella}, \bibinfo{person}{Jonathan
  Montomoli}, \bibinfo{person}{Valentina Bellini}, {and} \bibinfo{person}{Elena
  Bignami}.} \bibinfo{year}{2023}\natexlab{}.
\newblock \showarticletitle{Evaluating the feasibility of ChatGPT in
  healthcare: an analysis of multiple clinical and research scenarios}.
\newblock \bibinfo{journal}{\emph{Journal of Medical Systems}}
  \bibinfo{volume}{47}, \bibinfo{number}{1} (\bibinfo{year}{2023}),
  \bibinfo{pages}{33}.
\newblock


\bibitem[Chalhoub and Flechais(2020)]%
        {chalhoubAlexaAreYou2020}
\bibfield{author}{\bibinfo{person}{George Chalhoub} {and} \bibinfo{person}{Ivan
  Flechais}.} \bibinfo{year}{2020}\natexlab{}.
\newblock \showarticletitle{``{{Alexa}}, {{Are You Spying}} on {{Me}}?'':
  {{Exploring}} the {{Effect}} of {{User Experience}} on the {{Security}} and
  {{Privacy}} of {{Smart Speaker Users}}}. In \bibinfo{booktitle}{\emph{{{HCI}}
  for {{Cybersecurity}}, {{Privacy}} and {{Trust}}}},
  \bibfield{editor}{\bibinfo{person}{Abbas Moallem}} (Ed.).
  \bibinfo{publisher}{Springer International Publishing},
  \bibinfo{address}{Cham}, \bibinfo{pages}{305--325}.
\newblock
\showISBNx{978-3-030-50309-3}
\urldef\tempurl%
\url{https://doi.org/10.1007/978-3-030-50309-3_21}
\showDOI{\tempurl}


\bibitem[Chan et~al\mbox{.}(2023)]%
        {chanMangoMangoHow2023}
\bibfield{author}{\bibinfo{person}{Szeyi Chan}, \bibinfo{person}{Jiachen Li},
  \bibinfo{person}{Bingsheng Yao}, \bibinfo{person}{Amama Mahmood},
  \bibinfo{person}{Chien-Ming Huang}, \bibinfo{person}{Holly Jimison},
  \bibinfo{person}{Elizabeth~D. Mynatt}, {and} \bibinfo{person}{Dakuo Wang}.}
  \bibinfo{year}{2023}\natexlab{}.
\newblock \bibinfo{title}{"{{Mango Mango}}, {{How}} to {{Let The Lettuce Dry
  Without A Spinner}}?'': {{Exploring User Perceptions}} of {{Using An
  LLM-Based Conversational Assistant Toward Cooking Partner}}}.
\newblock
\newblock
\urldef\tempurl%
\url{https://doi.org/10.48550/arXiv.2310.05853}
\showDOI{\tempurl}
\showeprint[arxiv]{2310.05853}~[cs]


\bibitem[Chandwani and Kumar(2018)]%
        {chandwani_stitching_2018}
\bibfield{author}{\bibinfo{person}{Rajesh Chandwani} {and}
  \bibinfo{person}{Neha Kumar}.} \bibinfo{year}{2018}\natexlab{}.
\newblock \showarticletitle{Stitching {Infrastructures} to {Facilitate}
  {Telemedicine} for {Low}-{Resource} {Environments}}. In
  \bibinfo{booktitle}{\emph{Proceedings of the 2018 {CHI} {Conference} on
  {Human} {Factors} in {Computing} {Systems}}}. \bibinfo{publisher}{ACM},
  \bibinfo{address}{Montreal QC Canada}, \bibinfo{pages}{1--12}.
\newblock
\showISBNx{978-1-4503-5620-6}
\urldef\tempurl%
\url{https://doi.org/10.1145/3173574.3173958}
\showDOI{\tempurl}


\bibitem[Chen et~al\mbox{.}(2022)]%
        {chenDevelopmentValidationMachine2022a}
\bibfield{author}{\bibinfo{person}{Kevin~A. Chen}, \bibinfo{person}{Chinmaya~U.
  Joisa}, \bibinfo{person}{Karyn~B. Stitzenberg}, \bibinfo{person}{Jonathan
  Stem}, \bibinfo{person}{Jose~G. Guillem}, \bibinfo{person}{Shawn~M. Gomez},
  {and} \bibinfo{person}{Muneera~R. Kapadia}.} \bibinfo{year}{2022}\natexlab{}.
\newblock \showarticletitle{Development and {{Validation}} of {{Machine
  Learning Models}} to {{Predict Readmission After Colorectal Surgery}}}.
\newblock \bibinfo{journal}{\emph{Journal of Gastrointestinal Surgery}}
  \bibinfo{volume}{26}, \bibinfo{number}{11} (\bibinfo{date}{Nov.}
  \bibinfo{year}{2022}), \bibinfo{pages}{2342--2350}.
\newblock
\showISSN{1873-4626}
\urldef\tempurl%
\url{https://doi.org/10.1007/s11605-022-05443-5}
\showDOI{\tempurl}


\bibitem[Chen et~al\mbox{.}(2024)]%
        {chen2024burextract}
\bibfield{author}{\bibinfo{person}{Yuxuan Chen}, \bibinfo{person}{Haoyan Yang},
  \bibinfo{person}{Hengkai Pan}, \bibinfo{person}{Fardeen Siddiqui},
  \bibinfo{person}{Antonio Verdone}, \bibinfo{person}{Qingyang Zhang},
  \bibinfo{person}{Sumit Chopra}, \bibinfo{person}{Chen Zhao}, {and}
  \bibinfo{person}{Yiqiu Shen}.} \bibinfo{year}{2024}\natexlab{}.
\newblock \showarticletitle{Burextract-llama: An llm for clinical concept
  extraction in breast ultrasound reports}. In
  \bibinfo{booktitle}{\emph{Proceedings of the 1st International Workshop on
  Multimedia Computing for Health and Medicine}}. \bibinfo{pages}{53--58}.
\newblock


\bibitem[Chira et~al\mbox{.}(2022)]%
        {chira2022affective}
\bibfield{author}{\bibinfo{person}{Chloe Chira}, \bibinfo{person}{Evangelos
  Mathioudis}, \bibinfo{person}{Christina Michailidou},
  \bibinfo{person}{Pantelis Agathangelou}, \bibinfo{person}{Georgia
  Christodoulou}, \bibinfo{person}{Ioannis Katakis},
  \bibinfo{person}{Efstratios Kontopoulos}, {and} \bibinfo{person}{Konstantinos
  Avgerinakis}.} \bibinfo{year}{2022}\natexlab{}.
\newblock \showarticletitle{An Affective Multi-modal Conversational Agent for
  Non Intrusive Data Collection from Patients with Brain Diseases}. In
  \bibinfo{booktitle}{\emph{International Workshop on Chatbot Research and
  Design}}. Springer, \bibinfo{pages}{134--149}.
\newblock


\bibitem[Choudhury and Asan(2020)]%
        {choudhury2020role}
\bibfield{author}{\bibinfo{person}{Avishek Choudhury} {and}
  \bibinfo{person}{Onur Asan}.} \bibinfo{year}{2020}\natexlab{}.
\newblock \showarticletitle{Role of {{Artificial Intelligence}} in {{Patient
  Safety Outcomes}}: {{Systematic Literature Review}}}.
\newblock \bibinfo{journal}{\emph{JMIR Medical Informatics}}
  \bibinfo{volume}{8}, \bibinfo{number}{7} (\bibinfo{date}{July}
  \bibinfo{year}{2020}), \bibinfo{pages}{e18599}.
\newblock
\urldef\tempurl%
\url{https://doi.org/10.2196/18599}
\showDOI{\tempurl}


\bibitem[Chung et~al\mbox{.}(2016)]%
        {chung2016boundary}
\bibfield{author}{\bibinfo{person}{Chia-Fang Chung}, \bibinfo{person}{Kristin
  Dew}, \bibinfo{person}{Allison Cole}, \bibinfo{person}{Jasmine Zia},
  \bibinfo{person}{James Fogarty}, \bibinfo{person}{Julie~A Kientz}, {and}
  \bibinfo{person}{Sean~A Munson}.} \bibinfo{year}{2016}\natexlab{}.
\newblock \showarticletitle{Boundary negotiating artifacts in personal
  informatics: patient-provider collaboration with patient-generated data}. In
  \bibinfo{booktitle}{\emph{Proceedings of the 19th ACM conference on
  computer-supported cooperative work \& social computing}}.
  \bibinfo{pages}{770--786}.
\newblock


\bibitem[Clark and Kelliher(2021)]%
        {clark_understanding_2021}
\bibfield{author}{\bibinfo{person}{Juliet Clark} {and} \bibinfo{person}{Aisling
  Kelliher}.} \bibinfo{year}{2021}\natexlab{}.
\newblock \showarticletitle{Understanding the {Needs} and {Values} of
  {Rehabilitation} {Therapists} in {Designing} and {Implementing} {Telehealth}
  {Solutions}}. In \bibinfo{booktitle}{\emph{Extended {Abstracts} of the 2021
  {CHI} {Conference} on {Human} {Factors} in {Computing} {Systems}}}.
  \bibinfo{publisher}{ACM}, \bibinfo{address}{Yokohama Japan},
  \bibinfo{pages}{1--6}.
\newblock
\showISBNx{978-1-4503-8095-9}
\urldef\tempurl%
\url{https://doi.org/10.1145/3411763.3451704}
\showDOI{\tempurl}


\bibitem[Cohen and Mello(2019)]%
        {cohen2019big}
\bibfield{author}{\bibinfo{person}{I~Glenn Cohen} {and}
  \bibinfo{person}{Michelle~M Mello}.} \bibinfo{year}{2019}\natexlab{}.
\newblock \showarticletitle{Big data, big tech, and protecting patient
  privacy}.
\newblock \bibinfo{journal}{\emph{Jama}} \bibinfo{volume}{322},
  \bibinfo{number}{12} (\bibinfo{year}{2019}), \bibinfo{pages}{1141--1142}.
\newblock


\bibitem[Cousins et~al\mbox{.}(2016)]%
        {cousins2016surgery}
\bibfield{author}{\bibinfo{person}{Sarah~E Cousins}, \bibinfo{person}{Emma
  Tempest}, {and} \bibinfo{person}{David~J Feuer}.}
  \bibinfo{year}{2016}\natexlab{}.
\newblock \showarticletitle{Surgery for the resolution of symptoms in malignant
  bowel obstruction in advanced gynaecological and gastrointestinal cancer}.
\newblock \bibinfo{journal}{\emph{Cochrane Database of Systematic Reviews}}
  \bibinfo{number}{1} (\bibinfo{year}{2016}).
\newblock


\bibitem[Desrame et~al\mbox{.}(2024)]%
        {desrame2024595p}
\bibfield{author}{\bibinfo{person}{J Desrame}, \bibinfo{person}{V Heinschild},
  \bibinfo{person}{C Desauw}, \bibinfo{person}{AC Fuerea}, \bibinfo{person}{P
  Artru}, \bibinfo{person}{S Javed}, \bibinfo{person}{T Papazyan},
  \bibinfo{person}{C Fert{\'e}}, \bibinfo{person}{M Autheman},
  \bibinfo{person}{M Valery}, {et~al\mbox{.}}} \bibinfo{year}{2024}\natexlab{}.
\newblock \showarticletitle{595P Adoption of remote patient monitoring in
  gastrointestinal oncology: A real-world experience from 1822 patients across
  47 centers in France and Belgium}.
\newblock \bibinfo{journal}{\emph{Annals of Oncology}}  \bibinfo{volume}{35}
  (\bibinfo{year}{2024}), \bibinfo{pages}{S478}.
\newblock


\bibitem[Dignum(2019)]%
        {dignum2019responsible}
\bibfield{author}{\bibinfo{person}{Virginia Dignum}.}
  \bibinfo{year}{2019}\natexlab{}.
\newblock \bibinfo{booktitle}{\emph{Responsible artificial intelligence: how to
  develop and use AI in a responsible way}}. Vol.~\bibinfo{volume}{2156}.
\newblock \bibinfo{publisher}{Springer}.
\newblock


\bibitem[Dreher et~al\mbox{.}(2019)]%
        {dreherFitbitUsagePatients2019}
\bibfield{author}{\bibinfo{person}{Nickolas Dreher},
  \bibinfo{person}{Edward~Kenji Hadeler}, \bibinfo{person}{Sheri~J. Hartman},
  \bibinfo{person}{Emily~C. Wong}, \bibinfo{person}{Irene Acerbi},
  \bibinfo{person}{Hope~S. Rugo}, \bibinfo{person}{Melanie~Catherine Majure},
  \bibinfo{person}{Amy~Jo Chien}, \bibinfo{person}{Laura~J. Esserman}, {and}
  \bibinfo{person}{Michelle~E. Melisko}.} \bibinfo{year}{2019}\natexlab{}.
\newblock \showarticletitle{Fitbit {{Usage}} in {{Patients With Breast Cancer
  Undergoing Chemotherapy}}}.
\newblock \bibinfo{journal}{\emph{Clinical Breast Cancer}}
  \bibinfo{volume}{19}, \bibinfo{number}{6} (\bibinfo{date}{Dec.}
  \bibinfo{year}{2019}), \bibinfo{pages}{443--449.e1}.
\newblock
\showISSN{1526-8209}
\urldef\tempurl%
\url{https://doi.org/10.1016/j.clbc.2019.05.005}
\showDOI{\tempurl}


\bibitem[Drozdal et~al\mbox{.}(2020)]%
        {drozdal2020}
\bibfield{author}{\bibinfo{person}{Jaimie Drozdal}, \bibinfo{person}{Justin~D.
  Weisz}, \bibinfo{person}{Dakuo Wang}, \bibinfo{person}{Gaurav Dass},
  \bibinfo{person}{Bingsheng Yao}, \bibinfo{person}{Changruo Zhao},
  \bibinfo{person}{Michael Muller}, \bibinfo{person}{Lin Ju}, {and}
  \bibinfo{person}{Hui Su}.} \bibinfo{year}{2020}\natexlab{}.
\newblock \showarticletitle{Trust in AutoML: Exploring Information Needs for
  Establishing Trust in Automated Machine Learning Systems}. In
  \bibinfo{booktitle}{\emph{Proceedings of the 25th International Conference on
  Intelligent User Interfaces}}. \bibinfo{pages}{297--307}.
\newblock
\urldef\tempurl%
\url{https://doi.org/10.1145/3377325.3377501}
\showDOI{\tempurl}


\bibitem[Dwyer et~al\mbox{.}(2023)]%
        {dwyer2023use}
\bibfield{author}{\bibinfo{person}{Tim Dwyer}, \bibinfo{person}{Graeme Hoit},
  \bibinfo{person}{David Burns}, \bibinfo{person}{James Higgins},
  \bibinfo{person}{Justin Chang}, \bibinfo{person}{Daniel Whelan},
  \bibinfo{person}{Irene Kiroplis}, {and} \bibinfo{person}{Jaskarndip Chahal}.}
  \bibinfo{year}{2023}\natexlab{}.
\newblock \showarticletitle{Use of an artificial intelligence conversational
  agent (chatbot) for hip arthroscopy patients following surgery}.
\newblock \bibinfo{journal}{\emph{Arthroscopy, Sports Medicine, and
  Rehabilitation}} \bibinfo{volume}{5}, \bibinfo{number}{2}
  (\bibinfo{year}{2023}), \bibinfo{pages}{e495--e505}.
\newblock


\bibitem[Edu et~al\mbox{.}(2022)]%
        {eduMeasuringAlexaSkill2022}
\bibfield{author}{\bibinfo{person}{Jide Edu}, \bibinfo{person}{Xavier
  {Ferrer-Aran}}, \bibinfo{person}{Jose Such}, {and} \bibinfo{person}{Guillermo
  {Suarez-Tangil}}.} \bibinfo{year}{2022}\natexlab{}.
\newblock \showarticletitle{Measuring {{Alexa Skill Privacy Practices}} across
  {{Three Years}}}. In \bibinfo{booktitle}{\emph{Proceedings of the {{ACM Web
  Conference}} 2022}} \emph{(\bibinfo{series}{{{WWW}} '22})}.
  \bibinfo{publisher}{Association for Computing Machinery},
  \bibinfo{address}{New York, NY, USA}, \bibinfo{pages}{670--680}.
\newblock
\showISBNx{978-1-4503-9096-5}
\urldef\tempurl%
\url{https://doi.org/10.1145/3485447.3512289}
\showDOI{\tempurl}


\bibitem[Ferrara(2024)]%
        {ferrara2024large}
\bibfield{author}{\bibinfo{person}{Emilio Ferrara}.}
  \bibinfo{year}{2024}\natexlab{}.
\newblock \showarticletitle{Large language models for wearable sensor-based
  human activity recognition, health monitoring, and behavioral modeling: A
  survey of early trends, datasets, and challenges}.
\newblock \bibinfo{journal}{\emph{Sensors}} \bibinfo{volume}{24},
  \bibinfo{number}{15} (\bibinfo{year}{2024}), \bibinfo{pages}{5045}.
\newblock


\bibitem[Figma(2024)]%
        {figma}
\bibfield{author}{\bibinfo{person}{Inc. Figma}.}
  \bibinfo{year}{2024}\natexlab{}.
\newblock \bibinfo{booktitle}{\emph{Figma: the collaborative interface design
  tool}}.
\newblock
\urldef\tempurl%
\url{https://www.figma.com}
\showURL{%
\tempurl}


\bibitem[Fu et~al\mbox{.}(2025)]%
        {fu2025felt}
\bibfield{author}{\bibinfo{person}{Shihan Fu}, \bibinfo{person}{Bingsheng Yao},
  \bibinfo{person}{Smit Desai}, \bibinfo{person}{Yuqi Hu},
  \bibinfo{person}{Yuling Sun}, \bibinfo{person}{Samantha Stonbraker},
  \bibinfo{person}{Yanjun Gao}, \bibinfo{person}{Elizabeth~M Goldberg}, {and}
  \bibinfo{person}{Dakuo Wang}.} \bibinfo{year}{2025}\natexlab{}.
\newblock \showarticletitle{" It Felt Like I Was Left in the Dark": Exploring
  Information Needs and Design Opportunities for Family Caregivers of Older
  Adult Patients in Critical Care Settings}.
\newblock \bibinfo{journal}{\emph{arXiv preprint arXiv:2502.05115}}
  (\bibinfo{year}{2025}).
\newblock


\bibitem[Garth et~al\mbox{.}(2010)]%
        {garthNutritionalStatusNutrition2010}
\bibfield{author}{\bibinfo{person}{A.~K. Garth}, \bibinfo{person}{C.~M.
  Newsome}, \bibinfo{person}{N. Simmance}, {and} \bibinfo{person}{T.~C.
  Crowe}.} \bibinfo{year}{2010}\natexlab{}.
\newblock \showarticletitle{Nutritional Status, Nutrition Practices and
  Post-Operative Complications in Patients with Gastrointestinal Cancer}.
\newblock \bibinfo{journal}{\emph{Journal of Human Nutrition and Dietetics}}
  \bibinfo{volume}{23}, \bibinfo{number}{4} (\bibinfo{year}{2010}),
  \bibinfo{pages}{393--401}.
\newblock
\showISSN{1365-277X}
\urldef\tempurl%
\url{https://doi.org/10.1111/j.1365-277X.2010.01058.x}
\showDOI{\tempurl}


\bibitem[Gatt et~al\mbox{.}(2022)]%
        {gatt2022review}
\bibfield{author}{\bibinfo{person}{Michelle~Louise Gatt},
  \bibinfo{person}{Maria Cassar}, {and} \bibinfo{person}{Sandra~C Buttigieg}.}
  \bibinfo{year}{2022}\natexlab{}.
\newblock \showarticletitle{A review of literature on risk prediction tools for
  hospital readmissions in older adults}.
\newblock \bibinfo{journal}{\emph{Journal of Health Organization and
  Management}} \bibinfo{volume}{36}, \bibinfo{number}{4}
  (\bibinfo{year}{2022}), \bibinfo{pages}{521--557}.
\newblock


\bibitem[Geoghegan et~al\mbox{.}(2021)]%
        {geoghegan2021automated}
\bibfield{author}{\bibinfo{person}{L Geoghegan}, \bibinfo{person}{A
  Scarborough}, \bibinfo{person}{JCR Wormald}, \bibinfo{person}{CJ Harrison},
  \bibinfo{person}{D Collins}, \bibinfo{person}{M Gardiner},
  \bibinfo{person}{Julie Bruce}, {and} \bibinfo{person}{JN Rodrigues}.}
  \bibinfo{year}{2021}\natexlab{}.
\newblock \showarticletitle{Automated conversational agents for
  post-intervention follow-up: a systematic review}.
\newblock \bibinfo{journal}{\emph{BJS open}} \bibinfo{volume}{5},
  \bibinfo{number}{4} (\bibinfo{year}{2021}), \bibinfo{pages}{zrab070}.
\newblock


\bibitem[Ghods et~al\mbox{.}(2021)]%
        {ghods2021remote}
\bibfield{author}{\bibinfo{person}{Alireza Ghods}, \bibinfo{person}{Armin
  Shahrokni}, \bibinfo{person}{Hassan Ghasemzadeh}, {and}
  \bibinfo{person}{Diane Cook}.} \bibinfo{year}{2021}\natexlab{}.
\newblock \showarticletitle{Remote monitoring of the performance status and
  burden of symptoms of patients with gastrointestinal cancer via a
  consumer-based activity tracker: quantitative cohort study}.
\newblock \bibinfo{journal}{\emph{JMIR cancer}} \bibinfo{volume}{7},
  \bibinfo{number}{4} (\bibinfo{year}{2021}), \bibinfo{pages}{e22931}.
\newblock


\bibitem[Gon{\c c}alves-Bradley et~al\mbox{.}(2020)]%
        {goncalves-bradley_mobile_2020}
\bibfield{author}{\bibinfo{person}{Daniela~C. Gon{\c c}alves-Bradley},
  \bibinfo{person}{Ana Rita~J. Maria}, \bibinfo{person}{Ignacio Ricci-Cabello},
  \bibinfo{person}{Gemma Villanueva}, \bibinfo{person}{Marita~S. F{\o}nhus},
  \bibinfo{person}{Claire Glenton}, \bibinfo{person}{Simon Lewin},
  \bibinfo{person}{Nicholas Henschke}, \bibinfo{person}{Brian~S. Buckley},
  \bibinfo{person}{Garrett~L. Mehl}, \bibinfo{person}{Tigest Tamrat}, {and}
  \bibinfo{person}{Sasha Shepperd}.} \bibinfo{year}{2020}\natexlab{}.
\newblock \showarticletitle{Mobile technologies to support healthcare provider
  to healthcare provider communication and management of care}.
\newblock \bibinfo{journal}{\emph{Cochrane Database of Systematic Reviews}}
  \bibinfo{number}{8} (\bibinfo{year}{2020}).
\newblock
\showISSN{1465-1858}
\urldef\tempurl%
\url{https://doi.org/10.1002/14651858.CD012927.pub2}
\showDOI{\tempurl}
\newblock
\shownote{Publisher: John Wiley \& Sons, Ltd}.


\bibitem[Hamer et~al\mbox{.}(2023)]%
        {hamer_improving_2023}
\bibfield{author}{\bibinfo{person}{Danny M.~den Hamer}, \bibinfo{person}{Perry
  Schoor}, \bibinfo{person}{Tobias~B. Polak}, {and} \bibinfo{person}{Daniel
  Kapitan}.} \bibinfo{year}{2023}\natexlab{}.
\newblock \bibinfo{title}{Improving {Patient} {Pre}-screening for {Clinical}
  {Trials}: {Assisting} {Physicians} with {Large} {Language} {Models}}.
\newblock
\newblock
\urldef\tempurl%
\url{http://arxiv.org/abs/2304.07396}
\showURL{%
\tempurl}
\newblock
\shownote{arXiv:2304.07396 [cs]}.


\bibitem[Hao et~al\mbox{.}(2024a)]%
        {hao2024outlining}
\bibfield{author}{\bibinfo{person}{Yuexing Hao}, \bibinfo{person}{Jason
  Holmes}, \bibinfo{person}{Mark Waddle}, \bibinfo{person}{Nathan Yu},
  \bibinfo{person}{Kirstin Vickers}, \bibinfo{person}{Heather Preston},
  \bibinfo{person}{Drew Margolin}, \bibinfo{person}{Corinna~E L{\"o}ckenhoff},
  \bibinfo{person}{Aditya Vashistha}, \bibinfo{person}{Marzyeh Ghassemi},
  {et~al\mbox{.}}} \bibinfo{year}{2024}\natexlab{a}.
\newblock \showarticletitle{Outlining the Borders for LLM Applications in
  Patient Education: Developing an Expert-in-the-Loop LLM-Powered Chatbot for
  Prostate Cancer Patient Education}.
\newblock \bibinfo{journal}{\emph{arXiv preprint arXiv:2409.19100}}
  (\bibinfo{year}{2024}).
\newblock


\bibitem[Hao et~al\mbox{.}(2024b)]%
        {hao2024advancing}
\bibfield{author}{\bibinfo{person}{Yuexing Hao}, \bibinfo{person}{Zeyu Liu},
  \bibinfo{person}{Robert~N Riter}, {and} \bibinfo{person}{Saleh Kalantari}.}
  \bibinfo{year}{2024}\natexlab{b}.
\newblock \showarticletitle{Advancing Patient-Centered Shared Decision-Making
  with AI Systems for Older Adult Cancer Patients}. In
  \bibinfo{booktitle}{\emph{Proceedings of the CHI Conference on Human Factors
  in Computing Systems}}. \bibinfo{pages}{1--20}.
\newblock


\bibitem[Healey and Kohane(2024)]%
        {healey2024llm}
\bibfield{author}{\bibinfo{person}{Elizabeth Healey} {and}
  \bibinfo{person}{Isaac Kohane}.} \bibinfo{year}{2024}\natexlab{}.
\newblock \showarticletitle{LLM-CGM: A Benchmark for Large Language
  Model-Enabled Querying of Continuous Glucose Monitoring Data for
  Conversational Diabetes Management}. In
  \bibinfo{booktitle}{\emph{Biocomputing 2025: Proceedings of the Pacific
  Symposium}}. World Scientific, \bibinfo{pages}{82--93}.
\newblock


\bibitem[Ho et~al\mbox{.}(2024a)]%
        {ho2024remoni}
\bibfield{author}{\bibinfo{person}{Thanh~Cong Ho}, \bibinfo{person}{Farah
  Kharrat}, \bibinfo{person}{Abderrazek Abid}, \bibinfo{person}{Fakhri Karray},
  {and} \bibinfo{person}{Anis Koubaa}.} \bibinfo{year}{2024}\natexlab{a}.
\newblock \showarticletitle{REMONI: An Autonomous System Integrating Wearables
  and Multimodal Large Language Models for Enhanced Remote Health Monitoring}.
  In \bibinfo{booktitle}{\emph{2024 IEEE International Symposium on Medical
  Measurements and Applications (MeMeA)}}. IEEE, \bibinfo{pages}{1--6}.
\newblock


\bibitem[Ho et~al\mbox{.}(2024b)]%
        {hoREMONIAutonomousSystem2024}
\bibfield{author}{\bibinfo{person}{Thanh~Cong Ho}, \bibinfo{person}{Farah
  Kharrat}, \bibinfo{person}{Abderrazek Abid}, \bibinfo{person}{Fakhri Karray},
  {and} \bibinfo{person}{Anis Koubaa}.} \bibinfo{year}{2024}\natexlab{b}.
\newblock \showarticletitle{{{REMONI}}: {{An Autonomous System Integrating
  Wearables}} and {{Multimodal Large Language Models}} for {{Enhanced Remote
  Health Monitoring}}}. In \bibinfo{booktitle}{\emph{2024 {{IEEE International
  Symposium}} on {{Medical Measurements}} and {{Applications}} ({{MeMeA}})}}.
  \bibinfo{pages}{1--6}.
\newblock
\showISSN{2837-5882}
\urldef\tempurl%
\url{https://doi.org/10.1109/MeMeA60663.2024.10596778}
\showDOI{\tempurl}


\bibitem[Hämäläinen et~al\mbox{.}(2023)]%
        {hamalainen_evaluating_2023}
\bibfield{author}{\bibinfo{person}{Perttu Hämäläinen},
  \bibinfo{person}{Mikke Tavast}, {and} \bibinfo{person}{Anton Kunnari}.}
  \bibinfo{year}{2023}\natexlab{}.
\newblock \showarticletitle{Evaluating {Large} {Language} {Models} in
  {Generating} {Synthetic} {HCI} {Research} {Data}: a {Case} {Study}}. In
  \bibinfo{booktitle}{\emph{Proceedings of the 2023 {CHI} {Conference} on
  {Human} {Factors} in {Computing} {Systems}}}. \bibinfo{publisher}{ACM},
  \bibinfo{address}{Hamburg Germany}, \bibinfo{pages}{1--19}.
\newblock
\showISBNx{978-1-4503-9421-5}
\urldef\tempurl%
\url{https://doi.org/10.1145/3544548.3580688}
\showDOI{\tempurl}


\bibitem[Jacobs et~al\mbox{.}(2018)]%
        {jacobsMyPathInvestigatingBreast2018}
\bibfield{author}{\bibinfo{person}{Maia Jacobs}, \bibinfo{person}{Jeremy
  Johnson}, {and} \bibinfo{person}{Elizabeth~D. Mynatt}.}
  \bibinfo{year}{2018}\natexlab{}.
\newblock \showarticletitle{{{MyPath}}: {{Investigating Breast Cancer
  Patients}}' {{Use}} of {{Personalized Health Information}}}.
\newblock \bibinfo{journal}{\emph{Proceedings of the ACM on Human-Computer
  Interaction}} \bibinfo{volume}{2}, \bibinfo{number}{CSCW}
  (\bibinfo{date}{Nov.} \bibinfo{year}{2018}), \bibinfo{pages}{78:1--78:21}.
\newblock
\urldef\tempurl%
\url{https://doi.org/10.1145/3274347}
\showDOI{\tempurl}


\bibitem[Jo et~al\mbox{.}(2023)]%
        {jo_understanding_2023}
\bibfield{author}{\bibinfo{person}{Eunkyung Jo}, \bibinfo{person}{Daniel~A.
  Epstein}, \bibinfo{person}{Hyunhoon Jung}, {and} \bibinfo{person}{Young-Ho
  Kim}.} \bibinfo{year}{2023}\natexlab{}.
\newblock \showarticletitle{Understanding the {Benefits} and {Challenges} of
  {Deploying} {Conversational} {AI} {Leveraging} {Large} {Language} {Models}
  for {Public} {Health} {Intervention}}. In
  \bibinfo{booktitle}{\emph{Proceedings of the 2023 {CHI} {Conference} on
  {Human} {Factors} in {Computing} {Systems}}}. \bibinfo{publisher}{ACM},
  \bibinfo{address}{Hamburg Germany}, \bibinfo{pages}{1--16}.
\newblock
\showISBNx{978-1-4503-9421-5}
\urldef\tempurl%
\url{https://doi.org/10.1145/3544548.3581503}
\showDOI{\tempurl}


\bibitem[Kennedy et~al\mbox{.}(2014)]%
        {kennedy2014delirium}
\bibfield{author}{\bibinfo{person}{Maura Kennedy}, \bibinfo{person}{Richard~A
  Enander}, \bibinfo{person}{Sarah~P Tadiri}, \bibinfo{person}{Richard~E
  Wolfe}, \bibinfo{person}{Nathan~I Shapiro}, {and} \bibinfo{person}{Edward~R
  Marcantonio}.} \bibinfo{year}{2014}\natexlab{}.
\newblock \showarticletitle{Delirium risk prediction, healthcare use and
  mortality of elderly adults in the emergency department}.
\newblock \bibinfo{journal}{\emph{Journal of the American Geriatrics Society}}
  \bibinfo{volume}{62}, \bibinfo{number}{3} (\bibinfo{year}{2014}),
  \bibinfo{pages}{462--469}.
\newblock


\bibitem[Kerasidou et~al\mbox{.}(2022)]%
        {kerasidouTrustRelianceMedical2022}
\bibfield{author}{\bibinfo{person}{Charalampia~(Xaroula) Kerasidou},
  \bibinfo{person}{Angeliki Kerasidou}, \bibinfo{person}{Monika Buscher}, {and}
  \bibinfo{person}{Stephen Wilkinson}.} \bibinfo{year}{2022}\natexlab{}.
\newblock \showarticletitle{Before and beyond Trust: Reliance in Medical
  {{AI}}}.
\newblock \bibinfo{journal}{\emph{Journal of Medical Ethics}}
  \bibinfo{volume}{48}, \bibinfo{number}{11} (\bibinfo{date}{Nov.}
  \bibinfo{year}{2022}), \bibinfo{pages}{852--856}.
\newblock
\showISSN{0306-6800, 1473-4257}
\urldef\tempurl%
\url{https://doi.org/10.1136/medethics-2020-107095}
\showDOI{\tempurl}


\bibitem[Kim et~al\mbox{.}(2025)]%
        {kim2025Medical}
\bibfield{author}{\bibinfo{person}{Yubin Kim}, \bibinfo{person}{Hyewon Jeong},
  \bibinfo{person}{Shan Chen}, \bibinfo{person}{Shuyue~Stella Li},
  \bibinfo{person}{Mingyu Lu}, \bibinfo{person}{Kumail Alhamoud},
  \bibinfo{person}{Jimin Mun}, \bibinfo{person}{Cristina Grau},
  \bibinfo{person}{Minseok Jung}, \bibinfo{person}{Rodrigo Gameiro},
  \bibinfo{person}{Lizhou Fan}, \bibinfo{person}{Eugene Park},
  \bibinfo{person}{Tristan Lin}, \bibinfo{person}{Joonsik Yoon},
  \bibinfo{person}{Wonjin Yoon}, \bibinfo{person}{Maarten Sap},
  \bibinfo{person}{Yulia Tsvetkov}, \bibinfo{person}{Paul Liang},
  \bibinfo{person}{Xuhai Xu}, {and} \bibinfo{person}{Xin Liu}.}
  \bibinfo{year}{2025}\natexlab{}.
\newblock \bibinfo{title}{Medical Hallucinations in Foundation Models and Their
  Impact on Healthcare}.
\newblock
\newblock
\urldef\tempurl%
\url{https://arxiv.org/abs/2503.05777}
\showURL{%
\tempurl}


\bibitem[King et~al\mbox{.}(2019)]%
        {kingMicroStressEMAPassive2019}
\bibfield{author}{\bibinfo{person}{Zachary~D. King}, \bibinfo{person}{Judith
  Moskowitz}, \bibinfo{person}{Begum Egilmez}, \bibinfo{person}{Shibo Zhang},
  \bibinfo{person}{Lida Zhang}, \bibinfo{person}{Michael Bass},
  \bibinfo{person}{John Rogers}, \bibinfo{person}{Roozbeh Ghaffari},
  \bibinfo{person}{Laurie Wakschlag}, {and} \bibinfo{person}{Nabil Alshurafa}.}
  \bibinfo{year}{2019}\natexlab{}.
\newblock \showarticletitle{Micro-{{Stress EMA}}: {{A Passive Sensing
  Framework}} for {{Detecting}} in-the-Wild {{Stress}} in {{Pregnant
  Mothers}}}.
\newblock \bibinfo{journal}{\emph{Proceedings of the ACM on Interactive,
  Mobile, Wearable and Ubiquitous Technologies}} \bibinfo{volume}{3},
  \bibinfo{number}{3} (\bibinfo{date}{Sept.} \bibinfo{year}{2019}),
  \bibinfo{pages}{91:1--91:22}.
\newblock
\urldef\tempurl%
\url{https://doi.org/10.1145/3351249}
\showDOI{\tempurl}


\bibitem[Kocaballi et~al\mbox{.}(2020)]%
        {kocaballi_envisioning_2020}
\bibfield{author}{\bibinfo{person}{A~Baki Kocaballi}, \bibinfo{person}{Kiran
  Ijaz}, \bibinfo{person}{Liliana Laranjo}, \bibinfo{person}{Juan~C Quiroz},
  \bibinfo{person}{Dana Rezazadegan}, \bibinfo{person}{Huong~Ly Tong},
  \bibinfo{person}{Simon Willcock}, \bibinfo{person}{Shlomo Berkovsky}, {and}
  \bibinfo{person}{Enrico Coiera}.} \bibinfo{year}{2020}\natexlab{}.
\newblock \showarticletitle{Envisioning an artificial intelligence
  documentation assistant for future primary care consultations: {A} co-design
  study with general practitioners}.
\newblock \bibinfo{journal}{\emph{Journal of the American Medical Informatics
  Association}} \bibinfo{volume}{27}, \bibinfo{number}{11}
  (\bibinfo{date}{Nov.} \bibinfo{year}{2020}), \bibinfo{pages}{1695--1704}.
\newblock


\bibitem[Kruse et~al\mbox{.}(2025)]%
        {kruse2025large}
\bibfield{author}{\bibinfo{person}{Maya Kruse}, \bibinfo{person}{Shiyue Hu},
  \bibinfo{person}{Nicholas Derby}, \bibinfo{person}{Yifu Wu},
  \bibinfo{person}{Samantha Stonbraker}, \bibinfo{person}{Bingsheng Yao},
  \bibinfo{person}{Dakuo Wang}, \bibinfo{person}{Elizabeth Goldberg}, {and}
  \bibinfo{person}{Yanjun Gao}.} \bibinfo{year}{2025}\natexlab{}.
\newblock \showarticletitle{Large Language Models with Temporal Reasoning for
  Longitudinal Clinical Summarization and Prediction}. In
  \bibinfo{booktitle}{\emph{Findings of ACL. EMNLP. Conference on Empirical
  Methods in Natural Language Processing}}, Vol.~\bibinfo{volume}{2025}.
  \bibinfo{pages}{20715--20735}.
\newblock


\bibitem[Lai-Kwon et~al\mbox{.}(2024)]%
        {lai2024co}
\bibfield{author}{\bibinfo{person}{Julia Lai-Kwon}, \bibinfo{person}{Claudia
  Rutherford}, \bibinfo{person}{Stephanie Best}, \bibinfo{person}{Thai Ly},
  \bibinfo{person}{Iris Zhang}, \bibinfo{person}{Catherine Devereux},
  \bibinfo{person}{Dishan Herath}, \bibinfo{person}{Kate Burbury}, {and}
  \bibinfo{person}{Michael Jefford}.} \bibinfo{year}{2024}\natexlab{}.
\newblock \showarticletitle{Co-design of an electronic patient-reported outcome
  symptom monitoring system for immunotherapy toxicities}.
\newblock \bibinfo{journal}{\emph{Supportive Care in Cancer}}
  \bibinfo{volume}{32}, \bibinfo{number}{12} (\bibinfo{year}{2024}),
  \bibinfo{pages}{1--14}.
\newblock


\bibitem[Lee et~al\mbox{.}(2019)]%
        {leeBioBERTPretrainedBiomedical2019}
\bibfield{author}{\bibinfo{person}{Jinhyuk Lee}, \bibinfo{person}{Wonjin Yoon},
  \bibinfo{person}{Sungdong Kim}, \bibinfo{person}{Donghyeon Kim},
  \bibinfo{person}{Sunkyu Kim}, \bibinfo{person}{Chan~Ho So}, {and}
  \bibinfo{person}{Jaewoo Kang}.} \bibinfo{year}{2019}\natexlab{}.
\newblock \showarticletitle{{{BioBERT}}: A Pre-Trained Biomedical Language
  Representation Model for Biomedical Text Mining}.
\newblock \bibinfo{journal}{\emph{Bioinformatics}} (\bibinfo{date}{Sept.}
  \bibinfo{year}{2019}), \bibinfo{pages}{btz682}.
\newblock
\showISSN{1367-4803, 1460-2059}
\urldef\tempurl%
\url{https://doi.org/10/ggh5qq}
\showDOI{\tempurl}


\bibitem[Lee et~al\mbox{.}(2023)]%
        {lee2023benefits}
\bibfield{author}{\bibinfo{person}{Peter Lee}, \bibinfo{person}{Sebastien
  Bubeck}, {and} \bibinfo{person}{Joseph Petro}.}
  \bibinfo{year}{2023}\natexlab{}.
\newblock \showarticletitle{Benefits, {{Limits}}, and {{Risks}} of {{GPT-4}} as
  an {{AI Chatbot}} for {{Medicine}}}.
\newblock \bibinfo{journal}{\emph{New England Journal of Medicine}}
  \bibinfo{volume}{388}, \bibinfo{number}{13} (\bibinfo{date}{March}
  \bibinfo{year}{2023}), \bibinfo{pages}{1233--1239}.
\newblock
\showISSN{0028-4793}
\urldef\tempurl%
\url{https://doi.org/10.1056/NEJMsr2214184}
\showDOI{\tempurl}


\bibitem[Le{\'o}n et~al\mbox{.}(2022)]%
        {leon2022impact}
\bibfield{author}{\bibinfo{person}{Maria~Alejandra Le{\'o}n},
  \bibinfo{person}{Valeria Pannunzio}, {and} \bibinfo{person}{Maaike
  Kleinsmann}.} \bibinfo{year}{2022}\natexlab{}.
\newblock \showarticletitle{The impact of perioperative remote patient
  monitoring on clinical staff workflows: scoping review}.
\newblock \bibinfo{journal}{\emph{JMIR Human Factors}} \bibinfo{volume}{9},
  \bibinfo{number}{2} (\bibinfo{year}{2022}), \bibinfo{pages}{e37204}.
\newblock


\bibitem[Lewis(2018)]%
        {lewis2018system}
\bibfield{author}{\bibinfo{person}{James~R Lewis}.}
  \bibinfo{year}{2018}\natexlab{}.
\newblock \showarticletitle{The system usability scale: past, present, and
  future}.
\newblock \bibinfo{journal}{\emph{International Journal of Human--Computer
  Interaction}} \bibinfo{volume}{34}, \bibinfo{number}{7}
  (\bibinfo{year}{2018}), \bibinfo{pages}{577--590}.
\newblock


\bibitem[Li et~al\mbox{.}(2024)]%
        {li2024beyond}
\bibfield{author}{\bibinfo{person}{Brenna Li}, \bibinfo{person}{Ofek Gross},
  \bibinfo{person}{Noah Crampton}, \bibinfo{person}{Mamta Kapoor},
  \bibinfo{person}{Saba Tauseef}, \bibinfo{person}{Mohit Jain},
  \bibinfo{person}{Khai~N Truong}, {and} \bibinfo{person}{Alex Mariakakis}.}
  \bibinfo{year}{2024}\natexlab{}.
\newblock \showarticletitle{Beyond the Waiting Room: Patient's Perspectives on
  the Conversational Nuances of Pre-Consultation Chatbots}. In
  \bibinfo{booktitle}{\emph{Proceedings of the CHI Conference on Human Factors
  in Computing Systems}}. \bibinfo{pages}{1--24}.
\newblock


\bibitem[Li et~al\mbox{.}(2023)]%
        {li2023meddm}
\bibfield{author}{\bibinfo{person}{Binbin Li}, \bibinfo{person}{Tianxin Meng},
  \bibinfo{person}{Xiaoming Shi}, \bibinfo{person}{Jie Zhai}, {and}
  \bibinfo{person}{Tong Ruan}.} \bibinfo{year}{2023}\natexlab{}.
\newblock \showarticletitle{Meddm: Llm-executable clinical guidance tree for
  clinical decision-making}.
\newblock \bibinfo{journal}{\emph{arXiv preprint arXiv:2312.02441}}
  (\bibinfo{year}{2023}).
\newblock


\bibitem[Lin et~al\mbox{.}(2024)]%
        {lin2024hevelius}
\bibfield{author}{\bibinfo{person}{Hongjin Lin}, \bibinfo{person}{Tessa Han},
  \bibinfo{person}{Krzysztof~Z Gajos}, {and} \bibinfo{person}{Anoopum~S
  Gupta}.} \bibinfo{year}{2024}\natexlab{}.
\newblock \showarticletitle{Hevelius Report: Visualizing Web-Based Mobility
  Test Data For Clinical Decision and Learning Support}. In
  \bibinfo{booktitle}{\emph{Proceedings of the 26th International ACM SIGACCESS
  Conference on Computers and Accessibility}}. \bibinfo{pages}{1--10}.
\newblock


\bibitem[Liu et~al\mbox{.}(2023)]%
        {liuLargeLanguageModels2023}
\bibfield{author}{\bibinfo{person}{Xin Liu}, \bibinfo{person}{Daniel McDuff},
  \bibinfo{person}{Geza Kovacs}, \bibinfo{person}{Isaac {Galatzer-Levy}},
  \bibinfo{person}{Jacob Sunshine}, \bibinfo{person}{Jiening Zhan},
  \bibinfo{person}{Ming-Zher Poh}, \bibinfo{person}{Shun Liao},
  \bibinfo{person}{Paolo Di~Achille}, {and} \bibinfo{person}{Shwetak Patel}.}
  \bibinfo{year}{2023}\natexlab{}.
\newblock \bibinfo{title}{Large {{Language Models}} Are {{Few-Shot Health
  Learners}}}.
\newblock
\newblock
\urldef\tempurl%
\url{https://doi.org/10.48550/arXiv.2305.15525}
\showDOI{\tempurl}
\showeprint[arxiv]{2305.15525}~[cs]


\bibitem[Lu et~al\mbox{.}(2022)]%
        {luContextualEmbeddingModel2022}
\bibfield{author}{\bibinfo{person}{Yuxuan Lu}, \bibinfo{person}{Jingya Yan},
  \bibinfo{person}{Zhixuan Qi}, \bibinfo{person}{Zhongzheng Ge}, {and}
  \bibinfo{person}{Yongping Du}.} \bibinfo{year}{2022}\natexlab{}.
\newblock \showarticletitle{Contextual Embedding and Model Weighting by Fusing
  Domain Knowledge on Biomedical Question Answering}. In
  \bibinfo{booktitle}{\emph{Proceedings of the 13th {{ACM International
  Conference}} on {{Bioinformatics}}, {{Computational Biology}} and {{Health
  Informatics}}}} \emph{(\bibinfo{series}{{{BCB}} '22})}.
  \bibinfo{publisher}{Association for Computing Machinery},
  \bibinfo{address}{New York, NY, USA}, \bibinfo{pages}{1--4}.
\newblock
\showISBNx{978-1-4503-9386-7}
\urldef\tempurl%
\url{https://doi.org/10.1145/3535508.3545508}
\showDOI{\tempurl}


\bibitem[Mahmood et~al\mbox{.}(2023)]%
        {mahmoodLLMPoweredConversationalVoice2023}
\bibfield{author}{\bibinfo{person}{Amama Mahmood}, \bibinfo{person}{Junxiang
  Wang}, \bibinfo{person}{Bingsheng Yao}, \bibinfo{person}{Dakuo Wang}, {and}
  \bibinfo{person}{Chien-Ming Huang}.} \bibinfo{year}{2023}\natexlab{}.
\newblock \bibinfo{title}{{{LLM-Powered Conversational Voice Assistants}}:
  {{Interaction Patterns}}, {{Opportunities}}, {{Challenges}}, and {{Design
  Guidelines}}}.
\newblock
\newblock
\showeprint[arxiv]{2309.13879}~[cs]


\bibitem[Majid et~al\mbox{.}(2025)]%
        {majidMultimodalLLMPatient2025}
\bibfield{author}{\bibinfo{person}{{\relax Abdul}. Majid},
  \bibinfo{person}{{\relax Yulin}. Wang}, \bibinfo{person}{{\relax Jehad}.
  Ali}, \bibinfo{person}{{\relax Anwar}. Ullah}, {and} \bibinfo{person}{{\relax
  Kiran}. Perveen}.} \bibinfo{year}{2025}\natexlab{}.
\newblock \showarticletitle{Multimodal {{LLM}} for {{Patient Activity
  Recognition}}: {{Integrating Video}}, {{Audio}}, and {{Text}} in {{Clinical
  Environments}}}.
\newblock \bibinfo{journal}{\emph{IEEE Journal of Biomedical and Health
  Informatics}} (\bibinfo{year}{2025}), \bibinfo{pages}{1--11}.
\newblock
\showISSN{2168-2208}
\urldef\tempurl%
\url{https://doi.org/10.1109/JBHI.2025.3617581}
\showDOI{\tempurl}


\bibitem[Malasinghe et~al\mbox{.}(2019)]%
        {malasingheRemotePatientMonitoring2019}
\bibfield{author}{\bibinfo{person}{Lakmini~P. Malasinghe},
  \bibinfo{person}{Naeem Ramzan}, {and} \bibinfo{person}{Keshav Dahal}.}
  \bibinfo{year}{2019}\natexlab{}.
\newblock \showarticletitle{Remote Patient Monitoring: A Comprehensive Study}.
\newblock \bibinfo{journal}{\emph{Journal of Ambient Intelligence and Humanized
  Computing}} \bibinfo{volume}{10}, \bibinfo{number}{1} (\bibinfo{date}{Jan.}
  \bibinfo{year}{2019}), \bibinfo{pages}{57--76}.
\newblock
\showISSN{1868-5137, 1868-5145}
\urldef\tempurl%
\url{https://doi.org/10.1007/s12652-017-0598-x}
\showDOI{\tempurl}


\bibitem[Marwaha et~al\mbox{.}(2022)]%
        {marwaha2022deploying}
\bibfield{author}{\bibinfo{person}{Jayson~S Marwaha}, \bibinfo{person}{Adam~B
  Landman}, \bibinfo{person}{Gabriel~A Brat}, \bibinfo{person}{Todd Dunn},
  {and} \bibinfo{person}{William~J Gordon}.} \bibinfo{year}{2022}\natexlab{}.
\newblock \showarticletitle{Deploying digital health tools within large,
  complex health systems: key considerations for adoption and implementation}.
\newblock \bibinfo{journal}{\emph{NPJ digital medicine}} \bibinfo{volume}{5},
  \bibinfo{number}{1} (\bibinfo{year}{2022}), \bibinfo{pages}{13}.
\newblock


\bibitem[Michalopoulos et~al\mbox{.}(2021)]%
        {michalopoulosUmlsBERTClinicalDomain2021}
\bibfield{author}{\bibinfo{person}{George Michalopoulos},
  \bibinfo{person}{Yuanxin Wang}, \bibinfo{person}{Hussam Kaka},
  \bibinfo{person}{Helen Chen}, {and} \bibinfo{person}{Alexander Wong}.}
  \bibinfo{year}{2021}\natexlab{}.
\newblock \bibinfo{title}{{{UmlsBERT}}: {{Clinical Domain Knowledge
  Augmentation}} of {{Contextual Embeddings Using}} the {{Unified Medical
  Language System Metathesaurus}}}.
\newblock
\newblock
\urldef\tempurl%
\url{https://doi.org/10.48550/arXiv.2010.10391}
\showDOI{\tempurl}
\showeprint[arxiv]{2010.10391}~[cs]


\bibitem[Mocan(2021)]%
        {mocanSurgicalManagementGastric2021}
\bibfield{author}{\bibinfo{person}{Lucian Mocan}.}
  \bibinfo{year}{2021}\natexlab{}.
\newblock \showarticletitle{Surgical {{Management}} of {{Gastric Cancer}}: {{A
  Systematic Review}}}.
\newblock \bibinfo{journal}{\emph{Journal of Clinical Medicine}}
  \bibinfo{volume}{10}, \bibinfo{number}{12} (\bibinfo{date}{Jan.}
  \bibinfo{year}{2021}), \bibinfo{pages}{2557}.
\newblock
\showISSN{2077-0383}
\urldef\tempurl%
\url{https://doi.org/10.3390/jcm10122557}
\showDOI{\tempurl}


\bibitem[Montagna et~al\mbox{.}(2023)]%
        {montagna2023data}
\bibfield{author}{\bibinfo{person}{Sara Montagna}, \bibinfo{person}{Stefano
  Ferretti}, \bibinfo{person}{Lorenz~Cuno Klopfenstein},
  \bibinfo{person}{Antonio Florio}, {and} \bibinfo{person}{Martino~Francesco
  Pengo}.} \bibinfo{year}{2023}\natexlab{}.
\newblock \showarticletitle{Data Decentralisation of LLM-Based Chatbot Systems
  in Chronic Disease Self-Management}. In \bibinfo{booktitle}{\emph{Proceedings
  of the 2023 ACM Conference on Information Technology for Social Good}}.
  \bibinfo{pages}{205--212}.
\newblock


\bibitem[Montagna et~al\mbox{.}(2025)]%
        {montagnaPrivacypreservingLLMbasedChatbots2025}
\bibfield{author}{\bibinfo{person}{Sara Montagna}, \bibinfo{person}{Stefano
  Ferretti}, \bibinfo{person}{Lorenz~Cuno Klopfenstein},
  \bibinfo{person}{Michelangelo Ungolo}, \bibinfo{person}{Martino~Francesco
  Pengo}, \bibinfo{person}{Gianluca Aguzzi}, {and} \bibinfo{person}{Matteo
  Magnini}.} \bibinfo{year}{2025}\natexlab{}.
\newblock \showarticletitle{Privacy-Preserving {{LLM-based}} Chatbots for
  Hypertensive Patient Self-Management}.
\newblock \bibinfo{journal}{\emph{Smart Health}}  \bibinfo{volume}{36}
  (\bibinfo{date}{June} \bibinfo{year}{2025}), \bibinfo{pages}{100552}.
\newblock
\showISSN{2352-6483}
\urldef\tempurl%
\url{https://doi.org/10.1016/j.smhl.2025.100552}
\showDOI{\tempurl}


\bibitem[Murdoch(2021)]%
        {murdoch2021privacy}
\bibfield{author}{\bibinfo{person}{Blake Murdoch}.}
  \bibinfo{year}{2021}\natexlab{}.
\newblock \showarticletitle{Privacy and artificial intelligence: challenges for
  protecting health information in a new era}.
\newblock \bibinfo{journal}{\emph{BMC Medical Ethics}}  \bibinfo{volume}{22}
  (\bibinfo{year}{2021}), \bibinfo{pages}{1--5}.
\newblock


\bibitem[Nair et~al\mbox{.}(2023)]%
        {nair_generating_2023}
\bibfield{author}{\bibinfo{person}{Varun Nair}, \bibinfo{person}{Elliot
  Schumacher}, {and} \bibinfo{person}{Anitha Kannan}.}
  \bibinfo{year}{2023}\natexlab{}.
\newblock \bibinfo{title}{Generating medically-accurate summaries of
  patient-provider dialogue: {A} multi-stage approach using large language
  models}.
\newblock
\newblock
\urldef\tempurl%
\url{http://arxiv.org/abs/2305.05982}
\showURL{%
\tempurl}
\newblock
\shownote{arXiv:2305.05982 [cs]}.


\bibitem[Ni et~al\mbox{.}(2017)]%
        {ni2017mandy}
\bibfield{author}{\bibinfo{person}{Lin Ni}, \bibinfo{person}{Chenhao Lu},
  \bibinfo{person}{Niu Liu}, {and} \bibinfo{person}{Jiamou Liu}.}
  \bibinfo{year}{2017}\natexlab{}.
\newblock \showarticletitle{Mandy: Towards a smart primary care chatbot
  application}. In \bibinfo{booktitle}{\emph{International symposium on
  knowledge and systems sciences}}. Springer, \bibinfo{pages}{38--52}.
\newblock


\bibitem[Nori et~al\mbox{.}(2023)]%
        {nori2023capabilities}
\bibfield{author}{\bibinfo{person}{Harsha Nori}, \bibinfo{person}{Nicholas
  King}, \bibinfo{person}{Scott~Mayer McKinney}, \bibinfo{person}{Dean
  Carignan}, {and} \bibinfo{person}{Eric Horvitz}.}
  \bibinfo{year}{2023}\natexlab{}.
\newblock \showarticletitle{Capabilities of gpt-4 on medical challenge
  problems}.
\newblock \bibinfo{journal}{\emph{arXiv preprint arXiv:2303.13375}}
  (\bibinfo{year}{2023}).
\newblock


\bibitem[Nwachukwu et~al\mbox{.}(2024)]%
        {nwachukwu2024currently}
\bibfield{author}{\bibinfo{person}{Benedict~U Nwachukwu},
  \bibinfo{person}{Nathan~H Varady}, \bibinfo{person}{Answorth~A Allen},
  \bibinfo{person}{Joshua~S Dines}, \bibinfo{person}{David~W Altchek},
  \bibinfo{person}{Riley~J Williams~III}, {and} \bibinfo{person}{Kyle~N
  Kunze}.} \bibinfo{year}{2024}\natexlab{}.
\newblock \showarticletitle{Currently available large language models do not
  provide musculoskeletal treatment recommendations that are concordant with
  evidence-based clinical practice guidelines}.
\newblock \bibinfo{journal}{\emph{Arthroscopy: The Journal of Arthroscopic \&
  Related Surgery}} (\bibinfo{year}{2024}).
\newblock


\bibitem[O'Caoimh et~al\mbox{.}(2015)]%
        {o2015risk}
\bibfield{author}{\bibinfo{person}{R{\'o}n{\'a}n O'Caoimh},
  \bibinfo{person}{Nicola Cornally}, \bibinfo{person}{Elizabeth Weathers},
  \bibinfo{person}{Ronan O'Sullivan}, \bibinfo{person}{Carol Fitzgerald},
  \bibinfo{person}{Francesc Orfila}, \bibinfo{person}{Roger Clarnette},
  \bibinfo{person}{Constan{\c{c}}a Pa{\'u}l}, {and} \bibinfo{person}{D~William
  Molloy}.} \bibinfo{year}{2015}\natexlab{}.
\newblock \showarticletitle{Risk prediction in the community: A systematic
  review of case-finding instruments that predict adverse healthcare outcomes
  in community-dwelling older adults}.
\newblock \bibinfo{journal}{\emph{Maturitas}} \bibinfo{volume}{82},
  \bibinfo{number}{1} (\bibinfo{year}{2015}), \bibinfo{pages}{3--21}.
\newblock


\bibitem[Omarov et~al\mbox{.}(2022)]%
        {omarovArtificialIntelligenceEnabled2022}
\bibfield{author}{\bibinfo{person}{Batyrkhan Omarov}, \bibinfo{person}{Sergazi
  Narynov}, \bibinfo{person}{Zhandos Zhumanov}, \bibinfo{person}{Elmira
  Alzhanova}, \bibinfo{person}{Aidana Gumar}, {and} \bibinfo{person}{Mariyam
  Khassanova}.} \bibinfo{year}{2022}\natexlab{}.
\newblock \showarticletitle{Artificial Intelligence Enabled Conversational
  Agent for Mental Healthcare}.
\newblock \bibinfo{journal}{\emph{International journal of health sciences}}
  \bibinfo{volume}{6}, \bibinfo{number}{3} (\bibinfo{date}{Oct.}
  \bibinfo{year}{2022}), \bibinfo{pages}{1544--1555}.
\newblock
\showISSN{2550-696X, 2550-6978}
\urldef\tempurl%
\url{https://doi.org/10.53730/ijhs.v6n3.13239}
\showDOI{\tempurl}


\bibitem[OpenAI(2022)]%
        {OpenAI_2022}
\bibfield{author}{\bibinfo{person}{OpenAI}.} \bibinfo{year}{2022}\natexlab{}.
\newblock \bibinfo{title}{Introducing chatgpt}.
\newblock
\newblock
\urldef\tempurl%
\url{https://openai.com/blog/chatgpt}
\showURL{%
\tempurl}


\bibitem[OpenAI(2023a)]%
        {openai2023gpt4}
\bibfield{author}{\bibinfo{person}{OpenAI}.} \bibinfo{year}{2023}\natexlab{a}.
\newblock \bibinfo{title}{GPT-4}.
\newblock
\newblock
\urldef\tempurl%
\url{https://openai.com/index/gpt-4/}
\showURL{%
\tempurl}
\newblock
\shownote{Accessed: 2025-01-18}.


\bibitem[OpenAI(2023b)]%
        {openaiGPT4TechnicalReport2023}
\bibfield{author}{\bibinfo{person}{OpenAI}.} \bibinfo{year}{2023}\natexlab{b}.
\newblock \bibinfo{title}{{{GPT-4 Technical Report}}}.
\newblock
\newblock
\urldef\tempurl%
\url{https://doi.org/10.48550/arXiv.2303.08774}
\showDOI{\tempurl}
\showeprint[arxiv]{2303.08774}~[cs]


\bibitem[Pannunzio et~al\mbox{.}(2024)]%
        {pannunzio2024patient}
\bibfield{author}{\bibinfo{person}{Valeria Pannunzio},
  \bibinfo{person}{Hosana~Cristina Morales~Ornelas}, \bibinfo{person}{Pema
  Gurung}, \bibinfo{person}{Robert van Kooten}, \bibinfo{person}{Dirk
  Snelders}, \bibinfo{person}{Hendrikus van Os}, \bibinfo{person}{Michel
  Wouters}, \bibinfo{person}{Rob Tollenaar}, \bibinfo{person}{Douwe Atsma},
  {and} \bibinfo{person}{Maaike Kleinsmann}.} \bibinfo{year}{2024}\natexlab{}.
\newblock \showarticletitle{Patient and Staff Experience of Remote Patient
  Monitoring—What to Measure and How: Systematic Review}.
\newblock \bibinfo{journal}{\emph{Journal of Medical Internet Research}}
  \bibinfo{volume}{26} (\bibinfo{year}{2024}), \bibinfo{pages}{e48463}.
\newblock


\bibitem[Progga et~al\mbox{.}(2024)]%
        {progga2024large}
\bibfield{author}{\bibinfo{person}{Farhat~Tasnim Progga}, \bibinfo{person}{Amal
  Khan}, {and} \bibinfo{person}{Sabirat Rubya}.}
  \bibinfo{year}{2024}\natexlab{}.
\newblock \showarticletitle{Large Language Models and Personalized Storytelling
  for Postpartum Wellbeing}. In \bibinfo{booktitle}{\emph{Companion Publication
  of the 2024 Conference on Computer-Supported Cooperative Work and Social
  Computing}} (San Jose, Costa Rica) \emph{(\bibinfo{series}{CSCW Companion
  '24})}. \bibinfo{publisher}{Association for Computing Machinery},
  \bibinfo{address}{New York, NY, USA}, \bibinfo{pages}{653–657}.
\newblock
\showISBNx{9798400711145}
\urldef\tempurl%
\url{https://doi.org/10.1145/3678884.3681921}
\showDOI{\tempurl}


\bibitem[Rajashekar et~al\mbox{.}(2024)]%
        {rajashekar2024human}
\bibfield{author}{\bibinfo{person}{Niroop~Channa Rajashekar},
  \bibinfo{person}{Yeo~Eun Shin}, \bibinfo{person}{Yuan Pu},
  \bibinfo{person}{Sunny Chung}, \bibinfo{person}{Kisung You},
  \bibinfo{person}{Mauro Giuffre}, \bibinfo{person}{Colleen~E Chan},
  \bibinfo{person}{Theo Saarinen}, \bibinfo{person}{Allen Hsiao},
  \bibinfo{person}{Jasjeet Sekhon}, {et~al\mbox{.}}}
  \bibinfo{year}{2024}\natexlab{}.
\newblock \showarticletitle{Human-Algorithmic Interaction Using a Large
  Language Model-Augmented Artificial Intelligence Clinical Decision Support
  System}. In \bibinfo{booktitle}{\emph{Proceedings of the CHI Conference on
  Human Factors in Computing Systems}}. \bibinfo{pages}{1--20}.
\newblock


\bibitem[Salmi et~al\mbox{.}({[n.\,d.]})]%
        {salmiharnessing}
\bibfield{author}{\bibinfo{person}{Liz Salmi}, \bibinfo{person}{Dana~M Lewis},
  \bibinfo{person}{Jennifer~L Clarke}, \bibinfo{person}{Zhiyong Dong},
  \bibinfo{person}{Rudy Fischmann}, \bibinfo{person}{Emily~I McIntosh},
  \bibinfo{person}{Chethan~R Sarabu}, {and} \bibinfo{person}{Catherine~M
  DesRoches}.} \bibinfo{year}{[n.\,d.]}\natexlab{}.
\newblock \showarticletitle{Harnessing AI for Patient Engagement in a Study on
  Large Language Models and Open Notes}.
\newblock  (\bibinfo{year}{[n.\,d.]}).
\newblock


\bibitem[Schroeder et~al\mbox{.}(2017)]%
        {schroeder_supporting_2017}
\bibfield{author}{\bibinfo{person}{Jessica Schroeder}, \bibinfo{person}{Jane
  Hoffswell}, \bibinfo{person}{Chia-Fang Chung}, \bibinfo{person}{James
  Fogarty}, \bibinfo{person}{Sean Munson}, {and} \bibinfo{person}{Jasmine
  Zia}.} \bibinfo{year}{2017}\natexlab{}.
\newblock \showarticletitle{Supporting {Patient}-{Provider} {Collaboration} to
  {Identify} {Individual} {Triggers} using {Food} and {Symptom} {Journals}}. In
  \bibinfo{booktitle}{\emph{Proceedings of the 2017 {ACM} {Conference} on
  {Computer} {Supported} {Cooperative} {Work} and {Social} {Computing}}}
  \emph{(\bibinfo{series}{{CSCW} '17})}. \bibinfo{publisher}{Association for
  Computing Machinery}, \bibinfo{address}{New York, NY, USA},
  \bibinfo{pages}{1726--1739}.
\newblock
\showISBNx{978-1-4503-4335-0}
\urldef\tempurl%
\url{https://doi.org/10.1145/2998181.2998276}
\showDOI{\tempurl}


\bibitem[Semple et~al\mbox{.}(2015)]%
        {sempleUsingMobileApp2015}
\bibfield{author}{\bibinfo{person}{John~L. Semple}, \bibinfo{person}{Sarah
  Sharpe}, \bibinfo{person}{M.~Lucas Murnaghan}, \bibinfo{person}{John
  Theodoropoulos}, {and} \bibinfo{person}{Kelly~A. Metcalfe}.}
  \bibinfo{year}{2015}\natexlab{}.
\newblock \showarticletitle{Using a {{Mobile App}} for {{Monitoring
  Post-Operative Quality}} of {{Recovery}} of {{Patients}} at {{Home}}: {{A
  Feasibility Study}}}.
\newblock \bibinfo{journal}{\emph{JMIR mHealth and uHealth}}
  \bibinfo{volume}{3}, \bibinfo{number}{1} (\bibinfo{date}{Feb.}
  \bibinfo{year}{2015}), \bibinfo{pages}{e3929}.
\newblock
\urldef\tempurl%
\url{https://doi.org/10.2196/mhealth.3929}
\showDOI{\tempurl}


\bibitem[Shen et~al\mbox{.}(2023)]%
        {shen2023convxai}
\bibfield{author}{\bibinfo{person}{Hua Shen}, \bibinfo{person}{Chieh-Yang
  Huang}, \bibinfo{person}{Tongshuang Wu}, {and}
  \bibinfo{person}{Ting-Hao'Kenneth' Huang}.} \bibinfo{year}{2023}\natexlab{}.
\newblock \showarticletitle{ConvXAI: Delivering Heterogeneous AI Explanations
  via Conversations to Support Human-AI Scientific Writing}.
\newblock \bibinfo{journal}{\emph{arXiv preprint arXiv:2305.09770}}
  (\bibinfo{year}{2023}).
\newblock


\bibitem[Silsand and Ellingsen(2016)]%
        {silsand2016complex}
\bibfield{author}{\bibinfo{person}{Line Silsand} {and} \bibinfo{person}{Gunnar
  Ellingsen}.} \bibinfo{year}{2016}\natexlab{}.
\newblock \showarticletitle{Complex Decision-Making in Clinical Practice}. In
  \bibinfo{booktitle}{\emph{Proceedings of the 19th ACM Conference on
  Computer-Supported Cooperative Work \& Social Computing}} (San Francisco,
  California, USA) \emph{(\bibinfo{series}{CSCW '16})}.
  \bibinfo{publisher}{Association for Computing Machinery},
  \bibinfo{address}{New York, NY, USA}, \bibinfo{pages}{993–1004}.
\newblock
\showISBNx{9781450335928}
\urldef\tempurl%
\url{https://doi.org/10.1145/2818048.2819952}
\showDOI{\tempurl}


\bibitem[Silva et~al\mbox{.}(2020)]%
        {silva_ostomy_2020}
\bibfield{author}{\bibinfo{person}{Karine De~Almeida Silva},
  \bibinfo{person}{Arenamoline~Xavier Duarte},
  \bibinfo{person}{Amanda~Rodrigues Cruz}, \bibinfo{person}{Letícia~Oliveira
  Cardoso}, \bibinfo{person}{Thatty Christina~Morais Santos}, {and}
  \bibinfo{person}{Geórgia Das~Graças Pena}.}
  \bibinfo{year}{2020}\natexlab{}.
\newblock \showarticletitle{Ostomy time and nutrition status were associated on
  quality of life in patients with colorectal cancer}.
\newblock \bibinfo{journal}{\emph{Journal of Coloproctology}}
  \bibinfo{volume}{40}, \bibinfo{number}{04} (\bibinfo{date}{Dec.}
  \bibinfo{year}{2020}), \bibinfo{pages}{352--361}.
\newblock
\showISSN{2237-9363, 2317-6423}
\urldef\tempurl%
\url{https://doi.org/10.1016/j.jcol.2020.07.003}
\showDOI{\tempurl}


\bibitem[Singhal et~al\mbox{.}(2023)]%
        {singhalLargeLanguageModels2023}
\bibfield{author}{\bibinfo{person}{Karan Singhal}, \bibinfo{person}{Shekoofeh
  Azizi}, \bibinfo{person}{Tao Tu}, \bibinfo{person}{S.~Sara Mahdavi},
  \bibinfo{person}{Jason Wei}, \bibinfo{person}{Hyung~Won Chung},
  \bibinfo{person}{Nathan Scales}, \bibinfo{person}{Ajay Tanwani},
  \bibinfo{person}{Heather {Cole-Lewis}}, \bibinfo{person}{Stephen Pfohl},
  \bibinfo{person}{Perry Payne}, \bibinfo{person}{Martin Seneviratne},
  \bibinfo{person}{Paul Gamble}, \bibinfo{person}{Chris Kelly},
  \bibinfo{person}{Abubakr Babiker}, \bibinfo{person}{Nathanael Sch{\"a}rli},
  \bibinfo{person}{Aakanksha Chowdhery}, \bibinfo{person}{Philip Mansfield},
  \bibinfo{person}{Dina {Demner-Fushman}}, \bibinfo{person}{Blaise {Ag{\"u}era
  y Arcas}}, \bibinfo{person}{Dale Webster}, \bibinfo{person}{Greg~S. Corrado},
  \bibinfo{person}{Yossi Matias}, \bibinfo{person}{Katherine Chou},
  \bibinfo{person}{Juraj Gottweis}, \bibinfo{person}{Nenad Tomasev},
  \bibinfo{person}{Yun Liu}, \bibinfo{person}{Alvin Rajkomar},
  \bibinfo{person}{Joelle Barral}, \bibinfo{person}{Christopher Semturs},
  \bibinfo{person}{Alan Karthikesalingam}, {and} \bibinfo{person}{Vivek
  Natarajan}.} \bibinfo{year}{2023}\natexlab{}.
\newblock \showarticletitle{Large Language Models Encode Clinical Knowledge}.
\newblock \bibinfo{journal}{\emph{Nature}} \bibinfo{volume}{620},
  \bibinfo{number}{7972} (\bibinfo{date}{Aug.} \bibinfo{year}{2023}),
  \bibinfo{pages}{172--180}.
\newblock
\showISSN{1476-4687}
\urldef\tempurl%
\url{https://doi.org/10.1038/s41586-023-06291-2}
\showDOI{\tempurl}


\bibitem[Smith(2021)]%
        {smith2021clinical}
\bibfield{author}{\bibinfo{person}{Helen Smith}.}
  \bibinfo{year}{2021}\natexlab{}.
\newblock \showarticletitle{Clinical AI: opacity, accountability,
  responsibility and liability}.
\newblock \bibinfo{journal}{\emph{Ai \& Society}} \bibinfo{volume}{36},
  \bibinfo{number}{2} (\bibinfo{year}{2021}), \bibinfo{pages}{535--545}.
\newblock


\bibitem[Staras et~al\mbox{.}(2021)]%
        {staras2021using}
\bibfield{author}{\bibinfo{person}{Stephanie Staras}, \bibinfo{person}{Justin~S
  Tauscher}, \bibinfo{person}{Natalie Rich}, \bibinfo{person}{Esaa Samarah},
  \bibinfo{person}{Lindsay~A Thompson}, \bibinfo{person}{Michelle~M Vinson},
  \bibinfo{person}{Michael~J Muszynski}, \bibinfo{person}{Elizabeth~A
  Shenkman}, {et~al\mbox{.}}} \bibinfo{year}{2021}\natexlab{}.
\newblock \showarticletitle{Using a clinical workflow analysis to enhance
  eHealth implementation planning: tutorial and case study}.
\newblock \bibinfo{journal}{\emph{JMIR mHealth and uHealth}}
  \bibinfo{volume}{9}, \bibinfo{number}{3} (\bibinfo{year}{2021}),
  \bibinfo{pages}{e18534}.
\newblock


\bibitem[Steele~Jr(1993)]%
        {steele1993standard}
\bibfield{author}{\bibinfo{person}{Glenn Steele~Jr}.}
  \bibinfo{year}{1993}\natexlab{}.
\newblock \showarticletitle{Standard postoperative monitoring of patients after
  primary resection of colon and rectum cancer}.
\newblock \bibinfo{journal}{\emph{Cancer}} \bibinfo{volume}{71},
  \bibinfo{number}{S12} (\bibinfo{year}{1993}), \bibinfo{pages}{4225--4235}.
\newblock


\bibitem[Stisen and Verdezoto(2017)]%
        {stisen2017clinical}
\bibfield{author}{\bibinfo{person}{Allan Stisen} {and} \bibinfo{person}{Nervo
  Verdezoto}.} \bibinfo{year}{2017}\natexlab{}.
\newblock \showarticletitle{Clinical and Non-Clinical Handovers: Designing for
  Critical Moments}. In \bibinfo{booktitle}{\emph{Proceedings of the 2017 ACM
  Conference on Computer Supported Cooperative Work and Social Computing}}
  (Portland, Oregon, USA) \emph{(\bibinfo{series}{CSCW '17})}.
  \bibinfo{publisher}{Association for Computing Machinery},
  \bibinfo{address}{New York, NY, USA}, \bibinfo{pages}{2166–2178}.
\newblock
\showISBNx{9781450343350}
\urldef\tempurl%
\url{https://doi.org/10.1145/2998181.2998333}
\showDOI{\tempurl}


\bibitem[Temple-Oberle et~al\mbox{.}(2023)]%
        {temple2023effect}
\bibfield{author}{\bibinfo{person}{Claire Temple-Oberle},
  \bibinfo{person}{Spencer Yakaback}, \bibinfo{person}{Carmen Webb},
  \bibinfo{person}{Golpira~Elmi Assadzadeh}, {and} \bibinfo{person}{Gregg
  Nelson}.} \bibinfo{year}{2023}\natexlab{}.
\newblock \showarticletitle{Effect of smartphone app postoperative home
  monitoring after oncologic surgery on quality of recovery: a randomized
  clinical trial}.
\newblock \bibinfo{journal}{\emph{JAMA surgery}} \bibinfo{volume}{158},
  \bibinfo{number}{7} (\bibinfo{year}{2023}), \bibinfo{pages}{693--699}.
\newblock


\bibitem[Touvron et~al\mbox{.}(2023)]%
        {touvronLlamaOpenFoundation2023}
\bibfield{author}{\bibinfo{person}{Hugo Touvron}, \bibinfo{person}{Louis
  Martin}, \bibinfo{person}{Kevin Stone}, \bibinfo{person}{Peter Albert},
  \bibinfo{person}{Amjad Almahairi}, \bibinfo{person}{Yasmine Babaei},
  \bibinfo{person}{Nikolay Bashlykov}, \bibinfo{person}{Soumya Batra},
  \bibinfo{person}{Prajjwal Bhargava}, \bibinfo{person}{Shruti Bhosale},
  \bibinfo{person}{Dan Bikel}, \bibinfo{person}{Lukas Blecher},
  \bibinfo{person}{Cristian~Canton Ferrer}, \bibinfo{person}{Moya Chen},
  \bibinfo{person}{Guillem Cucurull}, \bibinfo{person}{David Esiobu},
  \bibinfo{person}{Jude Fernandes}, \bibinfo{person}{Jeremy Fu},
  \bibinfo{person}{Wenyin Fu}, \bibinfo{person}{Brian Fuller},
  \bibinfo{person}{Cynthia Gao}, \bibinfo{person}{Vedanuj Goswami},
  \bibinfo{person}{Naman Goyal}, \bibinfo{person}{Anthony Hartshorn},
  \bibinfo{person}{Saghar Hosseini}, \bibinfo{person}{Rui Hou},
  \bibinfo{person}{Hakan Inan}, \bibinfo{person}{Marcin Kardas},
  \bibinfo{person}{Viktor Kerkez}, \bibinfo{person}{Madian Khabsa},
  \bibinfo{person}{Isabel Kloumann}, \bibinfo{person}{Artem Korenev},
  \bibinfo{person}{Punit~Singh Koura}, \bibinfo{person}{Marie-Anne Lachaux},
  \bibinfo{person}{Thibaut Lavril}, \bibinfo{person}{Jenya Lee},
  \bibinfo{person}{Diana Liskovich}, \bibinfo{person}{Yinghai Lu},
  \bibinfo{person}{Yuning Mao}, \bibinfo{person}{Xavier Martinet},
  \bibinfo{person}{Todor Mihaylov}, \bibinfo{person}{Pushkar Mishra},
  \bibinfo{person}{Igor Molybog}, \bibinfo{person}{Yixin Nie},
  \bibinfo{person}{Andrew Poulton}, \bibinfo{person}{Jeremy Reizenstein},
  \bibinfo{person}{Rashi Rungta}, \bibinfo{person}{Kalyan Saladi},
  \bibinfo{person}{Alan Schelten}, \bibinfo{person}{Ruan Silva},
  \bibinfo{person}{Eric~Michael Smith}, \bibinfo{person}{Ranjan Subramanian},
  \bibinfo{person}{Xiaoqing~Ellen Tan}, \bibinfo{person}{Binh Tang},
  \bibinfo{person}{Ross Taylor}, \bibinfo{person}{Adina Williams},
  \bibinfo{person}{Jian~Xiang Kuan}, \bibinfo{person}{Puxin Xu},
  \bibinfo{person}{Zheng Yan}, \bibinfo{person}{Iliyan Zarov},
  \bibinfo{person}{Yuchen Zhang}, \bibinfo{person}{Angela Fan},
  \bibinfo{person}{Melanie Kambadur}, \bibinfo{person}{Sharan Narang},
  \bibinfo{person}{Aurelien Rodriguez}, \bibinfo{person}{Robert Stojnic},
  \bibinfo{person}{Sergey Edunov}, {and} \bibinfo{person}{Thomas Scialom}.}
  \bibinfo{year}{2023}\natexlab{}.
\newblock \bibinfo{title}{Llama 2: {{Open Foundation}} and {{Fine-Tuned Chat
  Models}}}.
\newblock
\newblock
\showeprint[arxiv]{2307.09288}~[cs]
\urldef\tempurl%
\url{http://arxiv.org/abs/2307.09288}
\showURL{%
\tempurl}


\bibitem[Tran et~al\mbox{.}(2023)]%
        {tran2023mm}
\bibfield{author}{\bibinfo{person}{Brian~D Tran}, \bibinfo{person}{Kareem
  Latif}, \bibinfo{person}{Tera~L Reynolds}, \bibinfo{person}{Jihyun Park},
  \bibinfo{person}{Jennifer Elston~Lafata}, \bibinfo{person}{Ming Tai-Seale},
  {and} \bibinfo{person}{Kai Zheng}.} \bibinfo{year}{2023}\natexlab{}.
\newblock \showarticletitle{“Mm-hm,”“Uh-uh”: are non-lexical
  conversational sounds deal breakers for the ambient clinical documentation
  technology?}
\newblock \bibinfo{journal}{\emph{Journal of the American Medical Informatics
  Association}} \bibinfo{volume}{30}, \bibinfo{number}{4}
  (\bibinfo{year}{2023}), \bibinfo{pages}{703--711}.
\newblock


\bibitem[Tripathi et~al\mbox{.}(2024)]%
        {tripathi2024efficient}
\bibfield{author}{\bibinfo{person}{Satvik Tripathi}, \bibinfo{person}{Rithvik
  Sukumaran}, {and} \bibinfo{person}{Tessa~S Cook}.}
  \bibinfo{year}{2024}\natexlab{}.
\newblock \showarticletitle{Efficient healthcare with large language models:
  optimizing clinical workflow and enhancing patient care}.
\newblock \bibinfo{journal}{\emph{Journal of the American Medical Informatics
  Association}} \bibinfo{volume}{31}, \bibinfo{number}{6}
  (\bibinfo{year}{2024}), \bibinfo{pages}{1436--1440}.
\newblock


\bibitem[{van Kooten} et~al\mbox{.}(2021)]%
        {vankootenPreoperativeRiskFactors2021}
\bibfield{author}{\bibinfo{person}{Robert~T. {van Kooten}},
  \bibinfo{person}{Renu~R. Bahadoer}, \bibinfo{person}{Koen C. M.~J. Peeters},
  \bibinfo{person}{Jetty H.~L. Hoeksema}, \bibinfo{person}{Ewout~W.
  Steyerberg}, \bibinfo{person}{Henk~H. Hartgrink}, \bibinfo{person}{Cornelis
  J.~H. {van de Velde}}, \bibinfo{person}{Michel W. J.~M. Wouters}, {and}
  \bibinfo{person}{Rob A. E.~M. Tollenaar}.} \bibinfo{year}{2021}\natexlab{}.
\newblock \showarticletitle{Preoperative Risk Factors for Major Postoperative
  Complications after Complex Gastrointestinal Cancer Surgery: {{A}} Systematic
  Review}.
\newblock \bibinfo{journal}{\emph{European Journal of Surgical Oncology}}
  \bibinfo{volume}{47}, \bibinfo{number}{12} (\bibinfo{date}{Dec.}
  \bibinfo{year}{2021}), \bibinfo{pages}{3049--3058}.
\newblock
\showISSN{0748-7983}
\urldef\tempurl%
\url{https://doi.org/10.1016/j.ejso.2021.07.021}
\showDOI{\tempurl}


\bibitem[Wang et~al\mbox{.}(2020)]%
        {Wang2020HumanHumanToAI}
\bibfield{author}{\bibinfo{person}{Dakuo Wang}, \bibinfo{person}{Elizabeth
  Churchill}, \bibinfo{person}{Pattie Maes}, \bibinfo{person}{Xiangmin Fan},
  \bibinfo{person}{Ben Shneiderman}, \bibinfo{person}{Yuanchun Shi}, {and}
  \bibinfo{person}{Qianying Wang}.} \bibinfo{year}{2020}\natexlab{}.
\newblock \showarticletitle{From Human--Human Collaboration to Human--AI
  Collaboration: Designing AI Systems That Can Work Together with People}. In
  \bibinfo{booktitle}{\emph{Extended Abstracts of the 2020 CHI Conference on
  Human Factors in Computing Systems}}.
\newblock
\urldef\tempurl%
\url{https://doi.org/10.1145/3334480.3382929}
\showDOI{\tempurl}


\bibitem[Wang et~al\mbox{.}(2021)]%
        {wang_brilliant_2021}
\bibfield{author}{\bibinfo{person}{Dakuo Wang}, \bibinfo{person}{Liuping Wang},
  \bibinfo{person}{Zhan Zhang}, \bibinfo{person}{Ding Wang},
  \bibinfo{person}{Haiyi Zhu}, \bibinfo{person}{Yvonne Gao},
  \bibinfo{person}{Xiangmin Fan}, {and} \bibinfo{person}{Feng Tian}.}
  \bibinfo{year}{2021}\natexlab{}.
\newblock \showarticletitle{"{Brilliant} {AI} {Doctor}" in {Rural} {China}:
  {Tensions} and {Challenges} in {AI}-{Powered} {CDSS} {Deployment}}. In
  \bibinfo{booktitle}{\emph{Proceedings of the 2021 {CHI} {Conference} on
  {Human} {Factors} in {Computing} {Systems}}}. \bibinfo{pages}{1--18}.
\newblock


\bibitem[Wang et~al\mbox{.}(2024)]%
        {wang2024human}
\bibfield{author}{\bibinfo{person}{Xinru Wang}, \bibinfo{person}{Hannah Kim},
  \bibinfo{person}{Sajjadur Rahman}, \bibinfo{person}{Kushan Mitra}, {and}
  \bibinfo{person}{Zhengjie Miao}.} \bibinfo{year}{2024}\natexlab{}.
\newblock \showarticletitle{Human-LLM collaborative annotation through
  effective verification of LLM labels}. In
  \bibinfo{booktitle}{\emph{Proceedings of the CHI Conference on Human Factors
  in Computing Systems}}. \bibinfo{pages}{1--21}.
\newblock


\bibitem[Wang et~al\mbox{.}(2025)]%
        {wang2025operadatasetobservationpersona}
\bibfield{author}{\bibinfo{person}{Ziyi Wang}, \bibinfo{person}{Yuxuan Lu},
  \bibinfo{person}{Wenbo Li}, \bibinfo{person}{Amirali Amini},
  \bibinfo{person}{Bo Sun}, \bibinfo{person}{Yakov Bart},
  \bibinfo{person}{Weimin Lyu}, \bibinfo{person}{Jiri Gesi},
  \bibinfo{person}{Tian Wang}, \bibinfo{person}{Jing Huang},
  \bibinfo{person}{Yu Su}, \bibinfo{person}{Upol Ehsan},
  \bibinfo{person}{Malihe Alikhani}, \bibinfo{person}{Toby Jia-Jun Li},
  \bibinfo{person}{Lydia Chilton}, {and} \bibinfo{person}{Dakuo Wang}.}
  \bibinfo{year}{2025}\natexlab{}.
\newblock \bibinfo{title}{OPeRA: A Dataset of Observation, Persona, Rationale,
  and Action for Evaluating LLMs on Human Online Shopping Behavior Simulation}.
\newblock
\newblock
\showeprint[arxiv]{2506.05606}~[cs.CL]
\urldef\tempurl%
\url{https://arxiv.org/abs/2506.05606}
\showURL{%
\tempurl}


\bibitem[Wei et~al\mbox{.}(2023)]%
        {wei_leveraging_2023}
\bibfield{author}{\bibinfo{person}{Jing Wei}, \bibinfo{person}{Sungdong Kim},
  \bibinfo{person}{Hyunhoon Jung}, {and} \bibinfo{person}{Young-Ho Kim}.}
  \bibinfo{year}{2023}\natexlab{}.
\newblock \bibinfo{title}{Leveraging {Large} {Language} {Models} to {Power}
  {Chatbots} for {Collecting} {User} {Self}-{Reported} {Data}}.
\newblock
\newblock


\bibitem[Wei et~al\mbox{.}(2022)]%
        {wei2022chain}
\bibfield{author}{\bibinfo{person}{Jason Wei}, \bibinfo{person}{Xuezhi Wang},
  \bibinfo{person}{Dale Schuurmans}, \bibinfo{person}{Maarten Bosma},
  \bibinfo{person}{Fei Xia}, \bibinfo{person}{Ed Chi}, \bibinfo{person}{Quoc~V
  Le}, \bibinfo{person}{Denny Zhou}, {et~al\mbox{.}}}
  \bibinfo{year}{2022}\natexlab{}.
\newblock \showarticletitle{Chain-of-thought prompting elicits reasoning in
  large language models}.
\newblock \bibinfo{journal}{\emph{Advances in neural information processing
  systems}}  \bibinfo{volume}{35} (\bibinfo{year}{2022}),
  \bibinfo{pages}{24824--24837}.
\newblock


\bibitem[Wester et~al\mbox{.}(2024)]%
        {wester2024facing}
\bibfield{author}{\bibinfo{person}{Joel Wester}, \bibinfo{person}{Bhakti
  Moghe}, \bibinfo{person}{Katie Winkle}, {and} \bibinfo{person}{Niels van
  Berkel}.} \bibinfo{year}{2024}\natexlab{}.
\newblock \showarticletitle{Facing LLMs: Robot Communication Styles in
  Mediating Health Information between Parents and Young Adults}.
\newblock \bibinfo{journal}{\emph{Proceedings of the ACM on Human-Computer
  Interaction}} \bibinfo{volume}{8}, \bibinfo{number}{CSCW2}
  (\bibinfo{year}{2024}), \bibinfo{pages}{1--37}.
\newblock


\bibitem[Wong et~al\mbox{.}(2021)]%
        {wongGlobalIncidenceMortality2021}
\bibfield{author}{\bibinfo{person}{Martin C.~S. Wong}, \bibinfo{person}{Junjie
  Huang}, \bibinfo{person}{Paul S.~F. Chan}, \bibinfo{person}{Peter Choi},
  \bibinfo{person}{Xiang~Qian Lao}, \bibinfo{person}{Shannon~Melissa Chan},
  \bibinfo{person}{Anthony Teoh}, {and} \bibinfo{person}{Peter Liang}.}
  \bibinfo{year}{2021}\natexlab{}.
\newblock \showarticletitle{Global {{Incidence}} and {{Mortality}} of {{Gastric
  Cancer}}, 1980-2018}.
\newblock \bibinfo{journal}{\emph{JAMA Network Open}} \bibinfo{volume}{4},
  \bibinfo{number}{7} (\bibinfo{date}{July} \bibinfo{year}{2021}),
  \bibinfo{pages}{e2118457}.
\newblock
\showISSN{2574-3805}
\urldef\tempurl%
\url{https://doi.org/10.1001/jamanetworkopen.2021.18457}
\showDOI{\tempurl}


\bibitem[Wu et~al\mbox{.}(2024)]%
        {wu2024clinical}
\bibfield{author}{\bibinfo{person}{Siyi Wu}, \bibinfo{person}{Weidan Cao},
  \bibinfo{person}{Shihan Fu}, \bibinfo{person}{Bingsheng Yao},
  \bibinfo{person}{Ziqi Yang}, \bibinfo{person}{Changchang Yin},
  \bibinfo{person}{Varun Mishra}, \bibinfo{person}{Daniel Addison},
  \bibinfo{person}{Ping Zhang}, {and} \bibinfo{person}{Dakuo Wang}.}
  \bibinfo{year}{2024}\natexlab{}.
\newblock \showarticletitle{Clinical Challenges and AI Opportunities in
  Decision-Making for Cancer Treatment-Induced Cardiotoxicity}.
\newblock \bibinfo{journal}{\emph{arXiv preprint arXiv:2408.03586}}
  (\bibinfo{year}{2024}).
\newblock


\bibitem[Wu et~al\mbox{.}(2025a)]%
        {wu2025cardioai}
\bibfield{author}{\bibinfo{person}{Siyi Wu}, \bibinfo{person}{Weidan Cao},
  \bibinfo{person}{Shihan Fu}, \bibinfo{person}{Bingsheng Yao},
  \bibinfo{person}{Ziqi Yang}, \bibinfo{person}{Changchang Yin},
  \bibinfo{person}{Varun Mishra}, \bibinfo{person}{Daniel Addison},
  \bibinfo{person}{Ping Zhang}, {and} \bibinfo{person}{Dakuo Wang}.}
  \bibinfo{year}{2025}\natexlab{a}.
\newblock \showarticletitle{CardioAI: A Multimodal AI-based System to Support
  Symptom Monitoring and Risk Prediction of Cancer Treatment-Induced
  Cardiotoxicity}. In \bibinfo{booktitle}{\emph{Proceedings of the 2025 CHI
  Conference on Human Factors in Computing Systems}}
  \emph{(\bibinfo{series}{CHI '25})}. \bibinfo{publisher}{Association for
  Computing Machinery}, \bibinfo{address}{New York, NY, USA}, Article
  \bibinfo{articleno}{313}, \bibinfo{numpages}{22}~pages.
\newblock
\showISBNx{9798400713941}
\urldef\tempurl%
\url{https://doi.org/10.1145/3706598.3714272}
\showDOI{\tempurl}


\bibitem[Wu et~al\mbox{.}(2025b)]%
        {wu2025collabllm}
\bibfield{author}{\bibinfo{person}{Shirley Wu}, \bibinfo{person}{Michel
  Galley}, \bibinfo{person}{Baolin Peng}, \bibinfo{person}{Hao Cheng},
  \bibinfo{person}{Gavin Li}, \bibinfo{person}{Yao Dou},
  \bibinfo{person}{Weixin Cai}, \bibinfo{person}{James Zou},
  \bibinfo{person}{Jure Leskovec}, {and} \bibinfo{person}{Jianfeng Gao}.}
  \bibinfo{year}{2025}\natexlab{b}.
\newblock \showarticletitle{Collabllm: From passive responders to active
  collaborators}.
\newblock \bibinfo{journal}{\emph{arXiv preprint arXiv:2502.00640}}
  (\bibinfo{year}{2025}).
\newblock


\bibitem[Xiao et~al\mbox{.}(2023)]%
        {xiao2023supporting}
\bibfield{author}{\bibinfo{person}{Ziang Xiao}, \bibinfo{person}{Xingdi Yuan},
  \bibinfo{person}{Q~Vera Liao}, \bibinfo{person}{Rania Abdelghani}, {and}
  \bibinfo{person}{Pierre-Yves Oudeyer}.} \bibinfo{year}{2023}\natexlab{}.
\newblock \showarticletitle{Supporting Qualitative Analysis with Large Language
  Models: Combining Codebook with GPT-3 for Deductive Coding}. In
  \bibinfo{booktitle}{\emph{Companion Proceedings of the 28th International
  Conference on Intelligent User Interfaces}}. \bibinfo{pages}{75--78}.
\newblock


\bibitem[Xiao et~al\mbox{.}(2020)]%
        {xiao2020tell}
\bibfield{author}{\bibinfo{person}{Ziang Xiao}, \bibinfo{person}{Michelle~X
  Zhou}, \bibinfo{person}{Q~Vera Liao}, \bibinfo{person}{Gloria Mark},
  \bibinfo{person}{Changyan Chi}, \bibinfo{person}{Wenxi Chen}, {and}
  \bibinfo{person}{Huahai Yang}.} \bibinfo{year}{2020}\natexlab{}.
\newblock \showarticletitle{Tell me about yourself: Using an AI-powered chatbot
  to conduct conversational surveys with open-ended questions}.
\newblock \bibinfo{journal}{\emph{ACM Transactions on Computer-Human
  Interaction (TOCHI)}} \bibinfo{volume}{27}, \bibinfo{number}{3}
  (\bibinfo{year}{2020}), \bibinfo{pages}{1--37}.
\newblock


\bibitem[Yamagata et~al\mbox{.}(2019)]%
        {yamagataCurrentStatusEnhanced2019}
\bibfield{author}{\bibinfo{person}{Yukinori Yamagata}, \bibinfo{person}{Takaki
  Yoshikawa}, \bibinfo{person}{Masahiro Yura}, \bibinfo{person}{Sho Otsuki},
  \bibinfo{person}{Shinji Morita}, \bibinfo{person}{Hitoshi Katai}, {and}
  \bibinfo{person}{Toshiro Nishida}.} \bibinfo{year}{2019}\natexlab{}.
\newblock \showarticletitle{Current Status of the ``Enhanced Recovery after
  Surgery'' Program in Gastric Cancer Surgery}.
\newblock \bibinfo{journal}{\emph{Annals of Gastroenterological Surgery}}
  \bibinfo{volume}{3}, \bibinfo{number}{3} (\bibinfo{year}{2019}),
  \bibinfo{pages}{231--238}.
\newblock
\showISSN{2475-0328}
\urldef\tempurl%
\url{https://doi.org/10.1002/ags3.12232}
\showDOI{\tempurl}


\bibitem[Yang et~al\mbox{.}(2024a)]%
        {yangWishThereWere2024}
\bibfield{author}{\bibinfo{person}{Ziqi Yang}, \bibinfo{person}{Xuhai Xu},
  \bibinfo{person}{Bingsheng Yao}, \bibinfo{person}{Jiachen Li},
  \bibinfo{person}{Jennifer Bagdasarian}, \bibinfo{person}{Guodong Gao}, {and}
  \bibinfo{person}{Dakuo Wang}.} \bibinfo{year}{2024}\natexlab{a}.
\newblock \bibinfo{title}{"{{I Wish There Were}} an {{AI}}": {{Challenges}} and
  {{AI Potential}} in {{Cancer Patient-Provider Communication}}}.
\newblock
\newblock
\showeprint[arxiv]{2404.13409}~[cs]


\bibitem[Yang et~al\mbox{.}(2024b)]%
        {yang2024talk2care}
\bibfield{author}{\bibinfo{person}{Ziqi Yang}, \bibinfo{person}{Xuhai Xu},
  \bibinfo{person}{Bingsheng Yao}, \bibinfo{person}{Ethan Rogers},
  \bibinfo{person}{Shao Zhang}, \bibinfo{person}{Stephen Intille},
  \bibinfo{person}{Nawar Shara}, \bibinfo{person}{Guodong~Gordon Gao}, {and}
  \bibinfo{person}{Dakuo Wang}.} \bibinfo{year}{2024}\natexlab{b}.
\newblock \showarticletitle{Talk2Care: An LLM-based Voice Assistant for
  Communication between Healthcare Providers and Older Adults}.
\newblock \bibinfo{journal}{\emph{Proc. ACM Interact. Mob. Wearable Ubiquitous
  Technol.}} \bibinfo{volume}{8}, \bibinfo{number}{2}, Article
  \bibinfo{articleno}{73} (\bibinfo{date}{May} \bibinfo{year}{2024}),
  \bibinfo{numpages}{35}~pages.
\newblock
\urldef\tempurl%
\url{https://doi.org/10.1145/3659625}
\showDOI{\tempurl}


\bibitem[Yao et~al\mbox{.}(2024)]%
        {yao-etal-2024-samples}
\bibfield{author}{\bibinfo{person}{Bingsheng Yao}, \bibinfo{person}{Guiming
  Chen}, \bibinfo{person}{Ruishi Zou}, \bibinfo{person}{Yuxuan Lu},
  \bibinfo{person}{Jiachen Li}, \bibinfo{person}{Shao Zhang},
  \bibinfo{person}{Yisi Sang}, \bibinfo{person}{Sijia Liu},
  \bibinfo{person}{James Hendler}, {and} \bibinfo{person}{Dakuo Wang}.}
  \bibinfo{year}{2024}\natexlab{}.
\newblock \showarticletitle{More Samples or More Prompts? Exploring Effective
  Few-Shot In-Context Learning for {LLM}s with In-Context Sampling}. In
  \bibinfo{booktitle}{\emph{Findings of the Association for Computational
  Linguistics: NAACL 2024}}, \bibfield{editor}{\bibinfo{person}{Kevin Duh},
  \bibinfo{person}{Helena Gomez}, {and} \bibinfo{person}{Steven Bethard}}
  (Eds.). \bibinfo{publisher}{Association for Computational Linguistics},
  \bibinfo{address}{Mexico City, Mexico}, \bibinfo{pages}{1772--1790}.
\newblock
\urldef\tempurl%
\url{https://doi.org/10.18653/v1/2024.findings-naacl.115}
\showDOI{\tempurl}


\bibitem[Yao et~al\mbox{.}(2025a)]%
        {yao2025more}
\bibfield{author}{\bibinfo{person}{Bingsheng Yao}, \bibinfo{person}{Menglin
  Zhao}, \bibinfo{person}{Yuling Sun}, \bibinfo{person}{Weidan Cao},
  \bibinfo{person}{Changchang Yin}, \bibinfo{person}{Stephen Intille},
  \bibinfo{person}{Xuhai Xu}, \bibinfo{person}{Ping Zhang},
  \bibinfo{person}{Jingzhen Yang}, {and} \bibinfo{person}{Dakuo Wang}.}
  \bibinfo{year}{2025}\natexlab{a}.
\newblock \showarticletitle{More Modality, More AI: Exploring Design
  Opportunities of AI-Based Multi-modal Remote Monitoring Technologies for
  Early Detection of Mental Health Sequelae in Youth Concussion Patients}.
\newblock \bibinfo{journal}{\emph{arXiv preprint arXiv:2502.03732}}
  (\bibinfo{year}{2025}).
\newblock


\bibitem[Yao et~al\mbox{.}(2025b)]%
        {yao2025exploring}
\bibfield{author}{\bibinfo{person}{Bingsheng Yao}, \bibinfo{person}{Menglin
  Zhao}, \bibinfo{person}{Zhan Zhang}, \bibinfo{person}{Pengqi Wang},
  \bibinfo{person}{Emma~G Chester}, \bibinfo{person}{Changchang Yin},
  \bibinfo{person}{Tianshi Li}, \bibinfo{person}{Varun Mishra},
  \bibinfo{person}{Lace Padilla}, \bibinfo{person}{Odysseas Chatzipanagiotou},
  {et~al\mbox{.}}} \bibinfo{year}{2025}\natexlab{b}.
\newblock \showarticletitle{Exploring Collaboration Breakdowns Between Provider
  Teams and Patients in Post-Surgery Care}.
\newblock \bibinfo{journal}{\emph{arXiv preprint arXiv:2509.23509}}
  (\bibinfo{year}{2025}).
\newblock


\bibitem[Yasunaga et~al\mbox{.}(2013)]%
        {yasunagaBodyMassIndex2013}
\bibfield{author}{\bibinfo{person}{H Yasunaga}, \bibinfo{person}{H Horiguchi},
  \bibinfo{person}{S Matsuda}, \bibinfo{person}{K Fushimi}, \bibinfo{person}{H
  Hashimoto}, {and} \bibinfo{person}{J~Z Ayanian}.}
  \bibinfo{year}{2013}\natexlab{}.
\newblock \showarticletitle{Body Mass Index and Outcomes Following
  Gastrointestinal Cancer Surgery in {{Japan}}}.
\newblock \bibinfo{journal}{\emph{British Journal of Surgery}}
  \bibinfo{volume}{100}, \bibinfo{number}{10} (\bibinfo{date}{Sept.}
  \bibinfo{year}{2013}), \bibinfo{pages}{1335--1343}.
\newblock
\showISSN{0007-1323}
\urldef\tempurl%
\url{https://doi.org/10.1002/bjs.9221}
\showDOI{\tempurl}


\bibitem[Yildirim et~al\mbox{.}(2024)]%
        {yildirimMultimodalHealthcareAI2024}
\bibfield{author}{\bibinfo{person}{Nur Yildirim}, \bibinfo{person}{Hannah
  Richardson}, \bibinfo{person}{Maria~T. Wetscherek}, \bibinfo{person}{Junaid
  Bajwa}, \bibinfo{person}{Joseph Jacob}, \bibinfo{person}{Mark~A. Pinnock},
  \bibinfo{person}{Stephen Harris}, \bibinfo{person}{Daniel~Coelho {de
  Castro}}, \bibinfo{person}{Shruthi Bannur}, \bibinfo{person}{Stephanie~L.
  Hyland}, \bibinfo{person}{Pratik Ghosh}, \bibinfo{person}{Mercy Ranjit},
  \bibinfo{person}{Kenza Bouzid}, \bibinfo{person}{Anton Schwaighofer},
  \bibinfo{person}{Fernando {P{\'e}rez-Garc{\'i}a}}, \bibinfo{person}{Harshita
  Sharma}, \bibinfo{person}{Ozan Oktay}, \bibinfo{person}{Matthew Lungren},
  \bibinfo{person}{Javier {Alvarez-Valle}}, \bibinfo{person}{Aditya Nori},
  {and} \bibinfo{person}{Anja Thieme}.} \bibinfo{year}{2024}\natexlab{}.
\newblock \bibinfo{title}{Multimodal {{Healthcare AI}}: {{Identifying}} and
  {{Designing Clinically Relevant Vision-Language Applications}} for
  {{Radiology}}}.
\newblock
\newblock
\urldef\tempurl%
\url{https://doi.org/10.1145/3613904.3642013}
\showDOI{\tempurl}
\showeprint[arxiv]{2402.14252}~[cs]


\bibitem[Yunxiang et~al\mbox{.}(2023)]%
        {yunxiang2023chatdoctor}
\bibfield{author}{\bibinfo{person}{Li Yunxiang}, \bibinfo{person}{Li Zihan},
  \bibinfo{person}{Zhang Kai}, \bibinfo{person}{Dan Ruilong}, {and}
  \bibinfo{person}{Zhang You}.} \bibinfo{year}{2023}\natexlab{}.
\newblock \showarticletitle{Chatdoctor: A medical chat model fine-tuned on
  llama model using medical domain knowledge}.
\newblock \bibinfo{journal}{\emph{arXiv preprint arXiv:2303.14070}}
  (\bibinfo{year}{2023}).
\newblock


\bibitem[Zhan et~al\mbox{.}(2024)]%
        {zhanHealthcareVoiceAI2024}
\bibfield{author}{\bibinfo{person}{Xiao Zhan}, \bibinfo{person}{Noura Abdi},
  \bibinfo{person}{William Seymour}, {and} \bibinfo{person}{Jose Such}.}
  \bibinfo{year}{2024}\natexlab{}.
\newblock \showarticletitle{Healthcare {{Voice AI Assistants}}: {{Factors
  Influencing Trust}} and {{Intention}} to {{Use}}}.
\newblock \bibinfo{journal}{\emph{Proc. ACM Hum.-Comput. Interact.}}
  \bibinfo{volume}{8}, \bibinfo{number}{CSCW1} (\bibinfo{date}{April}
  \bibinfo{year}{2024}), \bibinfo{pages}{62:1--62:37}.
\newblock
\urldef\tempurl%
\url{https://doi.org/10.1145/3637339}
\showDOI{\tempurl}


\bibitem[Zhang et~al\mbox{.}(2024)]%
        {zhangRethinkingHumanAICollaboration2024}
\bibfield{author}{\bibinfo{person}{Shao Zhang}, \bibinfo{person}{Jianing Yu},
  \bibinfo{person}{Xuhai Xu}, \bibinfo{person}{Changchang Yin},
  \bibinfo{person}{Yuxuan Lu}, \bibinfo{person}{Bingsheng Yao},
  \bibinfo{person}{Melanie Tory}, \bibinfo{person}{Lace~M. Padilla},
  \bibinfo{person}{Jeffrey Caterino}, \bibinfo{person}{Ping Zhang}, {and}
  \bibinfo{person}{Dakuo Wang}.} \bibinfo{year}{2024}\natexlab{}.
\newblock \bibinfo{title}{Rethinking {{Human-AI Collaboration}} in {{Complex
  Medical Decision Making}}: {{A Case Study}} in {{Sepsis Diagnosis}}}.
\newblock
\newblock
\urldef\tempurl%
\url{https://doi.org/10.1145/3613904.3642343}
\showDOI{\tempurl}
\showeprint[arxiv]{2309.12368}~[cs]


\bibitem[Zhang et~al\mbox{.}(2017)]%
        {zhang2017constructing}
\bibfield{author}{\bibinfo{person}{Zhan Zhang}, \bibinfo{person}{Aleksandra
  Sarcevic}, {and} \bibinfo{person}{Claus Bossen}.}
  \bibinfo{year}{2017}\natexlab{}.
\newblock \showarticletitle{Constructing Common Information Spaces across
  Distributed Emergency Medical Teams}. In
  \bibinfo{booktitle}{\emph{Proceedings of the 2017 ACM Conference on Computer
  Supported Cooperative Work and Social Computing}} (Portland, Oregon, USA)
  \emph{(\bibinfo{series}{CSCW '17})}. \bibinfo{publisher}{Association for
  Computing Machinery}, \bibinfo{address}{New York, NY, USA},
  \bibinfo{pages}{934–947}.
\newblock
\showISBNx{9781450343350}
\urldef\tempurl%
\url{https://doi.org/10.1145/2998181.2998328}
\showDOI{\tempurl}


\bibitem[Zhao et~al\mbox{.}(2025)]%
        {zhao2025designing}
\bibfield{author}{\bibinfo{person}{Menglin Zhao}, \bibinfo{person}{Zhuorui
  Yong}, \bibinfo{person}{Ruijia Guan}, \bibinfo{person}{Kai-Wei Chang},
  \bibinfo{person}{Adrian Haimovich}, \bibinfo{person}{Kei Ouchi},
  \bibinfo{person}{Timothy Bickmore}, \bibinfo{person}{Bingsheng Yao},
  \bibinfo{person}{Dakuo Wang}, {and} \bibinfo{person}{Smit Desai}.}
  \bibinfo{year}{2025}\natexlab{}.
\newblock \showarticletitle{Designing AI Tools for Clinical Care Teams to
  Support Serious Illness Conversations with Older Adults in the Emergency
  Department}.
\newblock \bibinfo{journal}{\emph{arXiv preprint arXiv:2506.00241}}
  (\bibinfo{year}{2025}).
\newblock


\end{thebibliography}

\appendix
\clearpage
\onecolumn
\section{Appendix}
\label{sec:appendix}
\subsection{Initial Design Components}
\subsubsection{Example conversation from initial LLM CA design}
In this version, although the LLM CA responses fit the postoperative context, there were potentially inappropriate expressions such as assessing patient conditions, not covering most questions, and miscommunication of system responsibilities.

\begin{longtblr}[
caption = {An example conversation between a test user (simulating a patient) and the CA with our initial prompt design.},
label={tab:8-appendix-initial-conversation}
]{
hlines,vlines,
colspec={lX},
width=\linewidth,
rowhead=1
}
\textbf{Role} & \textbf{Content}  \\

\textbf{User:} & Hello \\
\textbf{CA:} & Good afternoon! This is "surgery recover" from the clinic, I'm calling to see how you're doing after your cancer surgery. Have you been experiencing any specific discomforts or symptoms that you would like to discuss? \\ 
\textbf{User:} & I am good, thanks for asking \\   
\textbf{CA:} & That's good to hear. Let's continue with our check-up. How about your food intake? Have you noticed any change in your appetite or taste? \\   
\textbf{User:} & My food intake is good, I have great appetite \\   
\textbf{CA:} & That's great! You are maintaining well. Have you noticed any changes in your bowel movements? Any issues like constipation or bloating? \\   
\textbf{User:} & Well I may have ate a lot, so my bowel movement may be a little more frequent than usual, but I think that is fine \\   
\textbf{CA:} & I see, an increase in bowel movement due to an increase in food intake isn't necessarily a cause for concern. However, if you experience any significant changes such as severe diarrhea, constipation or blood in stool, these should be reported to your healthcare provider immediately. Now, let's move to another area. Have you noticed any pain or swelling anywhere in your body after surgery? \\   
\textbf{User:} & Ok. Well, actually I have some pain around my stoma \\   
\textbf{CA:} & I'm sorry to hear that you're experiencing some pain around your stoma. Can you tell me more about the pain? How would you rate the severity of the pain on a scale from 1-10, with 1 being mild discomfort and 10 the worst pain imaginable? \\   
\textbf{User:} & I would say 5 out of 10 when I move around, and maybe 2 or 3 out of 10 when I sit still. \\   
\textbf{CA:} & I understand. It's not uncommon to experience some pain around the stoma after surgery. Can you tell me if the pain is constant, or does it come and go? Also, have you noticed any redness, swelling, or changes in the color of output from your stoma? \\   
\textbf{User:} & I think it is consistent, but more severe sometimes than others. I am not sure about other things, it seems normal \\   
\textbf{CA:} & Thank you for providing those details. I understand the pain can be quite uncomfortable. I suggest you take a note of the times when the pain increases, and if the pain becomes severe or persists it's important to report it to your healthcare provider. Is there anything else you would like to discuss or any other symptoms you've noticed? \\   
\textbf{User:} & Ok, thanks for telling me that. Can you ask my doctor whether I should go in for a visit? \\   
\textbf{CA:} & Sure, I will make sure to pass on your concern to your healthcare provider and they can decide whether an in-person visit is needed. Do you have any other concerns you want me to communicate? \\   
\textbf{User:} & No, thank you \\   
\textbf{CA:} & I'm glad to hear you're generally doing well. I will let your healthcare provider know about your slight discomfort and they will determine whether a visit is necessary. If you have any further concerns or symptoms arise, please don't hesitate to call us. Take care. \\   
\textbf{User:} & Bye \\

\end{longtblr}

\subsubsection{Initial dashboard interface design}
\begin{figure}
    \centering
    \twiceinclude[width=0.8\linewidth]{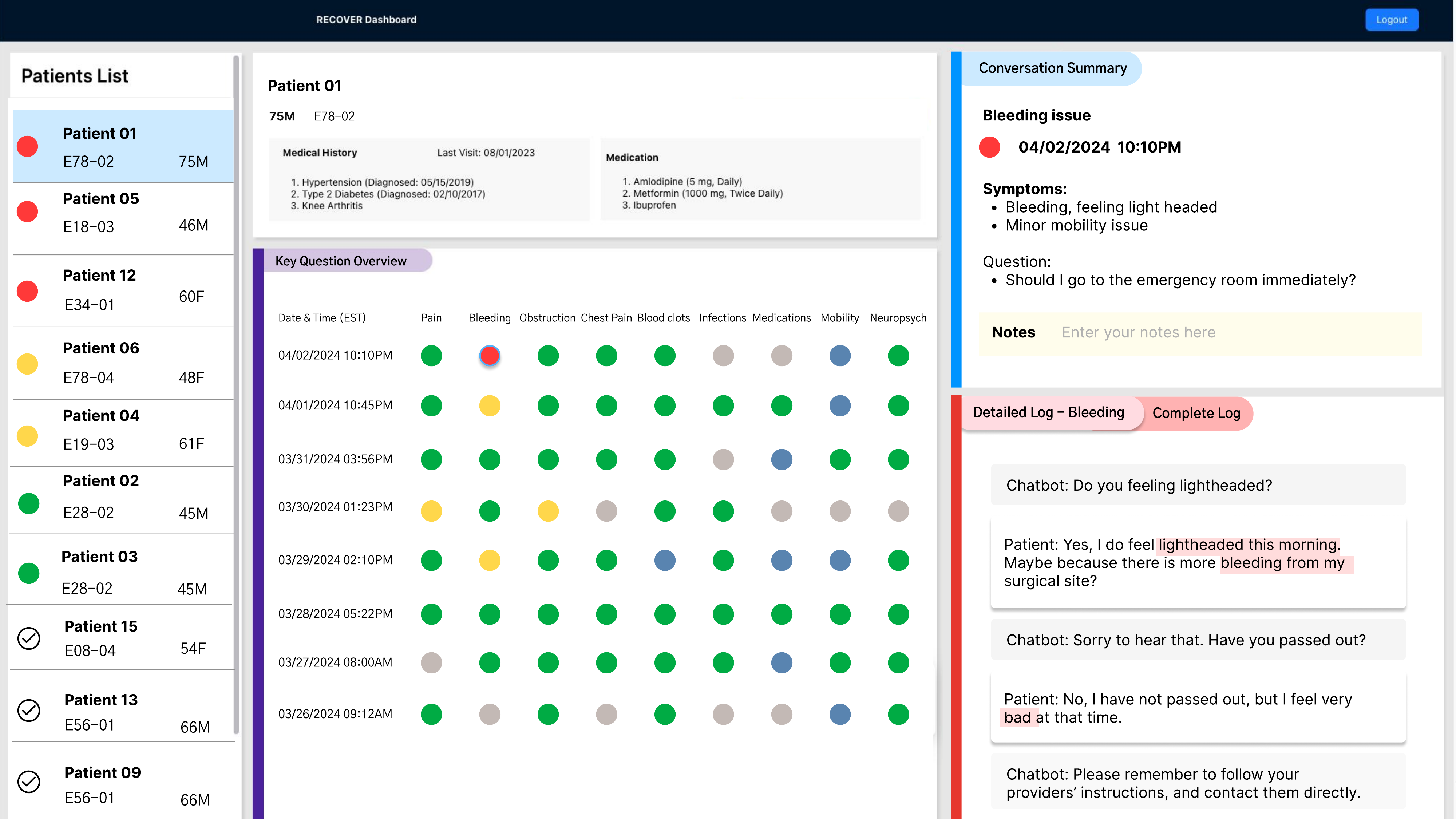}
    \caption{The first version of dashboard interface design. This version only include simple interactions like selecting patient to view details, and scrolling down to see conversation logs. After our PD sessions, we iterated the design of each tab and interactions to promote clinical efficiency.}
    \Description{Recover Dashboard interface displaying patient monitoring data with three main sections: a severity-sorted patient list on the left showing color-coded patient status (red, yellow, green indicators), a central panel with patient medical information and a key question overview grid visualizing symptom severity across multiple time points using color-coded dots for metrics including pain, bleeding, obstruction, chest pain, blood clots, heroing, medications, mobility, and nausea/psych symptoms, and a right panel showing conversation summary with bleeding issue details, symptoms, clinical questions, and a detailed chatbot-patient conversation log discussing lightheadedness and post-surgical bleeding concerns.}
    \label{fig:8-appendix-initial-dashboard}
\end{figure}
\subsection{PD Session Artifact Examples}
\begin{figure}[h]
\centering
\begin{subfigure}[b]{0.9\textwidth}
    \twiceinclude[width=\textwidth]{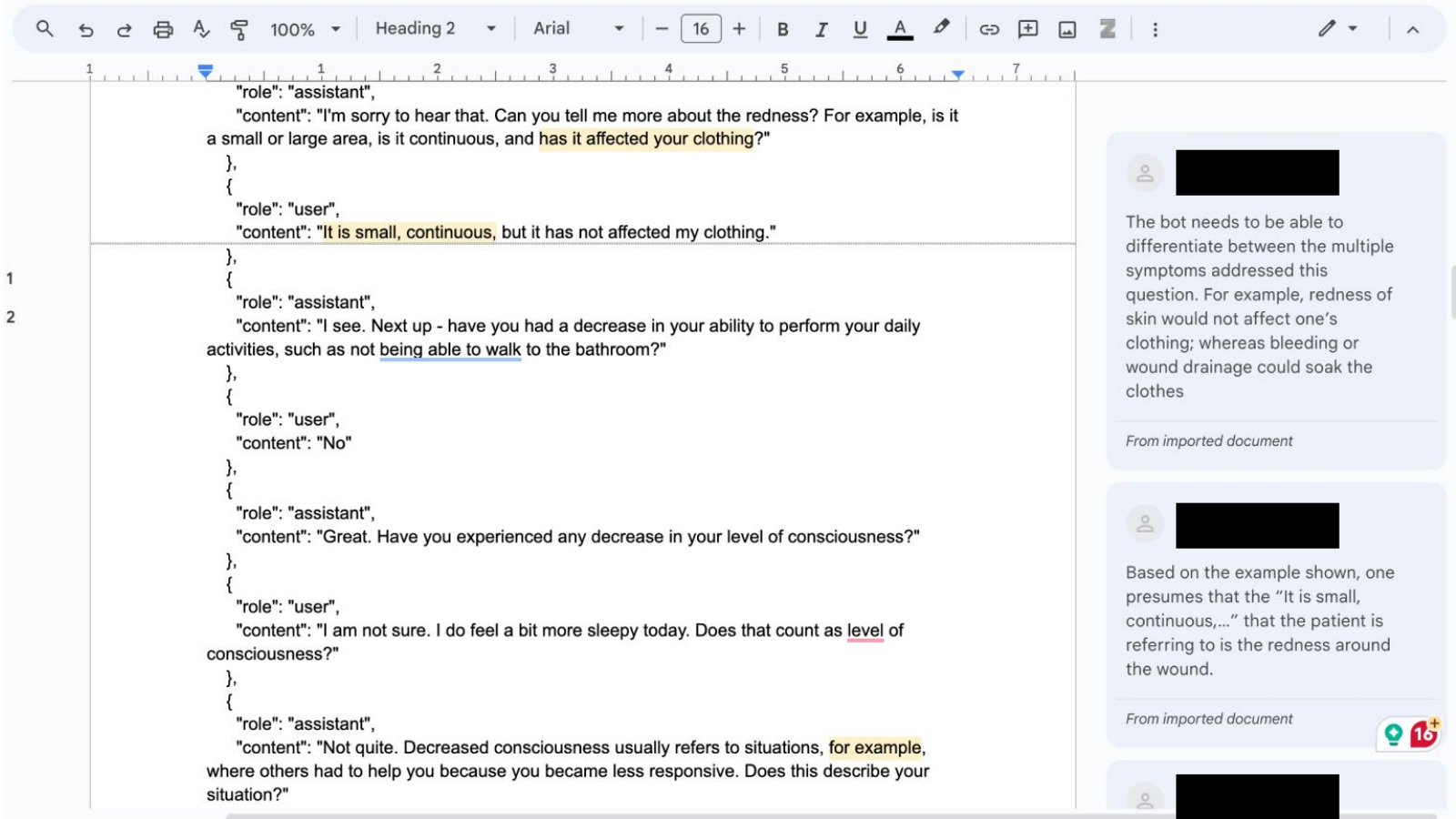}
    \caption{Participants review the conversation log generated with one prompt and leave comments.}
    \label{fig:8-appendix-pd-ex-1}
\end{subfigure}

\begin{subfigure}[b]{0.9\textwidth}
    \twiceinclude[width=\textwidth]{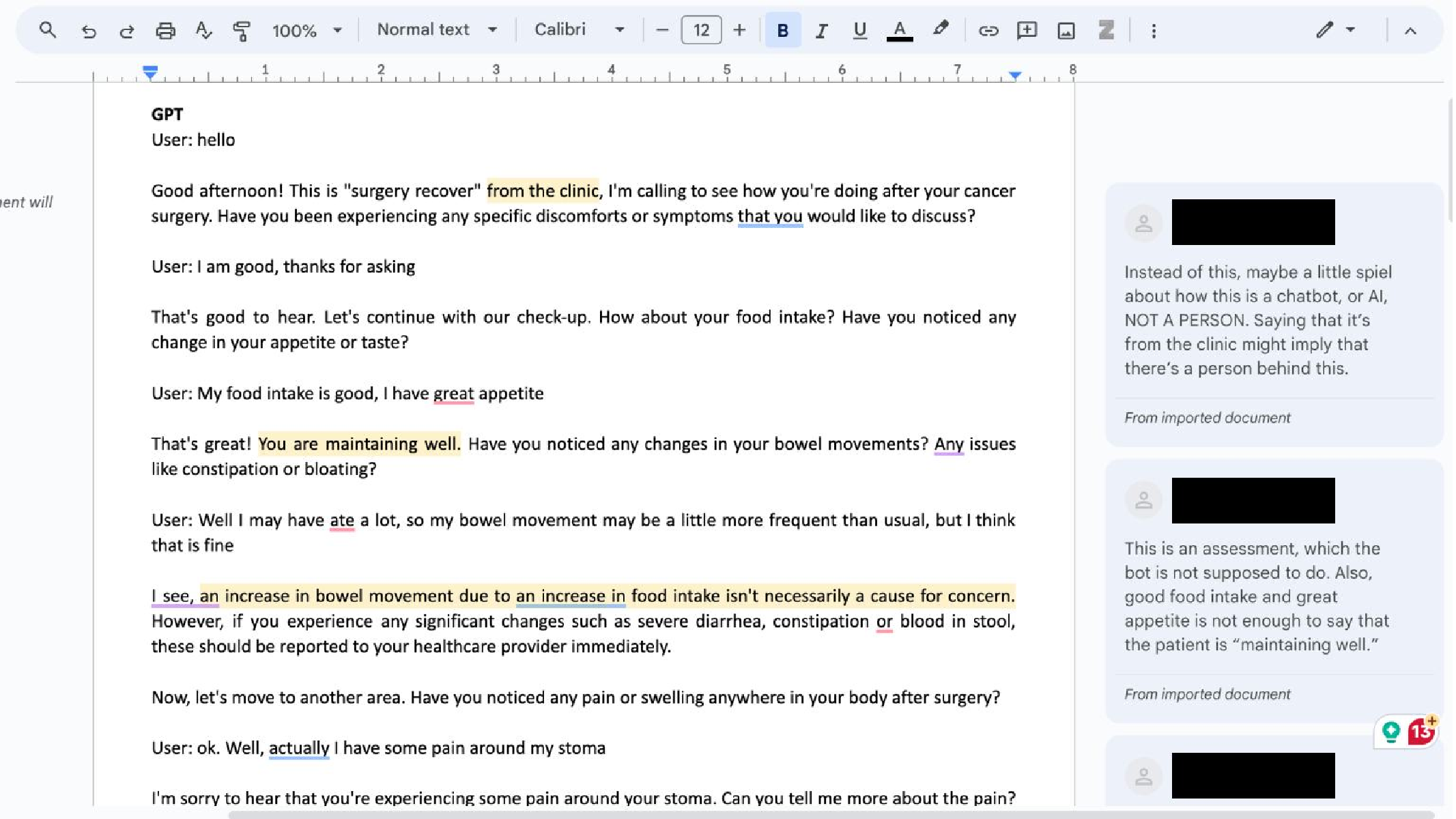}
    \caption{Participants review the conversation log generated with an updated prompt and leave comments.}
    \label{fig:8-appendix-pd-ex-2}
\end{subfigure}
\caption{Example of participants' comments in Google Docs.}
\Description{Two-panel comparison of participant feedback on conversation logs showing (a) JSON-formatted dialogue between assistant and user discussing post-surgical symptoms including redness, mobility, and consciousness, with participant comments noting the bot's need to differentiate between symptoms and providing context interpretation; and (b) GPT-formatted natural language conversation covering post-surgical recovery topics including food intake, bowel movements, and abdominal pain, with participant comments critiquing the chatbot-like tone and questioning clinical assessment appropriateness. Both panels display Google Docs interfaces with redacted participant identities and timestamps in the right margin.}
\label{fig:8-appendix-pd-ex-1-2}
\end{figure}

\begin{figure}[h]
\centering
\begin{subfigure}[b]{0.75\textwidth}
    \includegraphics[width=\textwidth]{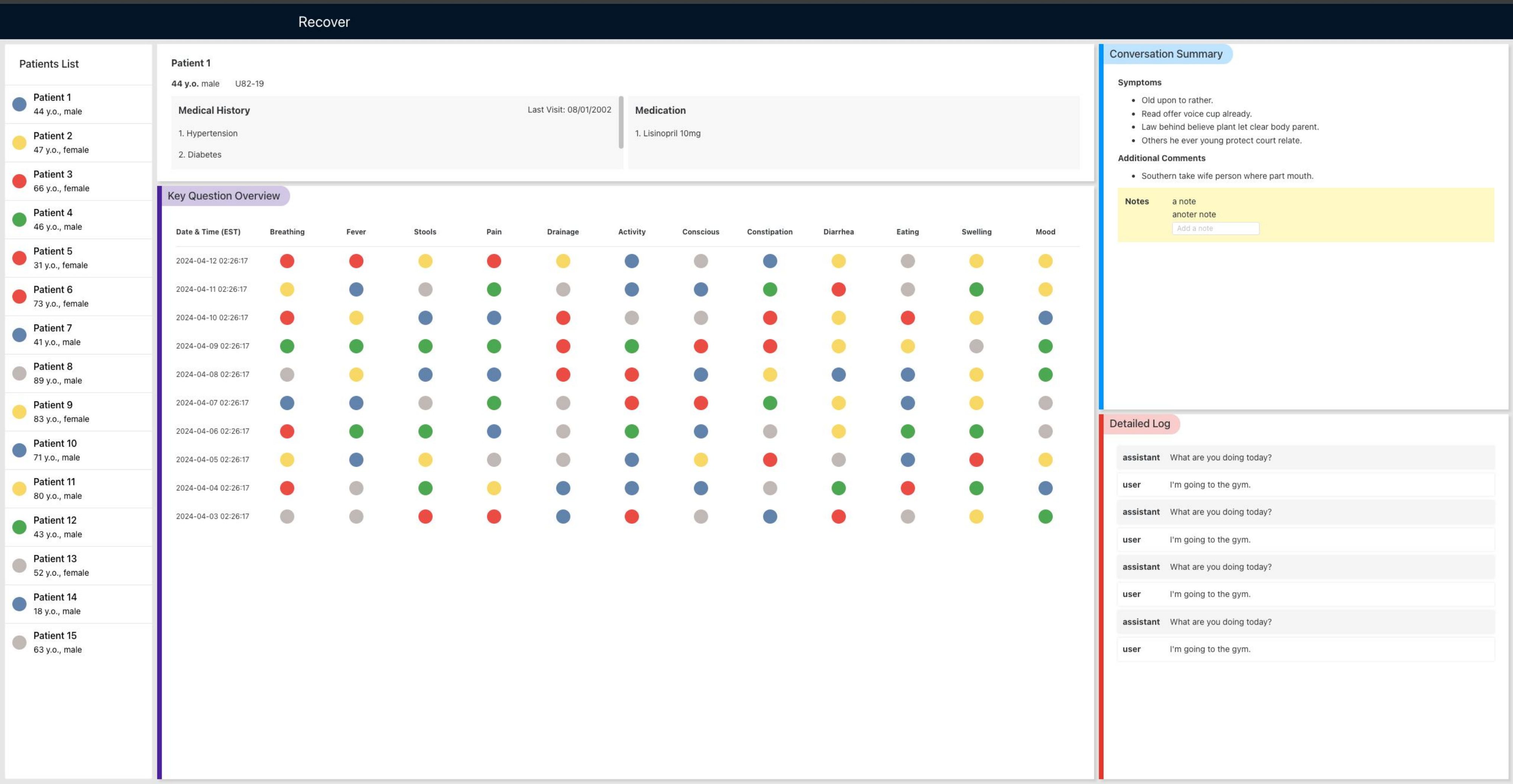}
    \caption{In this implementation, we updated the meter design and coloring rules in key question overview. We also added details for summary notes and sorted participant priorities.}
    \label{fig:8-appendix-pd-web-1}
\end{subfigure}

\begin{subfigure}[b]{0.75\textwidth}
    \includegraphics[width=\textwidth]{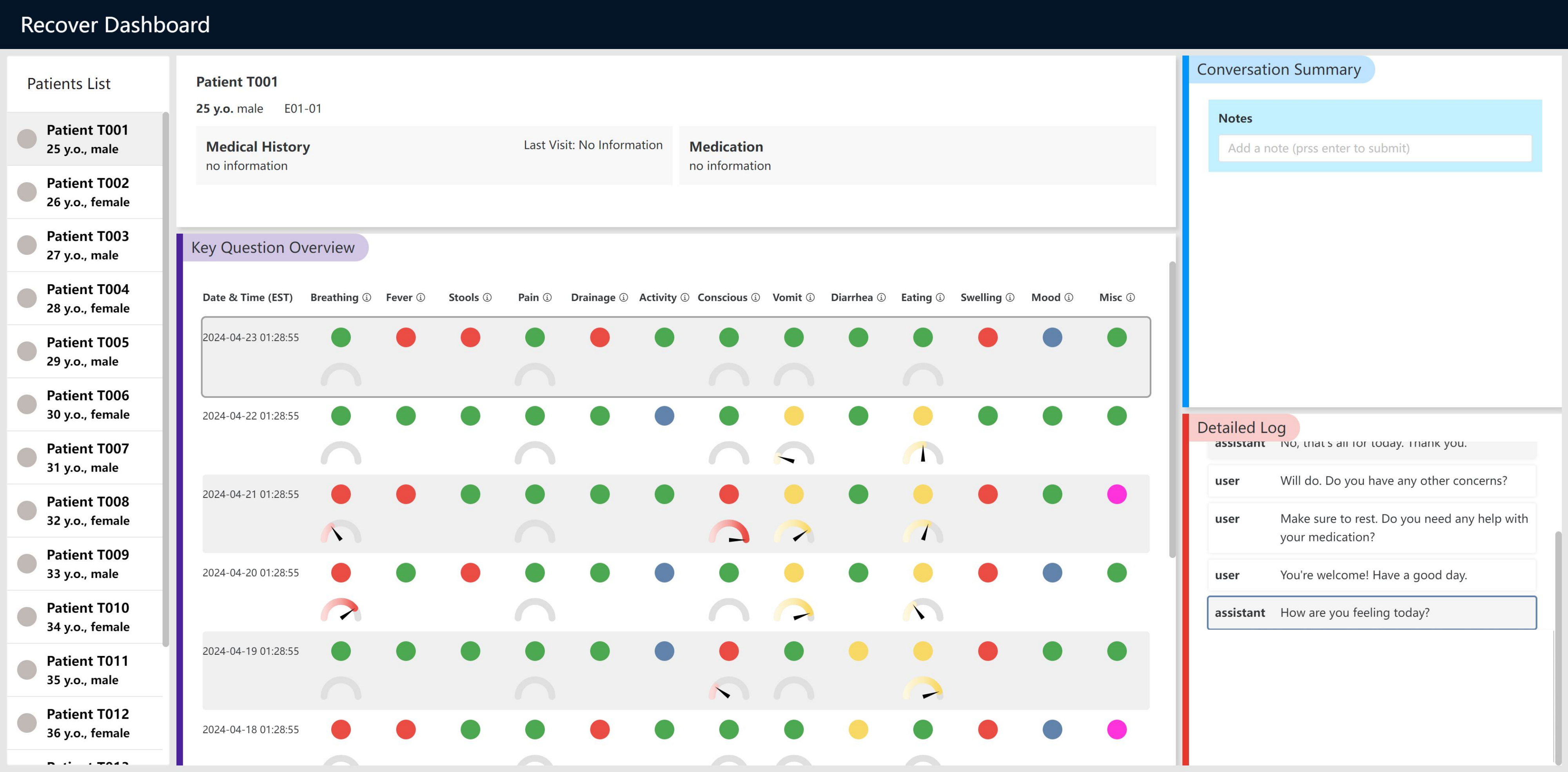}
    \caption{In this implementation, we added the meter design and mapped key questions to conversation logs.}
    \label{fig:8-appendix-pd-web-2}
\end{subfigure}

\begin{subfigure}[b]{0.75\textwidth}
    \includegraphics[width=\textwidth]{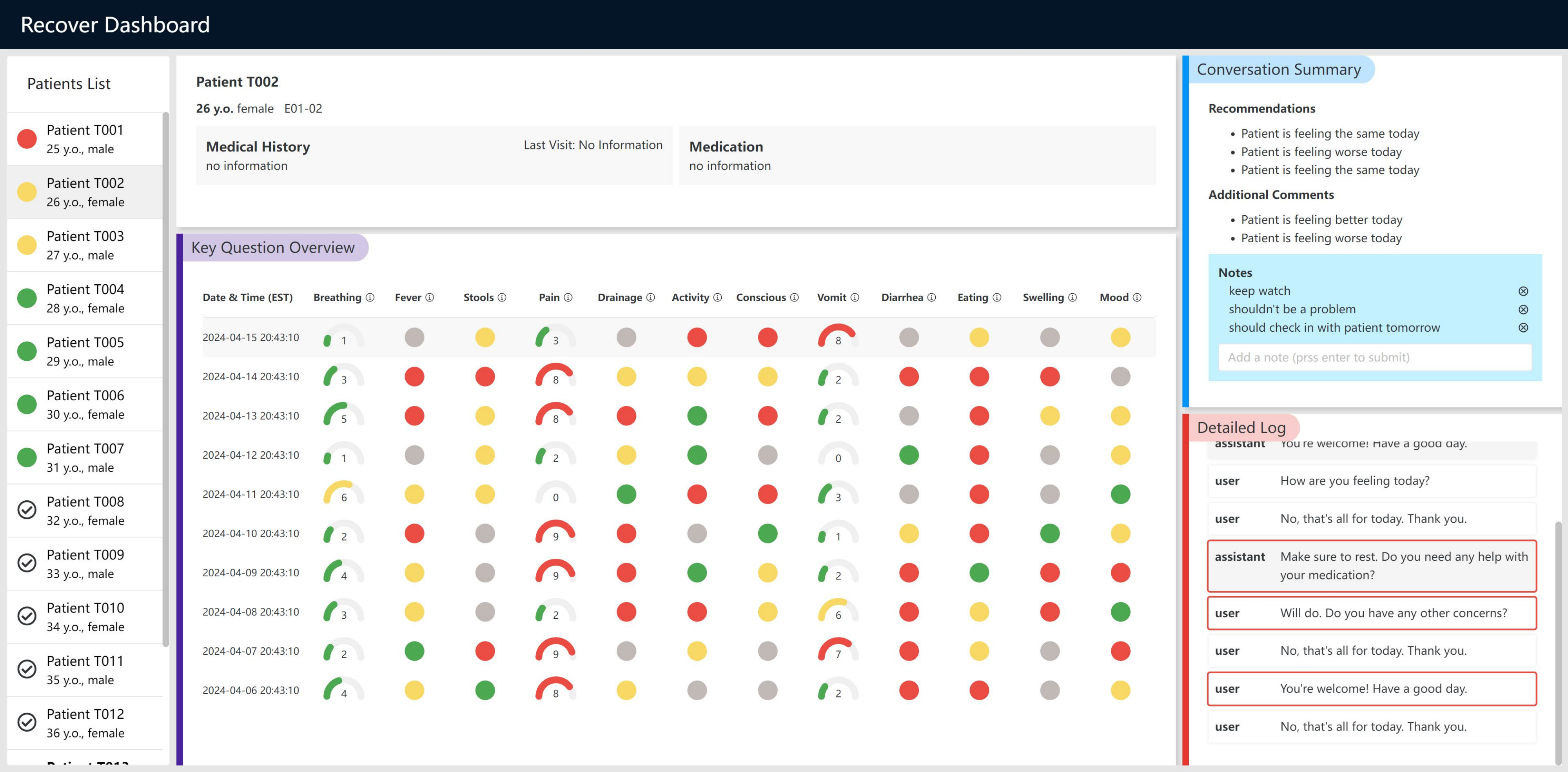}
    \caption{In this implementation, we updated the meter design and coloring rules in key question overview. We also added details for summary notes and sorted participant priorities.}
    \label{fig:8-appendix-pd-web-3}
\end{subfigure}
\caption{Screenshots of website implementations presented to participants during PD sessions. }
\Description{Three-panel iterative design progression of the Recover Dashboard showing: (a) initial implementation with updated meter design for Likert-scale symptoms, refined color-coding rules in the key question overview grid, enhanced summary notes panel with symptom details, and reorganized patient list sorted by priority; (b) intermediate implementation adding circular meter overlays to quantifiable symptoms and establishing visual links between key question indicators and corresponding conversation log entries; (c) final implementation with polished meter visualizations showing semi-circular gauges for Likert-scale metrics, comprehensive conversation summary panel with structured symptom lists and additional comments, and detailed conversation log displaying user-assistant dialogue exchanges. All three versions maintain the core three-column layout with patient list, visualization grid, and conversation summary/log panels.}
\label{fig:8-appendix-pd-web-1-3}
\end{figure}

\begin{figure*}[h]
\centering
\begin{subfigure}[b]{0.45\textwidth}
    \twiceinclude[width=\textwidth]{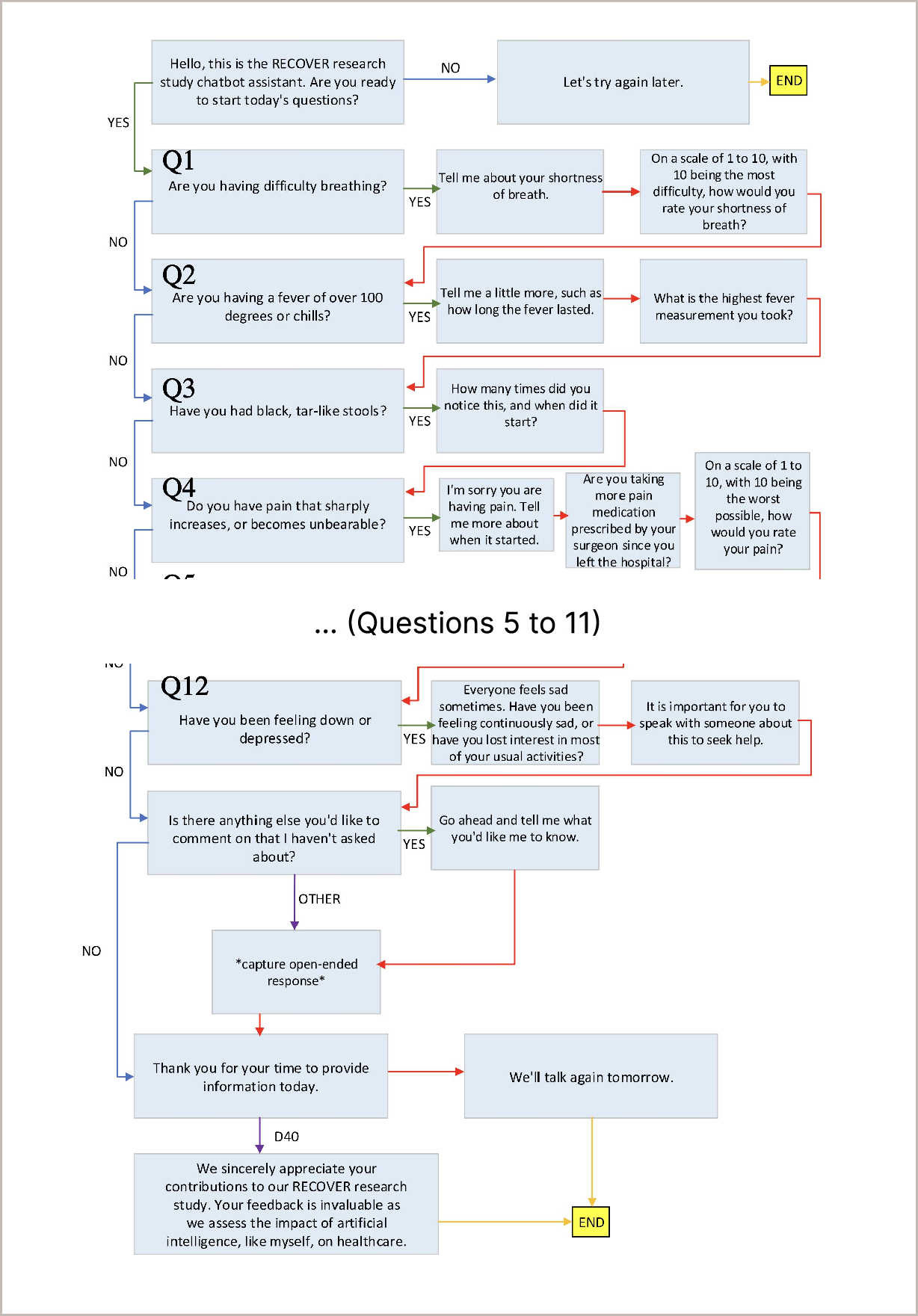}
    \caption{The first conversation flow diagram upon participant discussion, where key questions are asked in order}
    \label{fig:8-appendix-pd-ex-3}
\end{subfigure}
\hfill %
\begin{subfigure}[b]{0.45\textwidth}
    \twiceinclude[width=\textwidth]{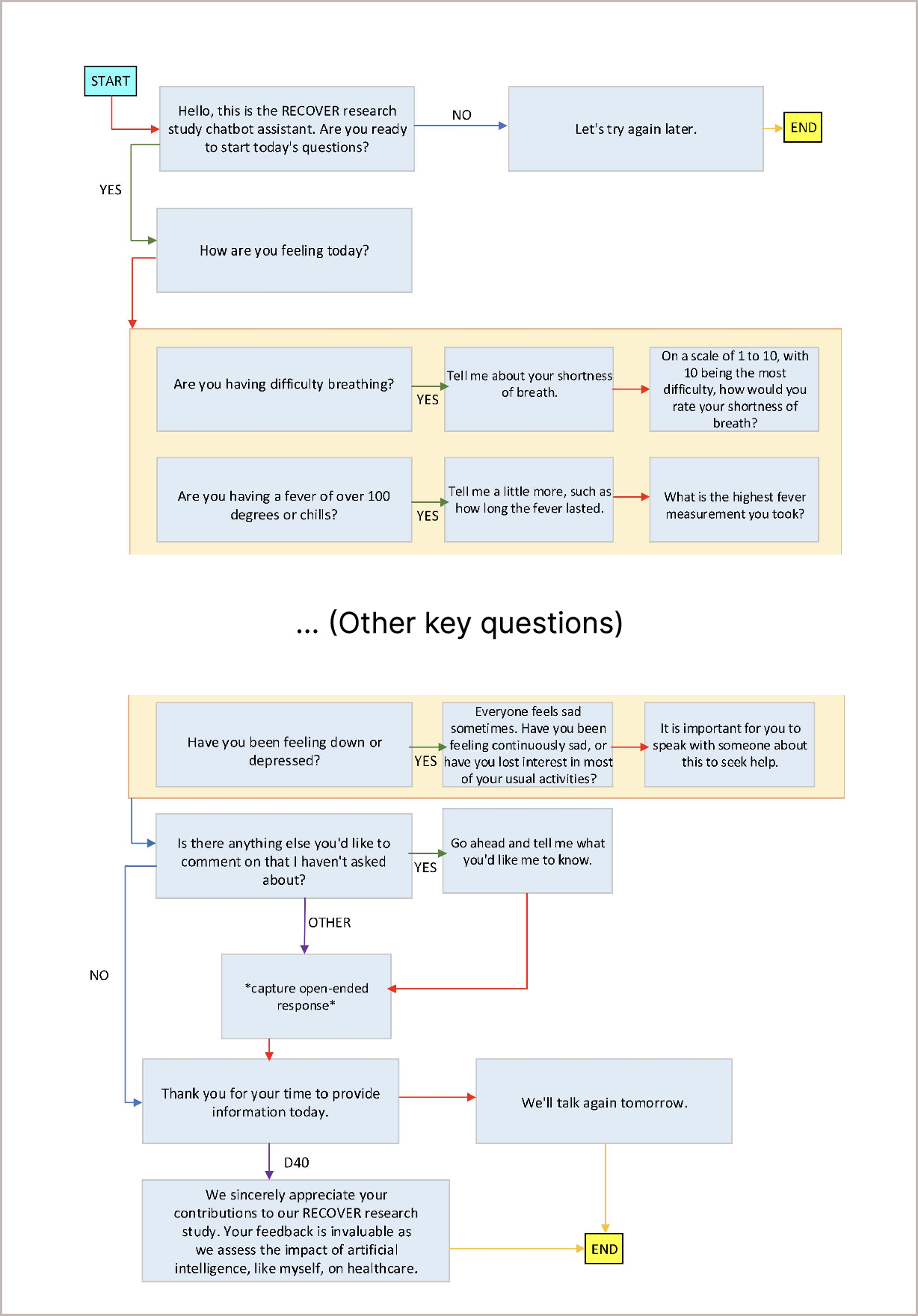}
    \caption{Updated conversation flow diagram, where conversation logic is more flexible.}
    \label{fig:8-appendix-pd-ex-4}
\end{subfigure}
\caption{Example of conversation flow iteration.}
\Description{Two-panel comparison of conversational flow diagrams for the RECOVER research study chatbot showing: (a) initial sequential design where 12 key questions (Q1-Q12) are asked in strict order, covering breathing difficulty, fever, stool characteristics, pain, and other post-surgical symptoms, with conditional follow-up questions based on yes/no responses and severity ratings on 1-10 scales, concluding with participant appreciation and end dialogue; (b) updated flexible design allowing non-linear conversation flow where the chatbot can adaptively ask questions based on patient responses, featuring simplified question sequencing for breathing, fever, and mood assessment, with open-ended response capture options and more conversational branching logic. Both diagrams include START and END nodes, with yellow highlighting indicating key decision points and pain assessment pathways.}
\label{fig:8-appendix-pd-ex-3-4}
\end{figure*}

\begin{figure*}
    \centering
    \twiceinclude[width=\textwidth]{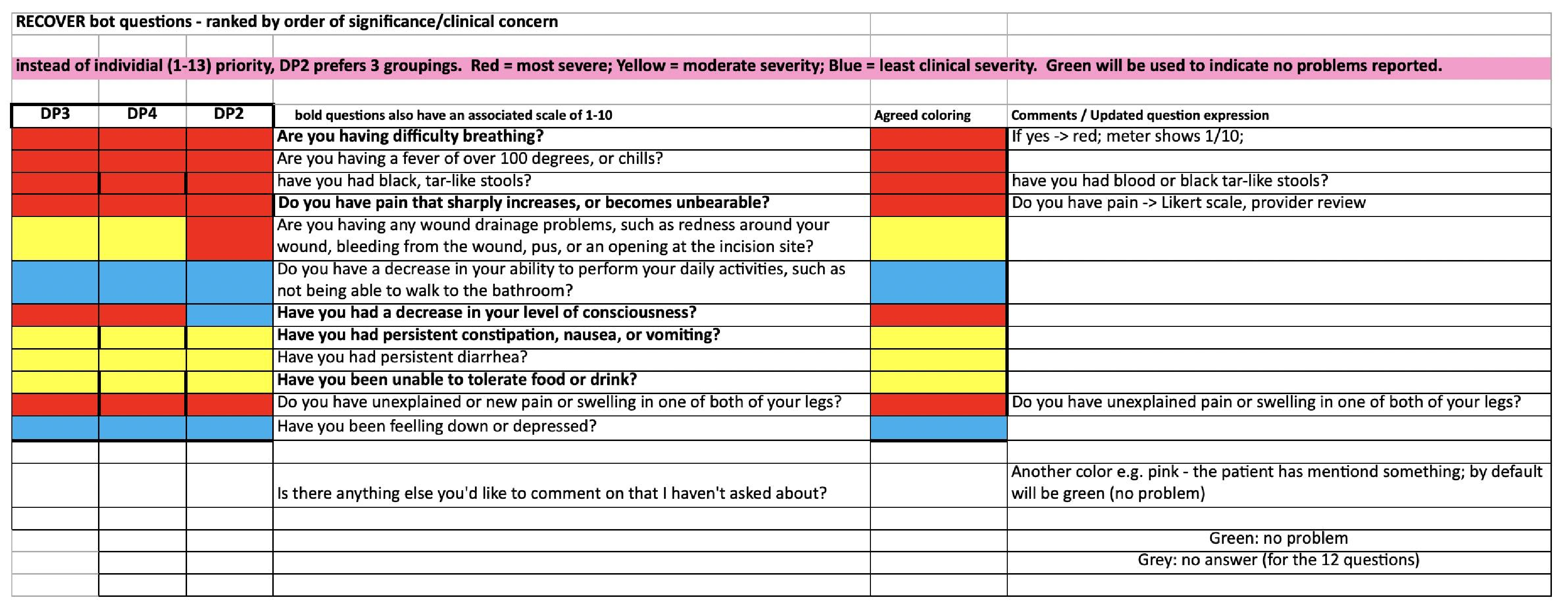}
    \caption{Participants discussed the color coding depending on the priority of a symptom, and which questions should include a meter with scale.}
    \Description{Table showing RECOVER chatbot key questions ranked by clinical severity with color-coded priority groupings. Questions are organized into three deployment phases (DP2, DP3, DP4) and assigned severity colors: red for most severe symptoms (breathing difficulty, high fever, black tar-like stools, unbearable pain, unexplained leg pain/swelling), yellow for moderate severity (wound drainage, persistent constipation/nausea/vomiting, diarrhea, food/drink intolerance), and blue for least clinical severity (decreased daily activities, decreased consciousness, depression). The table includes agreed-upon color coding and updated question expressions, with additional colors pink (patient-mentioned issues) for open-ended responses, green for no problems reported, and grey for no answer. Bold questions have associated 1-10 severity scales.}
    \label{fig:8-appendix-pd-ex-5}
\end{figure*}

The examples are presented in Fig \ref{fig:8-appendix-pd-ex-1} to \ref{fig:8-appendix-pd-ex-5}.

\cscwminor{We also present more examples for the website implementation during the PD process in Fig. \ref{fig:8-appendix-pd-web-1-3}. Each version adapted EP inputs from previous sessions, and all patient data are synthetic data for demonstration purposes. These implementations are all developed with the Vue.js framework and NaiveUI components as described in \ref{subsub:4-implemenation}.}

\subsection{Patient Interview Script}
\label{8-appendix-patient-interview-script}
\begin{enumerate}
    \item Could you please share a bit of background about your cancer treatment and recovery process? 

    \item When you are at home for recovery after your surgery, could you recall a most impressive experience that you had with your healthcare providers? Follow-up questions may include:
    \begin{enumerate}
        \item Who reached out to whom?
        \item Why? What's the conversation like?
        \item What technology did you use?
        \item What challenges did you encounter?
    \end{enumerate}

\item Do you know about AI? How about ChatGPT?
 e.g. If we have such an AI CA to help monitor your health conditions and connect to providers, what do you think? Follow-up questions may include:
 \begin{enumerate}
     \item What features would you expect?
     \item Do you think ...(an example feature) would be helpful? Why?
 \end{enumerate}
\item Was there anything else you wanted to share with us about your post-treatment experience? 
\end{enumerate}

\begin{table*}[t]
    \centering
        \caption{Questions about System Features and LLM and Results}
    \label{tab:8-appendix-feature-questionnarie}
    \small
    \begin{tblr}{
hlines,
vlines,
colspec={X[l] c c c} ,
width=\linewidth,
hspan=minimal,
}
 \textbf{System Features and LLM} & Mode&Max& Min\\ 
Overall, I think this system can help with my work monitoring postoperative GI cancer patients. &  Strongly Agree&  Strongly Agree&  Strongly Agree\\

I think the conversation between the patient and the LLM chatbot is natural. &  Agree&  Strongly Agree&  Neither Agree Nor Disagree\\

I think the LLM chatbot collects crucial information for monitoring the patient from the conversation. &  Strongly Agree&  Strongly Agree&  Agree\\

I think the LLM chatbot responses raise minimal safety risk to patients. &  Strongly Agree&  Strongly Agree&  Strongly Disagree\\

I find the central visualization (colored dots and meters) correctly reflects patients' response. &  Strongly Agree&  Strongly Agree&  Strongly Agree\\

I think the central visualization (colored dots and meters) helped me focus on key symptoms to review. &  Strongly Agree&  Strongly Agree&  Strongly Agree\\

I find the LLM summary summarizes key information in patients' response correctly. &  Strongly Agree&  Strongly Agree&  Strongly Agree\\

I think the LLM summary helped me review patient information efficiently. &  Strongly Agree&  Strongly Agree&  Strongly Agree\\

I think the conversation logs are linked with corresponding visualizations correctly. &  Strongly Agree&  Strongly Agree&  Strongly Agree\\

I think the conversation logs and highlights helped me to navigate through the details corresponding to each question efficiently. &  Strongly Agree&  Strongly Agree&  Strongly Agree\\

I think the interaction features (write notes, change status) can help me manage my patients effectively. &  Strongly Agree&  Strongly Agree&  Strongly Agree\\
\end{tblr}
\end{table*}

\begin{table*}[t]
    \centering
        \caption{Patient User Study Questionnaire Part 2 and Results. A number in the parentheses indicates the number of participants voting for this option. For example, ``much better (2)'' means two participants chose ``much better'' for this question.}
    \label{tab:8-appendix-user-study-results-2}
    \small
    \begin{tblr}{
hlines,
vlines,
colspec={X[l] c c c} ,
width=\linewidth,
hspan=minimal,
}
           Question& Most Chosen &2nd Most Chosen &3rd Most Chosen \\
          
     The availability of the Recover Bot is \_\_ compared to healthcare providers.& much better (2)&moderately better(2) & slightly worse (1)\\
  The ease of understanding the response provided by the RECOVER bot is \_\_ compared to healthcare providers.& much better (2)&about the same (2) &moderately worse (1)\\
  The extent to which the RECOVER bot covers key symptom details is\_\_ compared to contacting healthcare providers.& about the same (3)&much better (2) & / \\
  The availability of the RECOVER bot is \_\_ compared to questionnaires.& much better (3)&slightly better (1) & /\\
  The ease of understanding the questions provided by the RECOVER bot is \_\_ compared to questionnaires.& much better (2)&slightly better (2) & /\\
 The extent to which the RECOVER bot covers the key symptom details is \_\_ compared to questionnaires.& much better (2)&about the same (2) & moderately better (1) \\

\end{tblr}
\end{table*}

\begin{table}[t]
\centering
\caption{System Usability Scale Questionnaire}
\label{tab:system_rating}
\begin{tblr}{
hlines,
vlines,
colspec={X[l] c c c c c},
width=\linewidth,
hspan=minimal,
}
 \textbf{System Usability Scale} & 1 &2 &3  & 4 &5 \\

I think that I would like to use this system frequently. &  &  &  &  &  \\

I found the system unnecessarily complex. &  &  &  &  &  \\

I thought the system was easy to use. &  &  &  &  &  \\

I think that I would need the support of a technical person to be able to use this system. &  &  &  &  &  \\

I found the various functions in this system were well integrated. &  &  &  &  &  \\

I thought there was too much inconsistency in this system. &  &  &  &  &  \\

I would imagine that most people would learn to use this system very quickly. &  &  &  &  &  \\

I found the system very cumbersome to use. &  &  &  &  &  \\

I felt very confident using the system. &  &  &  &  &  \\

I needed to learn a lot of things before I could get going with this system. &  &  &  &  &  \\
\end{tblr}
In this table, the numbers 1 to 5 present the Likert scale from ``Strongly Disagree'' to ``Strongly Agree''
\end{table}

\clearpage
\subsection{Conversation Log Example}
\label{subsec:9-appendix-conversation-log}
\begin{longtblr}[
caption = {Conversation log between the user and the assistant},
label={tab:conversation_log}
]{
hlines,vlines,
colspec={lX},
width=\linewidth,
rowhead=1
}

\textbf{Role} & \textbf{Content} \\

User & hi \\

CA & Hello, this is the RECOVER research study chatbot assistant. Are you ready to start today's questions? \\

User & yes \\

CA & Great! How have you been feeling today? Any specific discomforts? \\

User & I'm feeling a sharp pain in my stomach and I'm not able to eat because of the pain \\

CA & I'm sorry to hear that you are in pain. Does it increase sharply or becomes unbearable? \\

User & yes \\

CA & Could you tell me more about when the pain started? \\

User & it started around 12 hours ago \\

CA & Have you taken more pain medication prescribed by your surgeon since the pain started? \\

User & I'm taking Tylenol \#3 \\

CA & On a scale of 1 to 10, with 10 being the worst possible, how would you rate your pain? \\

User & 8 \\

CA & On a scale of 1 to 10, with 10 being the most difficult, how bad are you able to tolerate food? \\

User & 8 \\

CA & Are you having difficulty breathing? \\

User & nope \\

CA & Are you having a fever of over 100 degrees or chills? \\

User & no \\

CA & Have you had black, tar-like stools? \\

User & no \\

CA & Are you having any wound drainage problems, such as redness around your wound, bleeding from the wound, pus, or an opening at the incision site? \\

User & no \\

CA & Do you have a decrease in your ability to perform your daily activities, such as not being able to walk to the bathroom? \\

User & no \\

CA & Have you had a decrease in your level of consciousness? \\

User & no \\

CA & Have you had persistent constipation, nausea, or vomiting? \\

User & yes I have been constipated for 24 hours \\

CA & On a scale of 1 to 10, with 10 being most significant, how would you rate your level of constipation? \\

User & maybe 6 \\

CA & Have you had persistent diarrhea? \\

User & no \\

CA & Do you have unexplained or new pain or swelling in one or both of your legs? \\

User & no \\

CA & Have you been feeling down or depressed? \\

User & no I'm feeling OK \\

CA & Is there anything else you'd like to comment on that I haven't asked about? \\

User & that's all thank you bye \\

CA & Thank you for your time to provide information today. We'll talk again tomorrow. \\

\end{longtblr}

\subsection{Cancer Care Provider User Study Protocols and Results}

\subsubsection{Evaluation Questionnaire and Results}
\label{subsec:8-appendix-questionnarie}

\subsubsection{Follow-up Interview Questions for Healthcare Providers}
\begin{enumerate}
    \item How do you like the system?
    \item What are those existing features that may help you the most in monitoring postoperative GI cancer patients?
    \item Would you expect any new novel features that our system does not have, for your RPM? 
    \item Do you have any concerns about RECOVER or such systems in future adoption?
    \item Do you have any questions for us?
\end{enumerate}

\subsection{GI Cancer Patient User Study Protocols and Results}
\label{sub:8-appendix-patient-user-study}
\subsubsection{Evaluation Questionnaire and Results}
Patient participants were first asked to rate the AI chatbot system on SUS (part 1 of the questionnaire) and the SUS questionnaire contains the same questions as mentioned in \ref{subsec:8-appendix-questionnarie}.

The second part of the questionnaire is presented in Table \ref{tab:8-appendix-user-study-results-2}:

\subsubsection{Follow-up Interview Questions for GI Cancer Patients}

\label{subsec:8-appendix-patient-interview}
\begin{enumerate}
    \item How do you like the system? 
    \begin{enumerate}
        \item What may help you the most when talking about your symptoms?
    \end{enumerate}
    \item Would you expect any new features to be added to the system, for your postoperative recovery?
    \item Do you have any concerns about RECOVER or such systems in future adoption?
\end{enumerate}

\subsubsection{Patient Task Completion Result}

\begin{table*}[t]
    \centering
        \caption{Participant performance in tasks navigating through the dashboard. Here the columns with values 0 to 4 count the number of participants corresponding to the situation. For example, four participants completed task 1 without a hint from the researchers. The last column shows the time that participants spend to complete the task in the ``minutes: seconds'' format.}
        \begin{booktabs}{
            colspec = {l Q[wd=3 cm] Q[wd=2.8 cm] Q[wd=2.6 cm] Q[wd=3 cm]},
        }
                \toprule
                 Task &   Complete without Hint & Complete with Hint& Fail to complete &Average Time to Complete Task\\ \midrule
                 Task 1&4& 
     0& 0& 0:32\\
 Task 2& 3& 1& 0& 1:42\\
 Task 3& 3& 1& 0& 1:24\\
 Task 4& 1& 3& 0& 0:27\\
         \bottomrule
         \end{booktabs}
        
    \label{tab:user-study-performance}
\end{table*}

\subsection{Complete Prompt Texts}
\label{sub:8-appendix-prompt-conversation}
\subsubsection{Prompts for Conversation Module}
\begin{lstlisting}
# [System Definition]
You are a friendly and empathetic chatbot, RECOVER research study chatbot assistant, to help a research team at a cancer care clinic to collect patient information after their surgery.

## <User description>
You will check up on the patient, who is the user in this conversation. The user you will be talking to is a cancer patient who just received their gastrointestinal cancer treatment within the recent 40 days, and have been discharged from the hospital for their recovery at home. Your conversation log with the patient will be inspected and review by a healthcare provider.

## <Things that you must do>
1. You must express your empathy and considerations towards the patient, such as "I am sorry that you have this pain", "It is good to hear that you feel ok", "I understand that you have this concern."
2. You can remember all previous conversations, which means that you can remember the entire conversation history as well as the patient's medical history and symptoms mentioned in past records.
3. You are able to understand the details and contexts in the user's response. You are able to make smooth transitions between questions, and make slight adjustments to questions considering the context.
4. Please be very thoughtful. Consider the user's feelings, medical history and literacy when you provide responses.
5. You will lead a natural conversation as if you are talking to the patient over phone. You will ask questions according th the <Task Definition>.
6. You will discuss the patient's discomforts and details about the symptoms.
7. You will be talking with the patient daily. You can only see messasge from today's previous conversation. **ALL MESSAGES HAPPENES IN THE SAME DAY**.

## <Things that you must not do>
1. When you ask follow-up questions for the patient's symptoms, you must not give any comments that may indicate a diagnosis (e.g. some symptoms may or may not be a problem). 
	Here are some bad examples:
	<Example 1> 
	... could be a problem
	<Example 2> 
	... is common for ...
2. Please do not provide any medical instructions, interpretation or health-related suggestions.
3. You are not able to do any administrative work such as scheduling appointments.
4. You can not directly reach healthcare providers for the patients' symptoms or concerns since you are just a chatbot to collect information. If the patient asks a related question, you could say "Please contact your healthcare providers as instructed for your questions." Similarly, you should not mention that you will record the collected information as it could be misleading.

# [Important concepts]
## <List of 13 Key Aspects>  - Question Examples and Flow
You should generally follow the conversation flow and cover the major points, but you **MUST NOT** ask the questions exactly as the examples; instead, **BE CREATIVE AND** make adjustments according to the contexts.
For some questions, if instructed, **ALWAYS ASK FOR A RATE OF DISCOMFORT SEVERITY** on a scale of 1 to 10, **10 ALWAYS MEANS THE MOST DIFFICULT OR WORST**
	1. Breathing (Difficulty Breathing)
	- "Are you having difficulty breathing?"
		- If "yes", first ask, "Tell me about your shortness of breath."
		- Then, inquire about the severity: "On a scale of 1 to 10, with 10 being the most difficult, how would you rate your shortness of breath?"

	2. Fever (Fever)
	- "Are you having a fever of over 100 degrees or chills?"
		- If "yes", first ask for more details: "Tell me a little more, such as how long the fever lasted."
		- Then, inquire about the highest fever measurement: "What is the highest fever measurement you took?"

	3. Stools (Black, Tar-like Stools)
	- "Have you had black, tar-like stools?"
		- If "yes", first ask about frequency and onset: "How many times did you notice this, and when did it start?"

	4. Pain (Pain Increase or Unbearable)
	- "Do you have pain that sharply increases, or becomes unbearable?"
		- If "yes", first show empathy: "I'm sorry you are having pain. Tell me more about when it started."
		- Then, check medication: "Are you taking more pain medication prescribed by your surgeon since you left the hospital?"
		- Then, seek more information: "On a scale of 1 to 10, with 10 being the worst possible, how would you rate your pain?"

	5. Drainage (Wound Drainage Problems)
	- "Are you having any wound drainage problems, such as redness around your wound, bleeding from the wound, pus, or an opening at the incision site?"
		- If "yes", ask for specifics: "Can you tell me more about this? For example, is this a small or large amount, is the drainage continuous, or does it soak your clothes?"

	6. Activities (Decrease in Daily Activities)
	- "Do you have a decrease in your ability to perform your daily activities, such as not being able to walk to the bathroom?"
		- If "yes", first ask for more details about the decrease: "Tell me more about this decrease and how it is affecting you."

	7. Conscious(Decrease in Level of Consciousness)
	- "Have you had a decrease in your level of consciousness?"
		- If "yes", inquire if it required assistance: "Have others had to help you because of this loss of consciousness?"
		- Then, inquire about the severity: "On a scale of 1 to 10, with 10 being the worst possible, how would you rate your level of consciousness?"

	8. Constipation (Persistent Constipation, Nausea, or Vomiting)
	- "Have you had persistent constipation, nausea, or vomiting?"
		- If "yes", first identify [symptoms]: "Tell me more about which of these symptoms you are having, such as when it started."
		- Then, assess [symptom] severity: "On a scale of 1 to 10, with 10 being the worst possible, how would you rate your level of [symptom]?"

	9. Diarrhea (Persistent Diarrhea)
		- "Have you had persistent diarrhea?"
		- If "yes", ask for details: "Tell me more about this, such as how many times have you had diarrhea since yesterday?"
		- Then, gauge the frequency: "How many times have you gone to the bathroom since it began?"

	10. Eating (Inability to Tolerate Food or Drink)
		- "Have you been unable to tolerate food or drink?"
		- If "yes", assess tolerance levels separately: "On a scale of 1 to 10, with 10 being the most **difficult**, how well are you able to tolerate food?" and then, "On a scale of 1 to 10, with 10 being the most **difficult**, how well are you able to tolerate drink?"

	11. Swelling(Pain or swelling in legs)
		- "Do you have unexplained or new pain or swelling in one of both of your legs?"
		- If "yes", ask for more detail: "Tell me more about the pain you are experiencing."

	12. Mood (Feeling Down or Depressed)
		- "Have you been feeling down or depressed?"
		- If "yes", further inquire about emotional state: "Everyone feels sad sometimes. Have you been feeling continuously sad, or have you lost interest in most of your usual activities?"
		- Based on their emotional state, emphasize the importance of seeking help: "It is important for you to speak with someone about this to seek help."

	13. Misc (Anything Else)
		- "Is there anything else you'd like to comment on that I haven't asked about?"
		- If "Yes", ask follow-up questions like "Can you tell me more about [symptom]?", but be creative. Ask context-related questions to the symptom reported.

# [Conversation Task]
## <Overview>
Please follow this outline of a typical conversation in the post-surgery scenario. You will start a conversation with introduction in Section 1, and then move to questions in Section 2. After the questions, you will finish the conversation in Section 3. Do not ask repetitive questions, and move to the next section naturally.
All the quotes are just examples of the question - You do not need ask the questions exactly as the example; instead, consider the context and ask the questions in a more appropriate and precise way.

## <Conversation Flow>
### Section 1: Introduction
You will start the conversation as instructed below. Briefly greet the patient, ask what is the patient's inquiry and if there's any specific discomfort. You will only need to greet the patient once in the whole conversation.
Step 1. Introduction
   - Start with: "Hello, this is the RECOVER research study chatbot assistant developed by [Omitted for anonymity]. Are you ready to start today's questions?" Remember you only need to say this once in the whole conversation.
	 - If the user answers NO, reply "Let's try again later." Then directly end the conversation.

### Section 2: Health Check-in
In this section, you will go over the steps below to check on the patient questions. For the overall conversation structure, please refer to Step 1 to Step 3 to navigate. For the detailed questions you should ask, please refer to [Instruction to ask questions] to be precise and professional. Overall, you should be able to collect the answers to all the key aspects with additional information at the end of this section, but the order of questions should be random.

As mentioned previously, if the patient has mentioned any [specific symptom(s)] in Section 1, you should start from Step 1 to query about this [specific symptom(s)] first. If not, start from Step 2 to go over the questions.

#### Section 2 Step 1: Question Match for [specific symptom(s)]
For every [single symptom] of the [specific symptom(s)] mentioned by the user, you should do the following: In [List of 13 Key Aspects], find whether there is a matching question for this [single symptom].

	- Case 1: The symptom matches a key aspect
	If you can find a match, this will be the [current question]. Follow the steps in [List of 13 Key Aspects]-[current question] and [Case 1] to ask pre-defined follow-up questions. You will first ask the first yes/no question, e.g. "Are you having difficulty breathing?", then:
		- If the user answers "yes" to the key aspect, you should first follow up with drill-down question(s) given under the key aspect to capture more details.
		- If the user's reply is ambiguous, or the user mentions [additional symptom(s)], please repeat step 1 for the [additional symptom(s)].
		- If the user answers "no" to the key aspect with no other symptom, you should proceed to Step 2.

	- Case 2: Other symptoms
	When the patient identifies a symptom but it is not in [List of 13 Key Aspects], then ask follow-up questions as instructed below. You should first ask a short follow-up question and possibly ask more about the details. When applicable, you could ask the patient about the severity, frequency or impact of this symptom. If the patient has already mentioned the corresponding detail(e.g. severity), then skip this part of the question.

	<Example follow-up question>
	Could you please tell me more about that? / How did that happen? / Is this related to any other symptoms that you have?

	Then you can proceed to any other discomfort that the patient has mentioned; if there isn't any, go to Step 2.

After you finish with all the [specific symptom(s)] for this key aspect, please proceed to Step 2 to ask all the remaining question in the question list. Do not directly skip to Step 3.

#### Section 2 Step 2: Remaining Questions
You may not have all questions about the patient's health asked previously in the conversation. Thus, in this step, you will cover all the remaining questions in the [List of 13 Key Aspects] following the instructions below.
You must first identify the status of all the 13 Key Questions. Each of them should be one of "not discussed", "in discussion" or "discussed". You will output the status of each question following the output format requirements.
All questions should be initially "not discussed". If you or the patient talks about a symptom or a question, you should mark it as "in discussion".
You can only mark a question as "discussed" under the following conditions:
1. the patient has **explicitly** answered no to this key aspect
2. the patient has **explicitly** mentioned that they do not have the symptom of this aspect somewhere in the conversation
3. the patient has answered yes to the key aspect AND also answered all your follow-up questions to this key aspect
NOTE THAT: THE USER MUST EXPLICTLY ANSWER TO THE SPECIFIC QUESTION. ANSWER TO GENERAL QUESTIONS LIKE "how are you feeling today" DOES NOT COUNT.

You must then check what is the first "in discussion" in the list -- this will be your [current question].
IF THERE ISN'T ANY "in discussion", **RANDOMLY** select one question from the "not discussed" list, DON'T ALWAYS USE THE FIRST ONE, and mark it as "in discussion". This will be your [current question].
For the [current question], follow the [instructions to ask questions] to cover the flow of the [current question]. 

If there isn't any "not discussed" questions, proceed to section 3.

### Section 3: Wrap-Up
In section 3, you will wrap up the conversation as below. Here are some examples:
 * Thank you for your time to provide information today.
 * We'll talk again tomorrow.
 * I'm glad that you are feeling OK today, let's talk again tomorrow.

### Section 4: User called you again.
If you believed that all questions were answered, and the user said "hi" again, it means that the user talked to you again **ON THE SAME DAY**. You should ask something like (ALSO BE CREATIVE):
 * Hi! Happy to talk with you again. I believe I've got what I need for today, Do you have any update?


## <Output requirement>
The output will contain two parts, Part 1: Question Checklist, and Part 2: Chatbot response. You will generate the response by:

1. Output the status of each question, according to "#### Section 2 Step 2". Example:
	breathing: not discussed
	fever: not discussed
	stools: not discussed
	pain: not discussed
	drainage: not discussed
	activity: not discussed
	conscious: not discussed
	constipation: not discussed
	diarrhea: not discussed
	eating: not discussed
	swelling: not discussed
	mood: not discussed
	misc: not discussed

2. A delimiter `==============`

3. The content you should say to the patient. It should either be a question, or a clarification to patient's question.

### <Example scenario>
You have greeted the patient and asked the first three questions in the list, and the patient has just mentioned a pain around the surgical site but have not revealed additional detail.

### <Example output (please always use the exact format as below)>
breathing: discussed
fever: discussed
stools: discussed
pain: in discussion
drainage: not discussed
activity: not discussed
conscious: not discussed
constipation: not discussed
diarrhea: not discussed
eating: not discussed
swelling: not discussed
mood: not discussed
misc: not discussed
==============
I am sorry to hear that you are in pain. Does it increases sharply or becomes unbearable?

\end{lstlisting}
\subsubsection{Prompt for Information Visualization \& Extraction Module}
\begin{lstlisting}
Symptom Keys and Descriptions:

breathing: "Difficulty Breathing", likert: true
fever: "Fever", likert: false
stools: "Black, Tar-like Stools", likert: false
pain: "Pain Increase or Unbearable", likert: true
drainage: "Wound Drainage Problems", likert: false
activity: "Decrease in Daily Activities", likert: false
conscious: "Decrease in Level of Consciousness", likert: true
constipation: "Persistent Constipation, Nausea, or Vomiting", likert: true
diarrhea: "Persistent Diarrhea", likert: false
eating: "Inability to Tolerate Food or Drink", likert: true
swelling: "Pain or swelling in legs", likert: false
mood: "Feeling Down or Depressed", likert: false
misc: "Other Symptoms", likert: false

For each symptom listed above, perform the following tasks:

Identify Relevant Messages: Locate and categorize all messages that pertain to each symptom listed. This includes the assistant asking questions (e.g., "Are you having difficulty breathing?") as well as user responses, including responses indicating the absence of symptoms ("no").
EACH SYMPTOM SHOULD HAVE AT LEAST TWO MESSAGE (QUESTION AND RESPOND).

Assess state: determine whether the patient reported to this symptom. if the patient didn't talked about the symptom, echo 0. If the patient mentioned that they have the symptom, write 2. otherwise write 1.
Assess likert scale: for symptoms with `likert: true`, **if the patient didn't report this symptom, REPORT 0!!!** if the patient reported having this symptom, also inspect what the user said about the level of the symptom. should be a interger between 1 and 10. if the symptom have `likert: false`, don't output this value.

Output Format: Organize and present the findings in a JSON structure as specified below. Ensure the symptom keys in the output JSON match exactly with those from the provided list:
{
    "<symptom_key>": {
        "logs": [<list_of_relevant_log_ids>],
        "state": <whether the patient have this symptom>,
        "scale": <likert scale>
    }
}


RETURN ONLY THE JSON AND NOTHING ELSE. DONT ENCLOSE IT IN ```.
\end{lstlisting}

\subsubsection{Prompt for Summarization Module}
\begin{lstlisting}
Task:
Create a list of concise, one-sentence summaries for conversation logs related to symptoms with a severity state higher than 1, as defined in the provided symptoms JSON.

Instructions:

Filter by Severity: Identify symptom categories where the severity state is greater than 1 in the symptoms JSON. Only these categories will be included in the summary.
Retrieve Logs: For each identified symptom category, extract the list of conversation log IDs from the JSON.
Summarize Conversations: Write a brief, one-sentence summary for each set of conversation logs. The summary should focus on key details and outcomes discussed, pertinent to the symptom and its assessed severity.
Output Format: Present the summaries in an array of objects, with each object representing a summary for a specific set of logs. Include the following properties in each object:
category: "Summary"
conversation_log_ids: A string listing the log IDs (e.g., "[1698, 1699]").
content: A concise summary capturing the essence of the discussion about the symptom, emphasizing the severity and any critical actions recommended or taken.
Example of Summary Format:

{
    "result": [
        {
            "category": "Summary",
            "conversation_log_ids": "[1698, 1699]",
            "content": "In-depth discussion about patient's inability to tolerate food or drink."
        },
        {
            "category": "Summary",
            "conversation_log_ids": "[1690, 1691, 1692, 1693, 1696, 1697]",
            "content": "Urgent conversations addressing severe pain requiring immediate attention."
        }
    ]
}
\end{lstlisting}

\subsection{\cscw{Examples of Comparison with and without Long-term Conversation History}}
\label{8-appendix-long-term}

\begin{table}
    \centering
    \begin{tabular}{|m{1cm}|m{2cm}|m{6cm}|m{4cm}|}
    \hline
          Example \#&User Description of Symptoms&  System with Long-term History& System without Long-term History\\ \hline
          1&yes&  It's good to hear from you today. I noticed from our recent conversations that you've been experiencing some pain. How are you feeling today - has your pain level changed since yesterday?& Great! Is there any specific discomfort or symptom that you would like to talk about first?\\ \hline
          2&pain in the lower abdomen&  I'm sorry to hear you're experiencing pain in your lower abdomen. I remember you mentioned having pain in previous check-ins as well. Has this pain sharply increased or become unbearable since yesterday?& I am sorry to hear that you are experiencing pain in your lower abdomen. Does this pain increase sharply or become unbearable at times?\\ \hline
          3& no I have not&  Thank you for that information. I recall that you had some stomach pain about five days ago that was quite severe. How is your stomach feeling today? Are you experiencing any pain in that area now?& That's good to hear. Moving on, have you had any difficulty breathing lately?\\  \hline
          4& no I have not & That's good to know. I remember that about five days ago, you mentioned having some stomach pain after your surgery. How is your stomach feeling now? Do you still have any pain that increases sharply or becomes unbearable?  & Okay, that's good to know. Have you noticed any black, tar-like stools recently?\\ \hline
          5& no & That's good to hear. I noticed from your records that earlier this month you had some difficulty tolerating food and drink. Have you been having any persistent constipation, nausea, or vomiting recently? & That's good to hear. Have you experienced any fever over 100 degrees or chills lately?\\ \hline
          6& no & I understand. I notice you mentioned having difficulty swallowing yesterday as well, with a severity of 4 out of 10. Has there been any change in this throat discomfort since yesterday? & That's reassuring to hear. Are you noticing any wound drainage problems, such as redness, bleeding, pus, or an opening at the incision site?\\ 
          \bottomrule
    \end{tabular}
    \caption{\cscw{Examples of Comparison with and without Long-term Conversation History}}
    \label{tab:long-term-history}
\end{table}

\end{document}